\documentclass{aa}
\usepackage{graphicx}
\begin{document}
   \title{Sources of carbon and the evolution of the abundance
 	of CNO elements} 
 
     \author{Y. C. Liang, G. Zhao 
          \and
          J. R. Shi
          }

   \offprints{G. Zhao, email: zg@orion.bao.ac.cn}

   \institute{Beijing Astronomical Observatory, 
                Chinese Academy of Sciences, Beijing 100012 China \\
            National Astronomical Observatories,  
                Chinese Academy of Sciences, Beijing 100012 China
             }

   \date{Received / Accepted }

 \abstract{
Using the standard infall model of Galactic chemical evolution, 
we explore the origin of carbon
and calculate the abundance evolution of CNO elements for  
8 different models of 
stellar nucleosynthesis yields. 
The results show that, in the early stage
of the Galaxy, massive stars are the main producer of carbon,
and that as our Galaxy evolves to the late stage,
the longer lived intermediate- and low-mass stars play an 
increasingly important role, while 
at the same time, 
metal-rich Wolf-Rayet stars eject a significant amount of
carbon into the ISM by radiative-driven stellar winds. 
However, from the present published nucleosynthesis yields we cannot distinguish whether 
the main source of carbon in the late Galactic stage is just 
the massive stars ($M>8M_{\odot}$) alone, or 
just the intermediate-, low-mass stars and 
$M\leq 40M_{\odot}$ massive stars that do not go through the Wolf-Rayet stage. 
The $^{12}$C$(\alpha,\gamma)^{16}$O reaction rate is very important in the stellar nucleosynthesis calculations: a lower rate will give a 
higher yield of carbon. 
The contribution to nitrogen is dominated by
intermediate- and low-mass stars,
and the secondary source of massive stars cannot explain the observed
[N/Fe] in metal-poor stars.
Most of oxygen is produced by massive stars. 
The fact that a higher O abundance in metal-poor stars is derived
from the \ion{O}{i} 7771$-$7775 \AA$~$ triplet than from
the forbidden [\ion{O}{i}] line at 6300 \AA$~$ poses a problem.
   }

  \titlerunning{Sources of carbon}
  \authorrunning{Y. C. Liang et al.}
   \maketitle

%

\section{Introduction}

The abundance trends of the CNO elements are crucial to study 
the Galactic chemical evolution (hereafter GCE).  
These elements play important
roles in the stellar interior as sources of opacity and energy production through the 
CNO cycle, and thus affect the star's lifetime, its position in the HR diagram, and its heavy-element yields.  Carbon is especially
important because it is the first
element that is synthesized out of the primordial elements (H,
He, Li). 

The main site of carbon synthesis has been a subject of argument for many years.  
Burbidge et al. (\cite{B2FH}) suggested that carbon is provided by mass loss from
red giants and supergiants. Iben \& Truran (\cite{Iben78}) concluded from thermally
pulsing models that intermediate-mass and high-mass stars contribute
carbon in roughly equal amounts.  
More recent theoretical and observational studies include the following.

Timmes et al. (\cite{timmes})
used
the carbon yields of Type {\sc II} supernova 
(hereafter SN\,II) explosion from Woosley \& Weaver (\cite{WW95}) 
(hereafter WW)
and the yields of 
intermediate-, low-mass stars (hereafter ILMS) 
from Renzini \& Voli (\cite{RV81}) (hereafter RV)
to calculate the abundance evolution of carbon.
They concluded that the ILMS
contribute a significant amount of carbon to the interstellar medium (hereafter ISM).
Chiappini et al. (\cite{CMG97}) obtained similar results with their two-infall model.
Oberhummer et al. (\cite{OCS20}) reviewed the ILMS
as the major producer of carbon.
Kobulnicky \& Skillman (\cite{KS98}) measured 
the C/O and N/O ratios of three metal-poor
galaxies with similar metallicities using HST ultraviolet 
and ground-based optical spectroscopy.
Their results implied that the mechanisms of C and N production are coupled,
and most of both C and N in galaxies originates in  
low- to intermediate-mass stars. 

Prantzos et al. (\cite{PVC94})
suggested that if the duration of the halo phase is as long as
$\sim 1-2$ Gyr, then massive stars with wind-driven mass loss could be the main
carbon source during the whole galactic evolution.
Gustafsson et
al. (1999) adopted an analytical GCE model 
to calculate and compare the carbon abundance evolution
in the Galactic disk ($-1.6<$ [Fe/H] $<+0.2$) 
and some irregular galaxies. They suggested that
carbon enrichment is caused mainly by superwinds from metal-rich,  massive stars.   
Karlsson et al. (\cite{Karlsson99}) reported similar results.
However, Gustafsson \& Ryde (\cite{Gus20}) wrote that the 
source of carbon is 
still not clear.
Henry et al. (\cite{henry20}) analyzed the behavior of 
the N/O and C/O abundance ratios 
as functions of metallicity 
in the Galactic and extragalactic
\ion{H}{ii} regions. Their results showed that C and N originate from
separate production sites and are decoupled from one another; and 
they suggested that massive stars
($M>8M_{\odot}$) dominate the production of carbon, while intermediate-mass stars
between 4 and 8$M_{\odot}$ dominate the nitrogen production.
Carigi (\cite{cari20}) adopted different sets of yields to calculate the C/O vs. Galactic
age relation, and suggested massive stars to be the main contributor  of carbon.

From calculated abundance gradients, Hou et al. (\cite{hou20}) argued that carbon comes from stellar winds of massive 
stars and ILMS. 
Goswami \& Prantzos (2000) using their new GCE model, concluded that the yields in WW
are not enough to explain the observed carbon abundance, 
and that some other sources, Wolf-Rayet (hereafter W-R) stars or 
ILMS, are needed
to explain the observed [C/Fe] vs. [Fe/H] curve.  
Garnett et al. (\cite{garnett99}) measured the gas-phase abundance ratio C/O
in six \ion{H}{ii} regions in the spiral galaxies M\,101 and NGC\,2403, based on 
ultraviolet spectroscopy from the HST.
These authors suggested that stellar winds in massive stars
have an important effect on carbon and oxygen, and 
their analysis showed that 
the source of carbon depended on the assumed UV reddening function. When a flatter  
UV reddening ($R_V=5$) was used, the data were consistent with the model 
that used the yields of massive stars from WW, 
while with a steeper value ($R_V=3$) the data
became more consistent with the model that used the yields
of Maeder (\cite{maeder92}) (hereafter M92). 

In summary, it is still not clear whether carbon comes from 
the SN\,II explosions of massive stars, from stellar winds of 
W-R stars or from ILMS. 
In this paper we try to address this question from 
two aspects: (1) by using GCE model calculations with different sets of nucleosynthesis yields;
(2) by using the observations of the large sample of dwarfs in our Galaxy as one of the main bservational constraints.

For the first aspect, 
GCE modelling is an excellent tool to 
study the evolution of elements of the whole Galaxy and other galaxies. 
The main parameters for setting up such a model  
include the form of the infall, the star formation rate (SFR), the initial mass function (IMF) and the nucleosynthesis yields.
The exponential form of infall 
(or Gaussian infall form: Chang et al. \cite{chang} and Prantzos \& Silk \cite{PS98}) has been found to be able to 
solve the G-dwarfs problem very well.
The SFR and IMF can 
be determined from the observed present-day mass function (PDMF) 
(Pagel \cite{pagel97}). 
The most uncertain factor of GCE model
is the nucleosynthesis yields, which largely decide the evolutionary trends of the elements.
Different research groups have calculated the yields of 
stellar nucleosynthesis,  
but the results are rather different due to the use of different 
input parameters. And different yields will result in 
different evolution trends. 

We set up a simple GCE model, like the one of Timmes et al. (\cite{timmes}), which 
fits some important observations. 
We choose the following different nucleosynthesis yields published recently 
in our calculations:
for the RV yields of intermediate- and low-mass single stars, 
van den Hoek \& Groenewegen (\cite{VG97}) (hereafter VG),
Marigo et al. (\cite{marigo96}) (hereafter MBC1), 
Marigo et al. (\cite{marigo98}) (hereafter MBC2) (hereafter, MBC represents
for combined MBC1 and MBC2) and Marigo
(\cite{M2K}) (hereafter M2K);
for the WW yields of massive single stars, Nomoto et al. (\cite{nomoto1}) (hereafter N97), M92 and
Portinari et al. (\cite{PCB98}) (hereafter PCB);
for the yields of binaries through SN\,Ia explosion, 
Nomoto et al. (\cite{nomoto2}).
Then we compared and analyzed 
the different evolutionary behavior of the CNO elements, particularly C, calculated on the basis of these 
different sets of yields.

In the early stage
of our Galaxy, massive stars are the main producers of carbon.
As our Galaxy evolves to a 
late stage, ILMS play an increasingly important role
in the enrichment of the ISM due to their longer lifetimes.  
At the same time, 
the high metallicity W-R stars eject significant amounts of carbon into the ISM 
by radiative-driven stellar wind (the metallicity influences the 
opacities in the outer layers of massive stars, hence their mass loss rate through winds).
Which is the main producer of carbon in the late stage of our 
Galaxy?
We attempt to answer in the following way.
We divide the massive stars
into two mass ranges, $M>40M_{\odot}$ and $M\leq 40M_{\odot}$. 
And assume that the former does, and the latter does not, go  
through the W-R stage.  
We then compare the contributions by the W-R stars and ILMS 
to carbon in the late Galactic stage.

In the second part of our work,
we pay particular attention to the observations of field dwarfs
in our Galaxy. Their atmospheres are assumed to 
represent perfect records of the chemical compositions of the ISM at the time of their formation, 
and we have available 
the abundances for a large sample of 
field dwarfs in the solar neighborhood.  
In addition, 
the metallicity distribution of G dwarfs is
very important, because the G dwarfs have lifetimes as long as the Galaxy, their metallicity distribution  
reflects the evolution of our Galaxy.
The age-metallicity relation is significant, too, 
for reflecting the Galactic evolution. 

This paper is organized as follows.
In Sect. 2, 
we analyze the observational data.
The GCE model and the relevant 
parameters are presented in Sect. 3.
Different nucleosynthesis yields are analyzed in Sect. 4.
Results of the analysis are given in Sect. 5$-$7. 
The main conclusions are summarized in Sect. 8.

\section{Observational data}

Observations of dwarf stars in the general field are very important 
to the evolution of  chemical abundance
as their atmospheres are thought to represent perfect
records of the chemical compositions of the ISM at the time they
formed.

\subsection {[C/Fe] vs. [Fe/H]}

Carbon abundance can be calculated from
the CH lines in the blue and ultraviolet (UV), the infrared 
CO emission, and a few
red and near-infrared \ion{C}{i} lines and [\ion{C}{i}] forbidden lines. 
The [\ion{C}{i}] line of low excitation potential being not so    
tightly dependent on NLTE considerations,
the values derived from it are less
dependent on the adopted stellar effective temperature than those from the
high-excitation \ion{C}{i} lines and CH lines. 
However, the small equivalent width (EW) of the [\ion{C}{i}] line 
makes the measurement difficult. 
And the [\ion{C}{i}] line at $\lambda$\,8727 \AA $~$is supposed to be 
blended with a weak \ion{Fe}{i} line.
Both the [\ion{C}{i}] and \ion{C}{i} lines become unmeasurably weak in metal-poor 
stars ([Fe/H] $<-1$). 

The CH line in the UV can be used to obtain the C abundance of halo and disk stars. 
Laird (\cite{laird85}) published an extensive study of the C abundances  of 116 field dwarfs,
10 faint field giants and 3 Hyades dwarfs 
in the range $-2.45<$ [Fe/H] $<+0.5$. They suggested that the [C/Fe]
were solar
in the whole [Fe/H] range of the samples. Tomkin et al. (\cite{TSL86}) 
analyzed the spectra
of 32 halo dwarfs, and derived [C/Fe] $\approx -0.2$, 
independent of metallicity for stars
with [Fe/H] $>-2$. And their data showed a rise of C abundance in more metal-poor stars:
[C/Fe] $\approx +0.2$ at [Fe/H] $\approx -2.5$. 
Carbon et al. (\cite{carbon87}) derived 
the C abundances of 83
halo dwarfs, which showed that [C/Fe] is constant over the range
of $-2.5<$ [Fe/H] $\leq -0.75$, and they found
an upturn at the very low metallicity value.
Wheeler et al. (\cite{wheeler}) reviewed these observed trends of 
[C/Fe] vs. [Fe/H].

Tomkin et al. (\cite{TLLS92}) derived the C abundances of halo and disk stars 
using the \ion{C}{i} $\lambda$\,9100 \AA $~$line.
Their results showed that the C abundances are +0.2 higher than
those derived from the CH lines.
Andersson \& Edvardsson (\cite{AE94}) (using [\ion{C}{i}] 8727 \AA), 
Tomkin et al. (\cite{TWLL95}) (using \ion{C}{i} 7100 \AA) and 
 Gustafsson et al. (\cite{Gus99}) (using [\ion{C}{i}] 8727 \AA) obtained the C abundances for large samples  
of disk dwarfs
(comprising 85, 105 and 80 stars respectively, with some in common).
Their results showed that 
[C/Fe] decreased with 
increasing [Fe/H] in the Galactic disk region.  
Using CH and C$_2$ lines, Carretta et al. (\cite{carretta20}) analyzed a sample of 19 stars, 
and reanalyzed the stars of Tomkin et al. (\cite{TLLS92}) and 
Edvardsson et al. (\cite{E93}). Their results showed that [C/Fe]
is roughly solar in the whole metallicity range of [Fe/H] $>-2.5$.

Considering the large scatter, we choose to take all the above-mentioned 
observations of C abundance
as observational constraints on our GCE calculations.

\subsection {[O/Fe] vs. [Fe/H]}

Oxygen is the third most abundant element in the universe, and it
is very important for the study of GCE.
In general, oxygen is produced by massive stars through 
SN\,II explosions, like other $\alpha$ elements. 
Thus, the observations of oxygen, especially in metal-poor region,
give a measure of the contribution by massive stars. 
Some parameters required in 
GCE calculations can be 
determined by the observed [O/Fe] in the stars
(Prantzos et al. \cite{PVC94}).

The oxygen abundance of stars can be derived from the 
\ion{O}{i} $\lambda$$\lambda$\,7771-7775 \AA $~$triplet, 
the [\ion{O}{i}] $\lambda$\,6300 \AA $~$forbidden line and OH lines
in the UV region.  
In main-sequence, metal-poor stars, the [\ion{O}{i}] lines are extremely weak
and are difficult to use for this purpose. 
Nonetheless, sensitive detectors, high-resolution, and long
exposures have enabled several teams (Spite \& Spite \cite{SS91}; 
Nissen \& Edvardsson \cite{NE92})
to measure the EWs of some dwarfs with [Fe/H] $\leq -0.1$. However, 
the [\ion{O}{i}] line
becomes too weak to measure at [Fe/H] $\leq -1.3$, and to provide useful
information on the trend of [O/Fe] vs. [Fe/H].
Thus, for the metal-poor stars,
the \ion{O}{i} triplet lines are the preferred choice in dwarfs because
they are of measurable strength for a wide range of stellar temperatures and
they lie in a fairly clear portion of the stellar spectrum.
Molecular OH lines in the near-ultraviolet at $\lambda$$\lambda$\,3060-3200 \AA, 
are observed in both giants and dwarfs with $T_\mathrm{eff}\leq$ 6500 K 
(OH lines are destroyed at higher temperatures), and
can be used to derive the O abundance.

An interesting result is that
there is a systematic difference of about 0.5 dex between 
the O abundances 
derived from the permitted and forbidden lines. 
The [\ion{O}{i}] 6300 line provides about 0.4$-$0.5 dex of 
the lower [O/Fe] values in metal-poor stars
(e.g. Spite \& Spite \cite{SS91}, Spiesman \& Wallerstein \cite{SW91},
Fulbright \& Kraft \cite{FK99} and Nissen et al. \cite{NPA20}
for dwarfs; 
Barbuy \cite{B88}, Barbuy \& Erdelyi-Mendes \cite{BE89} and
Sneden et al. \cite{sneden91} for giants).
The \ion{O}{i} triplet and OH lines, on the other hand, produce the higher [O/Fe] values, reaching +1.0 at
[Fe/H] $\approx -3$ (Abia \& Rebolo \cite{AR89}; King \cite{K94}; 
Nissen et al. \cite{NGEG94}; Israelian et al. \cite{I98};
Boesgaard et al. \cite{BKDV99}; Mishenina et al. \cite{MKKP20}).
Perhaps the higher O abundance obtained from the \ion{O}{i} triplet
was caused by some NLTE effect (Mishenina et al. \cite{MKKP20} and references therein). 
Kiselman (\cite{K91}) 
has performed a NLTE analysis of \ion{O}{i} $\lambda$\,7774 \AA $~$and  found significant 
NLTE corrections, up to 0.4 dex. 
However, Takeda (\cite{takeda}) found the NLTE effects to be small 
in metal-poor stars. 
Another source of discrepancy may be the fact that the O/Fe ratio is  very sensitive to
$T_\mathrm{eff}$ (Nissen \& Edvardsson \cite{NE92}).
King (\cite{K93}) determined a reduction in the temperature of subgiants of
150-200\,K, and obtained similar O abundances from the $\lambda$\,7774 \AA $~$and 
$\lambda$\,6300\,\AA $~$lines.
King \& Boesgaard (\cite{KB95}) obtained the O abundances from both the 
[\ion{O}{i}] $\lambda$\,6300 \AA $~$and \ion{O}{i} $\lambda$\,7774 \AA $~$lines 
for a sample of metal-rich F and G dwarfs. 
They found that for $T_\mathrm{eff}\leq 6200-6300$ K, no systematic
difference exists in the abundances determined from
the $\lambda$$\lambda$\,6300 and 7774 \AA$~$lines.  
For $T_\mathrm{eff}\geq 6200-6300$ K, 
however, the $\lambda$\,7774 \AA $~$abundance is substantially higher 
than the $\lambda$\,6300 \AA $~$abundance.

Thus, many papers have demonstrated that there is a 
systematic difference between the values of 
O abundance derived from the forbidden and permitted lines.
Possible reasons are as follows. The excitation potential of the \ion{O}{i} triplet is very
high (9.15 eV), the line is formed deep down in the atmosphere, and the abundance
determination has a large error; 
the near-UV 
OH lines at 3060-3200 \AA $~$are sensitive to the gas pressure
and electron density in the model atmosphere. 
In contrast, the [\ion{O}{i}] line
is roughly half as sensitive as the OH and \ion{O}{i} lines, 
and is insensitive to changes in the ratio of
$\alpha$-element to iron.
Also, the OH lines are very crowded in the UV region, 
which makes their measurement difficult. 

Nissen et al. (\cite{NPA20}) obtained the O abundances of 13 dwarfs and subgiants using 
data from the ESO VLT. They suggested that 
the [O/Fe] is nearly constant at $\approx$ +0.4 for [Fe/H] below $\approx -1.0$.
Maciel (\cite{maciel20}) suggested that [O/Fe] is not higher than
+0.4 according to the radial O/H abundance gradients in 
\ion{H}{ii} regions, hot stars and PNe, and [Fe/H]
gradients from open cluster stars.

From the foregoing discussion and considering the 
similar behavior
between O and $\alpha$ elements ([$\alpha$/Fe] is about +0.4 in
metal-poor stars, see Zhao \& Magain \cite{ZM90} and Fuhrmann \cite{Fu98}),  
we use the lower O abundances derived from the [\ion{O}{i}] line as 
observational constraints  
on our GCE calculations.
And we will specially discuss the higher abundances in metal-poor stars
obtained from the \ion{O}{i} triplet and OH lines in Sect. 8. 
Because the observed O abundances of
metal-poor dwarfs are not enough to reflect the abundance evolution,
we have supplemented our data with 
some metal-poor giants. 

\subsection {[N/Fe] vs. [Fe/H]}

The \ion{N}{i} line can be used in the abundance determination of nitrogen, but this can only be done for 
reasonably metal-rich stars. 
Clegg et al. (\cite{clegg}) obtained the N abundances of 20 disk stars 
($-0.9<$ [Fe/H] $<+0.4$)
using the \ion{N}{i} line, and suggested that [N/Fe] $\approx 0.0$.
For halo stars, some results have been derived 
from NH lines in the UV region. Tomkin \& Lambert (\cite{TL84}), 
Laird (\cite{laird85}) and Carbon et al. (\cite{carbon87}) obtained 
similar results, [N/Fe] $\approx 0.0$, 
irrespective of metallicity in the range $-2<$ [Fe/H] $<+0.3$. 
Indeed, there are still some difficulties 
in deriving reliable N abundance from the NH bands at 
$\lambda$$\lambda$\,3360 and 3370 \AA.
The difficulties are (a) uncertainties in the continuous opacities at
these wavelengths, (b) severe atomic-line contamination of the NH
features, (c) uncertainty in the dissociation energy of this molecule. 
(d) the crowdedness of the lines in the UV region.
Also, it is very difficult to derive the N abundance from the CN bands 
at $\lambda$$\lambda$\,4200 or 3800 \AA. Possible reasons are that
the dissociation energy of CN is not known to better than a
factor of 2, and that the determination of N abundance from CN depends on a 
prior determination of reliable C abundance from other features, 
and that the CN bands are extremely weak in very metal-poor stars. 
A detailed discussions 
can be found in Wheeler et al. (\cite{wheeler}).

\subsection {Table of observations}

Table 1 summarises the sources of the observations of carbon, nitrogen and oxygen.
Most of the objects are dwarfs and subgiants;
a few metal-poor giants provide additional O abundance data  
derived from the [\ion{O}{i}] $\lambda$\,6300 \AA $~$forbidden 
line. 

{
\begin{table*}
{\centering
 \caption[]{Observational data from different references and the line sources}
\begin{tabular}{llllll} \hline

References   &   C                     &    N                &   O                   & [Fe/H]      & Numbers of stars \\ 
             & line, wavelength (\AA) & line, wavelength (\AA) & line, wavelength (\AA) &              &                  \\ \hline 
 1   & \ion{C}{i}, [\ion{C}{i}], CH    & \ion{N}{i} 8683  &\ion{O}{i} 6158, [\ion{O}{i}] 6300 &[$-0.90,+0.40$] & 20             \\
 2   & [\ion{C}{i}] 8727              &                 &                  &[$-1.00,+0.25$] & 85         \\
 3   & [\ion{C}{i}] 8727              &                 &                  &[$-1.06,+0.26$] & 80          \\ 
 4   &  \ion{C}{i} 7100               &                 &                  &[$-0.80,+0.20$] & 105     \\
 5   &  \ion{C}{i} 9100               &                 &\ion{O}{i} 7770 triplet  &[$-3.00,-0.80$] & 34     \\
 6   &  CH  4300                     & NH 3360         &                  &[$-2.45,+0.50$] & 116    \\
 7   &  CH  4300                     &                 &                  &[$-2.60,-0.70$] & 32      \\
 8   &  CH  4300                     & NH 3360         &                  &[$-3.20,-1.50$] & 83  \\
 9   &  CH  C$_2$                    & CN              & \ion{O}{i}, [\ion{O}{i}]       &[$-2.61,+0.12$] & 19 (some giants)   \\
 10          &                       &  NH 3360        &                  & [$-2.30,-0.30$]& 14  \\
 11          &                       &                 &  [\ion{O}{i}] 6300       &[$-1.60,+0.30$] & 7      \\
 12          &                       &                 &  [\ion{O}{i}] 6300       & $-1.74,-1.43$  & 2  \\
 13          &                       &                 &  [\ion{O}{i}] 6300      &[$-0.80,+0.30$]  & 23  \\
 14          &                       &                 &  [\ion{O}{i}] 6300      & $-2.84,-2.31$  & 2   \\
 15          &                       &                 &  [\ion{O}{i}] 6300      &[$-1.80,-0.70$] & 13  \\
 16          &                       &                 & \ion{O}{i} 7770 triplet &[$-1.03,+0.26$] & 86 \\
 17          &                       &                 & \ion{O}{i} 7770 triplet &[$-1.00,+0.10$] & 90  \\
 18          &                       &                 & \ion{O}{i} 7770 triplet &$-1.84,-2.08$   & 2   \\
 19          &                       &                 & \ion{O}{i} 7770 triplet &[$-2.50,-0.50$] &  14 (some giants)  \\
 20          &                       &                 & \ion{O}{i} 7770 triplet &[$-2.72,-0.47$] &  14 \\
 21          &                       &                 & \ion{O}{i} 7770 triplet & $-3.00$        &  1 \\
 22          &                       &                 & \ion{O}{i} 7770 triplet &[$-3.50,-0.20$] & 30  \\
 23          &                       &                 & OH                & [$-3.00,-0.30$]& 24  \\
 24          &                       &                 & OH               &[$-3.02,-0.50$] &  24   \\
 25          &                       &                 & OH 3138-3155     & [$-3.20,-1.80$]& 9  \\
 26          &                       &                 &   OH , CO        & $-1.22$        & 1  \\
 27          &                       &                 & [\ion{O}{i}] 6300 &[$-1.20,-0.10$] & 24 (some giants)\\
 28          &                       &                 & [\ion{O}{i}] 6300 &[$-3.00,-1.10$] &  20 (giants) \\
 29          &                       &                 & [\ion{O}{i}] 6300 &[$-2.35,+0.37$] & 18  (giants) \\    
 30          &                       &                 & [\ion{O}{i}] 6300 &[$-2.88,-1.80$] & 10   (giants)  \\   \hline 

\end{tabular}
}

\vspace{0.2cm}
1. Clegg et al. \cite{clegg}; 2. Andersson \& Edvardsson \cite{AE94};
3. Gustafsson et al. \cite{Gus99}; 4. Tomkin et al. \cite{TWLL95};
5. Tomkin et al. \cite{TLLS92}; 6. Laird \cite{laird85};
7. Tomkin et al. \cite{TSL86}; 8. Carbon et al. \cite{carbon87};
9. Carretta et al. \cite{carretta20}; 10. Tomkin \& Lambert \cite{TL84};
11. Spite \& Spite \cite{SS91};
12. Spiesman \& Wallerstein \cite{SW91}; 13. Nissen \& Edvardsson \cite{NE92};
14. Fulbright \& Kraft \cite{FK99}; 15. Nissen et al. \cite{NPA20};
16. Edvardsson et al. \cite{E93}; 17. Chen et al. \cite{chen};
18. Beveridge \& Sneden \cite{BS94};
19. Mishenina et al. \cite{MKKP20}; 20. Boesgaard \& King \cite{BK93};
21. King \cite{K94}; 22. Abia \& Rebolo \cite{AR89}; 
23. Israelian et al. 1998; 24. Boesgaard et al. \cite{BKDV99};
25. Nissen et al. \cite{NGEG94}; 26. Balachandran \& Carney \cite{BC96};
27. Barbuy \& Erdelyi-Mendes \cite{BE89}; 28. Barbuy \cite{B88};
29. Gratton \& Ortolani \cite{GO86}; 30. Sneden et al. \cite{sneden91}
\end{table*} 
}

\subsection {The age-metallicity relation and the metallicity distribution of G dwarfs }

The age-metallicity relation for the solar neighborhood 
is one of the important observational constraints; it shows 
the [Fe/H]-represented 
metallicity as a function of the star's age. The observational data and
scatter are taken from Carlberg et al. (\cite{car85}), 
Meusinger et al. (\cite{meusinger}) and 
Edvardsson et al. (\cite{E93}) (their tables 14 and 15).

Another important observational constraint is the
metallicity distribution of long-lived G-type stars. 
Because these stars have long main-sequence lifetimes of about 15 Gyr, 
comparable to the estimated age of the Galaxy, they 
represent a sample of stars which have not been removed by stellar
evolution. The observational metallicity
distribution of G dwarfs is taken from Chang et al. (\cite{chang}). 
The solar abundances of CNO elements are
taken from Grevesse \& Sauval (\cite{GS98}).


\section{Model of galactic chemical evolution}

   A general, standard infall GCE model, namely,
a multizone model for GCE without mass exchange among the zones is adopted in our present study. 
The rate of change of the surface mass density of isotope $i$ in every zone in the 
gas at a given Galactocentric radius $r$ and time $t$ is 
\begin{eqnarray}
  & & {d\sigma_{g,i}(r,t) \over dt}  =  -X{_i}(r,t)\Psi (r,t)     \cr                         \cr
        & + &\int \limits_{M_l}^{M_{bl}} X{_i}(r,t- \tau_{M})\Psi (r,t- \tau_{M})\Phi(M)dM  \cr
        & + & \beta \int \limits_{M_{bl}}^{M_{bu}} \Phi(M_b)   
              \left [ \int \limits_{\mu_m}^{0.5} f(\mu)
               X{_i}(r,t-\tau_{M_2}) \Psi (r,t-\tau _{M_2})d{\mu} \right ] d{M_b}\cr
        & + & (1- \beta) \int \limits_{M_{bl}}^{M_{bu}}X{_i}(r,t- \tau_{M}) \Psi (r,t- \tau _{M})\Phi(M)dM \cr
        & + & \int  \limits_{M_{bu}}^{M_u} X{_i}(r,t- \tau_{M}) \Psi (r,t-\tau_{M})\Phi(M)dM         \cr
        & + &  \left [\frac {d \sigma{_i}(r,t)}{dt}\right ]_\mathrm{inf} ,
\end{eqnarray}

where the first term describes the disappearance of isotope $i$ due to new star 
formation, the second term describes the enrichment due to stellar wind mass
loss of low mass single stars, the third term represents the rate of enrichment 
due to binary systems that undergo SN\,Ia explosions.  The free parameter $\beta$ is
the amplitude factor of Type {\sc I}a supernova (hereafter SN\,Ia) explosion and 
$f(\mu)$ is the binary distribution
function.  The fourth term represents the enrichment due to single
stars or binary systems that do not undergo Type {\sc I}a events in the mass interval
$M_{bl}$ to $M_{bu}$.  The fifth term represents the enrichment rate due to
massive stars that become SN\,II explosions with masses between $M_{bu}$ to $M_{u}$.
The last term represents the infall of the primordial material.  $X{_i}(r,t)$ is the
elemental yield from nucleosynthesis (see Sect. 4).  More details have been given by 
Matteucci \& Fran\c{c}ois (\cite{MF89}) and Timmes et al. (\cite{timmes}).
   
{$\bullet$ \bf $\Psi (r,t)$ (SFR)~} 
Analytical prescriptions for the SFR can be
obtained from the observations. Schmidt (\cite{sch59}) suggested that the SFR is
related to the
surface mass density of interstellar gas $\sigma_g$. 
Subsequent studies showed
that the SFR 
depends not only on the surface mass density of gas but also on the total surface
mass density of the ISM (Talbot \& Arnett \cite{TA75}; Dopita \& Ryder \cite{DR94}; 
Timmes et al. \cite{timmes}; Chiappini et al. \cite{CMG97}; Dwek \cite{dwek98}).  

We adopt a similar formula to that of Timmes et al. (\cite{timmes}):  
\begin{equation} 
\Psi(r,t)=\nu \sigma_\mathrm{tot}(r,t)\left [\frac {\sigma_\mathrm{gas}(r,t)} 
           {\sigma_\mathrm{tot}(r,t)}
\right ]^{n} M_{\odot}{\rm pc^{-2}Gyr^{-1}}, 
\end{equation} 
where $n$ is in the range 1--2 (Timmes et al. \cite{timmes}), 
which we fix at 1.5 throughout our calculation,  
$\nu$ is the efficiency factor in unit Gyr$^{-1}$, 
which can be adjusted around the value 1.0.

{$\bullet$ \bf $\Phi (M)$ (IMF)~} 
For the solar neighborhood, the IMF is given in a series of papers. 
The first formula was applied by 
Salpeter (\cite{sal55}), who suggested the single power-law form: 
$\Phi(M)\sim M^{-(1+x)}$ ($x=1.35$),
normalized to 1 over a given mass range.
Then 
Scalo (\cite{scalo86}) suggested that the power law should become steeper 
for $M\geq 1M_{\odot}, x=1.7$. 
Following them, more authors suggested that
IMF should flatten at the low mass end, for example Kroupa et al. (\cite{KTG93}). 
Dwek (\cite{dwek98}) analyzed these results, 
and gave a similar formula as Kroupa et al. (\cite{KTG93}) (for three mass ranges).
An important issue 
for the IMF is it being ``top-heavy'', meaning that 
the formation frequency of massive stars in the early Galactic stage
was higher than at the present. 
Recently, the question of a changing IMF as our Galaxy evolves has been examined.
Chiappini et al. (\cite{CMP20}) found that                       
the metallicity distribution of G dwarfs is well explained by a constant IMF, rather than by  
a changing one.
Martinelli \& Matteucci (\cite{MM20})  
found that, with a changing IMF, even though
the metallicity distribution of G dwarfs could be reproduced, other important observations of [O/Fe] would then not fit so well.

In our calculation, we adopt the formula of Dwek (\cite{dwek98}):
\begin{eqnarray}
\Phi(M) & = & C_1~~~~~~~~ ~~~~~{\rm for}~ 0.1 \leq M/M_{\odot} \leq 0.3      \cr
        & = & C_2M^{-1.6}   ~~~~~{\rm for}~ 0.3 \leq M/M_{\odot} \leq 1  \cr
        & = & C_3M^{-2.6}   ~~~~~{\rm for}~ 1 \leq M/M_{\odot} \leq 100,
\end{eqnarray}
where $C_1, C_2$ and $C_3$, the normalized coefficients, are determined from the normalization 
$\int \limits_{M_l}^{M_u} M\Phi (M) dM=1$,
with ${M_l}=0.1M_{\odot}$ and ${M_u}=100M_{\odot}$.
In fact, the bulk of chemical enrichment is caused by the stars 
with $M \geq 1 M_{\odot}$. Thus,
it is meaningful to fix the fraction $\zeta$ of the total stellar mass distributed in
stars above $1M_{\odot}$ ($M_1$), which is equivalent to fixing $M_l$,
namely,
\begin{eqnarray}
\int \limits_{M_l}^{M_u} M\Phi (M) dM & = & 
   \int \limits_{M_l}^{M_1} M\Phi (M)dM + \int \limits_{M_1}^{M_u} M\Phi (M) dM \cr
& = & \int \limits_{M_l}^{M_1} M\Phi (M)dM + \zeta = 1.
\end{eqnarray}
Pagel (\cite{pagel97}) explained this method in detail.
In our calculations 
for different sets of yields, we adjust $\zeta$ so that the results give a good fit to 
the observations, especially the metallicity distribution of G dwarfs .

{$\bullet $\bf Infall } The infall rate is generally taken to follow an exponent law  
(Matteucci \& Fran\c{c}ois \cite{MF89}; Timmes et al. \cite{timmes}) 
or Gaussian form (Prantzos \& Silk \cite{PS98}; Chang et al. \cite{chang}).
We adopt the specific form  
of Timmes et al. (\cite{timmes}),
but with $\sigma_{\odot}=55 M_{\odot}$ pc$^{-2}$ (Sackett \cite{sackett}).

{$\bullet$\bf Main-sequence lifetimes $\tau_M~$} 
The main-sequence lifetimes of stars are
taken from Schaller et al. (\cite{schaller}), which 
have been used generally 
in GCE calculations,
such as Timmes et al. (\cite{timmes}) and Prantzos \& Silk (\cite{PS98}).

\section{Stellar yields and nucleosynthesis }

The stars are divided into binary and single stars.
We only consider SN\,Ia explosion for the binary stars,
and divide the single stars into
two groups: intermediate-, low-mass stars ($0.9-8M_{\odot}$,
or $0.9-6M_{\odot}$ in case of strong overshooting)
and massive stars ($M>8M_{\odot}$,
or $M>6M_{\odot}$ in case of strong overshooting). 

\subsection {Intermediate-, low-mass single stars}

The first detailed nucleosynthesis calculation for ILMS was made 
by RV ($M=$ $1-8M_{\odot}$, $Z=$ 0.004, 0.02). Recently,
VG extended the calculation to 
$M=$ $0.8-8.0M_{\odot}$ and $Z=$ 0.001, 0.004, 0.008, 0.02, 0.04.
Forestini \& Charbonnel (\cite{FC97}) calculated the evolution
of stars with $M=$ 3, 4, 5, 6, 7$M_{\odot}$ and $Z=$ 0.005, 0.02.
MBC calculated the evolution 
of stars with $M=0.819-5.0M_{\odot}$ and $Z=0.008, 0.02$ 
for the case of strong overshooting,
which means that the stars evolve rapidly so that the upper mass limit of
ILMS is changed from the standard value of 8$M_{\odot}$ to  6$M_{\odot}$. 
M2K gave a new set of stellar
yields (for $M=0.817 - 5.0M_{\odot}$ and $Z=$$0.004, 0.008, 0.02$) 
based on the results of an updated evolution calculation.

In our calculations, we choose the results of RV 
(their $\alpha=1.5, \eta=0.333$ case), 
VG, MBC and M2K as the yields
of ILMS for their wide mass ranges and metallicity-dependent results.
RV only gave results of $3.25M_{\odot}<M<8M_{\odot}$ for the case of
$\alpha=1.5, \eta=0.333$. In fact, the hot-bottom burning (HBB) process 
is not important for low
mass stars, but is very significant for 
intermediate mass stars above $4M_{\odot}$. 
So we choose the results with $\alpha=0.0$ for the lower mass stars.
M2K calculated three sets of yields with $\alpha=1.68, 2.00, 2.50$ 
for stars with $M\ge 3.5M_{\odot}$. $\alpha=2.50$ is too high to 
fit the observations (M2K), so we only consider the results for 
$\alpha=1.68, 2.00$.

\subsection {Binary stars through SN\,Ia explosions}

Type {\sc I}a SN explosions are assumed to occur in close binary systems 
(Whelan \& Iben \cite{WI73}). In this model, the explosion is caused by 
a carbon-deflagration of the material
accreting on the degenerate white dwarf (Nomoto et al. \cite{nomoto2}),
and the ejecta is dominated by the $^{56}$Fe isotope. 
In this paper, the nucleosynthesis yields of SN\,Ia explosions are taken 
from the classical W7-model of Nomoto et al. (\cite{nomoto2}), in which  $\sim 0.613M_{\odot}$ of Fe is produced.

\subsection {Massive single stars}

A star in the mass range of 8$-$10$M_{\odot}$ (or 6$-$8$M_{\odot}$  
in the case of strong overshooting) generally develops a degenerate O-Ne-Mg
core after C-burning and eventually explodes as an electron capture supernova,
leaving a neutron star of $1.3 M_{\odot}$ as a remnant and expelling a very small quantity of heavy elements. 

A star with a higher initial mass will end its life with an Fe-core 
collapse Type {\sc II} SN explosion. 
Some authors, including WW, N97, M92 and PCB, 
calculated the stellar nucleosynthesis yields 
for such stars either with or without mass loss through wind.
The results differ because of different choices of the 
parameters.
We have carefully compared their parameters and results in 
Liang \& Zhao (\cite{liang21}).
Some of the parameters adopted by them are listed in Table 2, including 
the stellar mass, metallicity, and $^{12}$C$(\alpha,\gamma)^{16}$O reaction rate.

Limongi et al. (\cite{lim20}) calculated the presupernova nucleosynthesis 
of massive stars with $M=13, 15, 20, 25M_{\odot}$ and $Z=0.02$,
and the explosive nucleosynthesis of stars with $Z=10^{-3}, 0$.
They did not give the yields of stars with $M>25M_{\odot}$, which
are very important for tracing the chemical evolution. And
they did not give the explosive nucleosynthesis of stars with $Z=0.02$.
So we did not use their results in our calculation of the abundance evolution of CNO elements in the present paper.

{
\begin{table*}
\centering
 \caption[]{Selected parameters of stellar evolution and nucleosynthesis of massive
             stars}
{\scriptsize
\begin{tabular}{lllll} \hline

            &       WW       & N97              &        M92  &  PCB     \\ \hline
upper mass  &  $40M_{\odot}$ & $70M_{\odot}$ &  120$M_{\odot}$      &  120$M_{\odot}$             \\
metallicity & $Z=1,0.1,0.01,10^{-4},0Z_{\odot}$   & $Z_{\odot}$  & $Z=0.001, 0.02$ & $Z=0.0004-0.05$  \\ 
stellar wind mass loss& No           & No             & Yes         & Yes         \\   
$^{12}$C$(\alpha,\gamma)^{16}$O rate& 1.7$\times$CF$\simeq$0.74$\times$CFHZ & CFHZ & CFHZ  &    CF         \\  \hline
CFHZ=Caughlan et al. (\cite{CFHZ}) & &                 &             &            \\ 
CF=Caughlan \& Fowler (\cite{CF88})& &                 &             &            \\

\end{tabular}
}
\end{table*} 
}

\subsubsection {Yields calculated by WW}

WW calculated, for a large mass range, the metallicity-dependent yields of SN\,II explosion, and the results 
have been widely used in GCE models (e.g. 
Timmes et al. \cite{timmes}; Chiappini et al. \cite{CMG97})

Timmes et al. (\cite{timmes}) suggested that the results of 
abundance evolution would be better 
if the Fe yields of WW is reduced by a factor of 2.
Samland (\cite{samland}) suggested that the actual Fe yield of SN\,II 
explosion should be $0.046M_{\odot}$,
approximately a factor of 2 less than the WW value.
Carigi (\cite{cari94})
used 0.075$M_{\odot}$ as the Fe yields of SN\,II explosion in their GCE model. 
Chieffi et al. (\cite{CLS98})
calculated the nucleosynthesis of SN\,II explosion for a $25M_{\odot}$ star, 
and got an 
Fe yield of about
0.075$M_{\odot}$. SN 1987A, for an initial main sequence mass of 20$M_{\odot}$,
ejected 0.075$M_{\odot}$ of Fe, while 
for the same initial mass, WW got 0.151$M_{\odot}$, 
again twice as large.
Because of these findings 
we reduced the results of WW by a factor of 2 in our calculations. 

\subsubsection {Yields given by N97}
 
N97 extended the calculations of Thielemann et al. (\cite{TNH96}) to include $13-70M_{\odot}$, solar metallicity stars.
We should note that
N97 used the evolution of the He core instead that of the entire star.
They used a relation 
$M(M_{\alpha})$ to transfer the result for a helium core of mass 
$M_{\alpha}$ to that of a star of initial mass $M$ (Sugimoto \& Nomoto \cite{SN80}). 
This can be understood since  
the total ejecta of star must be the sum of the contributions of both 
the He-core and the envelope of mass $M-M_{\alpha}$. 
(Thomas et al. \cite{thomas98}).

\subsubsection {Yields given by M92 and PCB}

Both M92 and PCB considered radiative-driven stellar wind in their
calculations, which was 
important for massive stars, especially the W-R stars.
The synthesized elements of C, N and O, in particular, C, can be ejected 
into the ISM in stellar wind.

How to choose the yields of W-R stars?  
Massey et al. (\cite{massey}) studied the massive stars both in the 
general field and associations
of the Magellanic Clouds.
Their data showed that stars with initial masses $M>30M_{\odot}$ evolved
through the W-R phase in the Large Magellanic Cloud (LMC); while the statistics of 
the Small Magellanic Cloud (SMC)  
are consistent with a somewhat higher mass limit of 
possibly
$50M_{\odot}$.
Conti (\cite{conti95}) suggested that the W-R stars are highly evolved, 
luminous, hot, and (mostly)
He-burning descendants of the most massive stars in the solar neighborhood, with masses $M\geq 35M_{\odot}$.
In the present paper, we follow    
the suggestions of Maeder \& Conti (\cite{MC94}) and the references therein and assume that the stars
with $M>40M_{\sun}$ will evolve through the W-R stage. 
So we divide the massive stars into two mass ranges,
$M\leq 40M_{\odot}$ and $M>40M_{\odot}$, in our discussion 
on the source of carbon.
For the upper mass limit of massive stars, we choose $100M_{\odot}$,
following the suggestion of Leitherer (\cite{lei95}), and  the usual practice with the upper mass limit of IMF used in many GCE models
(e.g. PCB, Prantzos \& Silk \cite{PS98})

M92 did not give Fe yields. When we use the M92 set of yields,
we choose the following values: 0.15$M_{\odot}$ 
for stars with initial main sequence mass $M\leq 14M_{\odot}$, 
0.075$M_{\odot}$ for $14M_{\odot}< M\leq 40M_{\odot}$,
and 0.15$M_{\odot}$ for $M>40M_{\odot}$. 
These choices are based on the following considerations.
In the case of SN 1987A (20$M_{\odot}$ during
the main-sequence stage), the light curve, powered by the decay of  $^{56}$Ni
and $^{56}$Co, gives a determination of the produced $^{56}$Fe of   
0.075$M_{\odot}$, and for SN 1993J (main-sequence mass 14$M_{\odot}$),
one of 0.15$M_{\odot}$.
(Thielemann et al. 1996). Carigi (1994)
adopted 0.075$M_{\odot}$ of Fe  
for a SN\,II explosion and 0.15$M_{\odot}$
for a SN\,Ib explosion. Since SN\,Ib takes place in W-R stars, and  
since we assume that the stars with $M>40M_{\odot}$ will undergo the W-R stage,
so we use the value 0.15$M_{\odot}$ in such cases.

M92 only gave the yield of N  ejected in 
wind, and not the final value. Thus the N yields used in 
our calculations
are lower limits. 

\section{ Results and analyses}

We set up a standard infall model of GCE
to map out 
the abundance evolution of carbon, nitrogen and oxygen in the solar
neighborhood.  
We choose the general formulae of SFR and IMF, and slightly adjust
them by changing $\nu$ and $\zeta$ respectively.
The 8 sets of nucleosynthesis yield, 
described in Sect. 4 are then used to set up 8 specific models.
For each model,
we try to fit the observational age-metallicity relation, the 
metallicity distribution of G dwarfs and the [O/Fe] vs. [Fe/H] relation.
The calculated results will then be discussed: 
the [C/Fe] vs. [Fe/H] relation, in this section. 
the [N/Fe] vs. [Fe/H] and [O/Fe] vs. [Fe/H] relations 
in Sect. 6 and Sect. 7, respectively.

\subsection{VG+WW}

In this model labelled (VG+WW), we use the yields of VG for ILMS,
and WW for massive stars (see Sect. 4).
The calculated age-metallicity relation, metallicity distribution of G dwarfs and [O/Fe] vs. [Fe/H]
relation can all fit the observations well (Figs. 1a, c, d).
The predicted abundance evolution of carbon is given in Fig. 1b; this 
shows that [C/Fe] vs. [Fe/H]
is approximately constant in time. The predicted trend agrees with the observations, including 
even 
the increasing positive [C/Fe] below [Fe/H] $<-1.5.$ 
The reason of positive [C/Fe] may be that for the massive stars, 
the C yields are high and the Fe yields are
relative low. 
In medium metal-poor region, [Fe/H] $>-1.0$, 
[C/Fe] slightly increases, showing the important  
C contribution from ILMS.
Up to the late evolutionary stage of the Galaxy, 
the ratio of SN\,Ia rate to SN\,II rate increases, therefore the [C/Fe] value
decreases with increasing [Fe/H] in the range of [Fe/H] $>-0.5$.
The results show that the combined contribution of carbon 
by VG and WW can 
explain the observed [C/Fe] vs. [Fe/H] relation.

\begin{figure*}
\centering
\input epsf
\vspace{0cm}
\hbox{\hspace{0cm}\epsfxsize=8.8cm \epsfbox{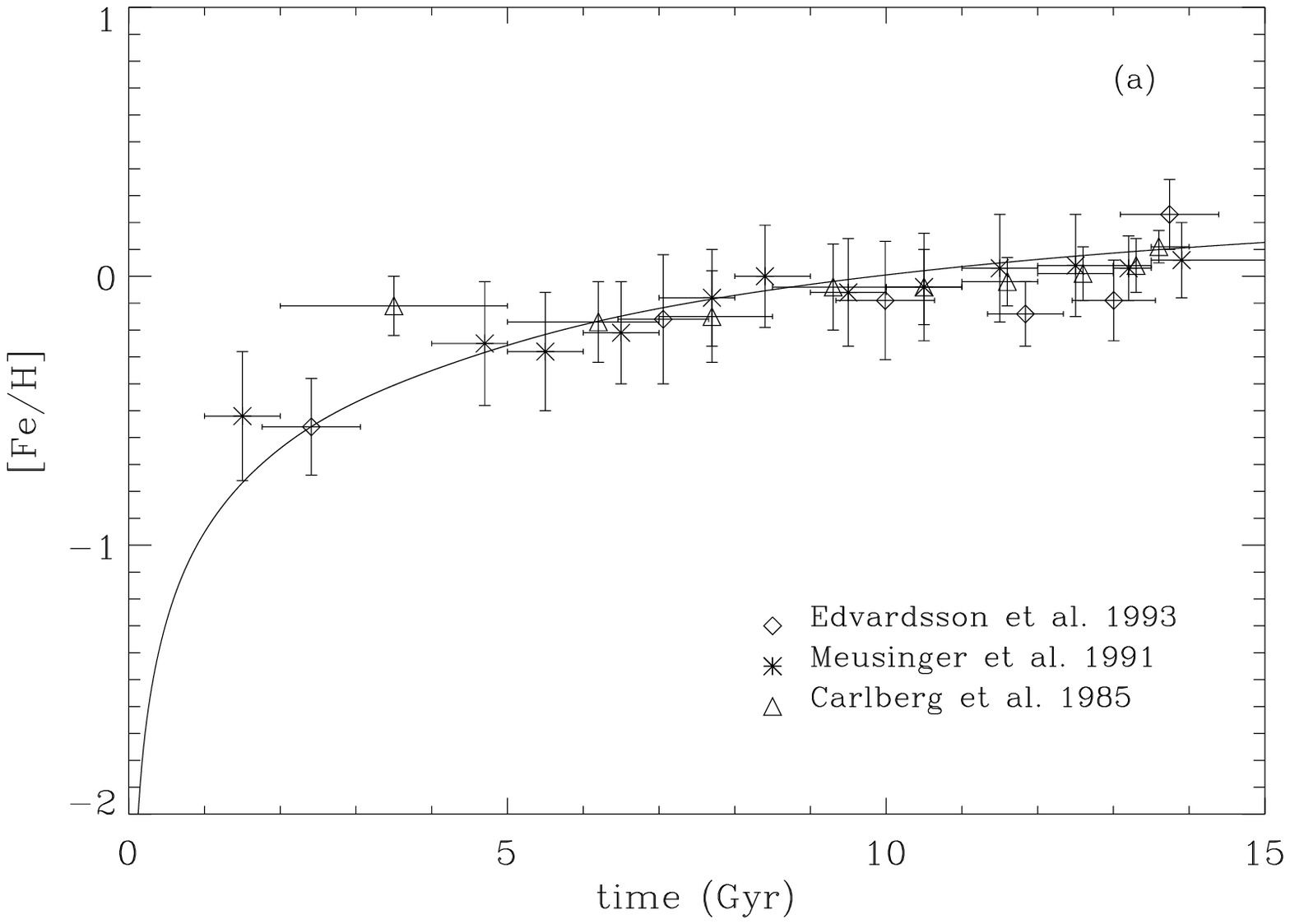}
\epsfxsize=8.8cm \epsfbox{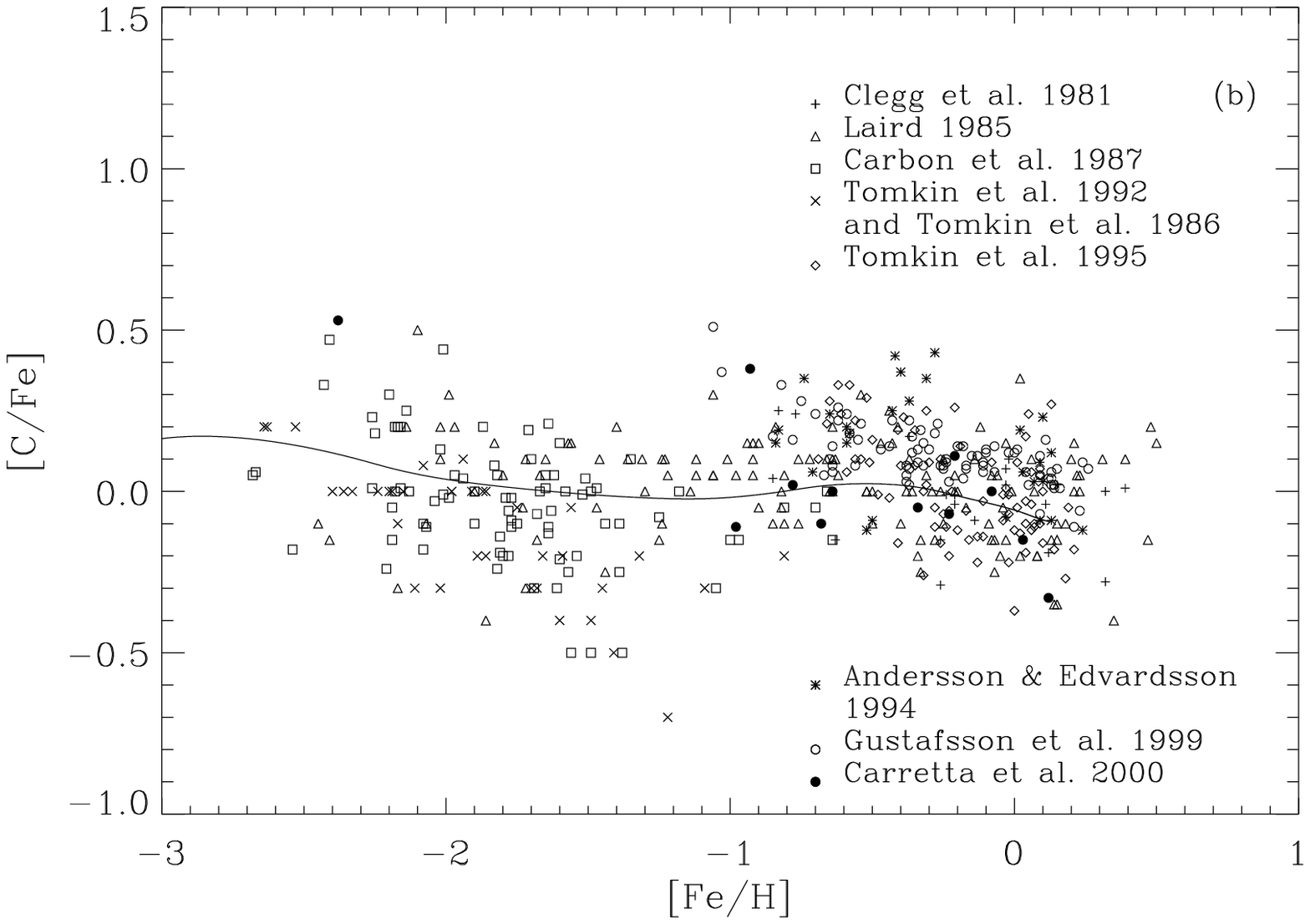}}
\vspace{0cm} 
\hbox{\hspace{0cm}\epsfxsize=8.8cm\epsfbox{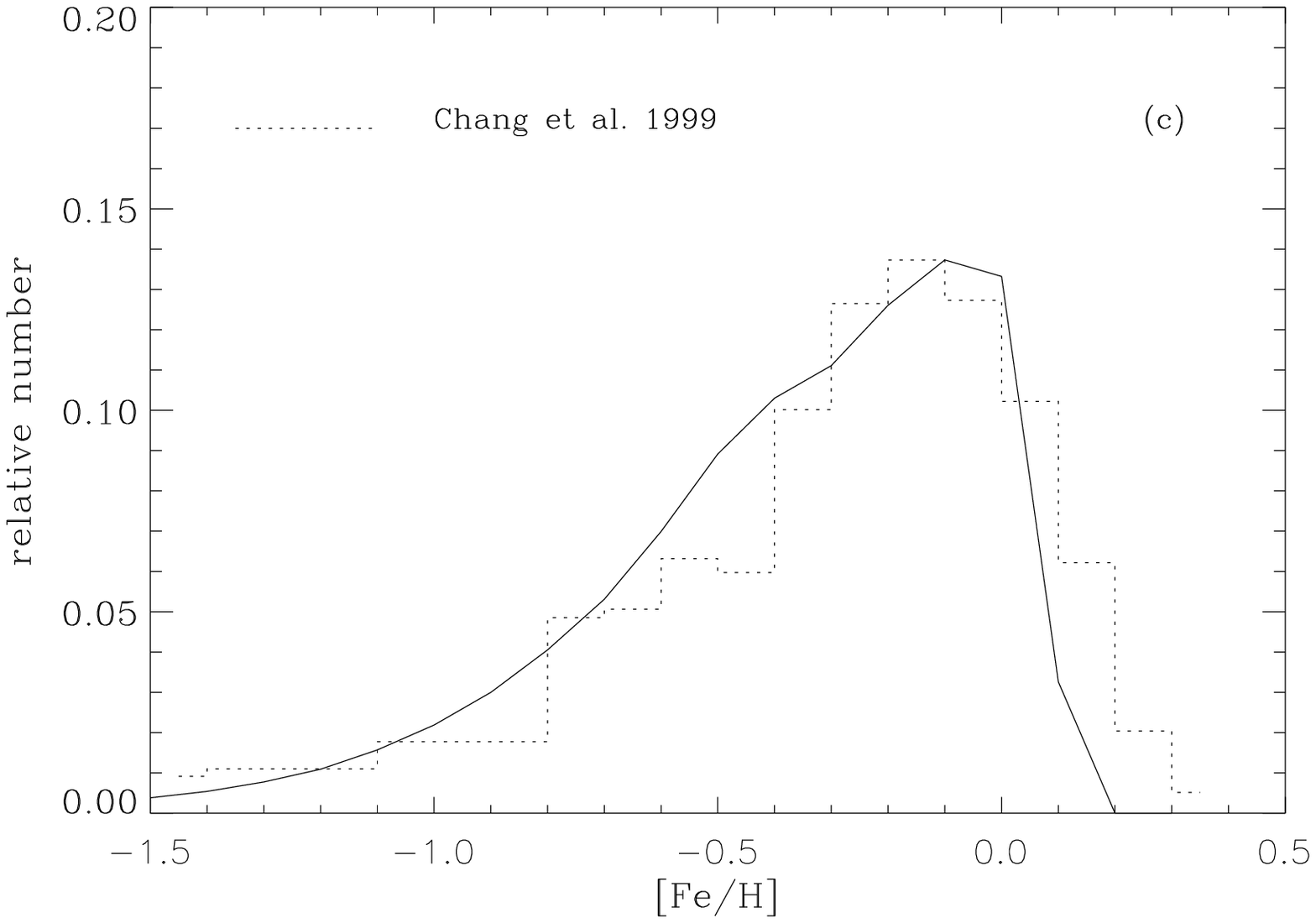}
\epsfxsize=8.8cm \epsfbox{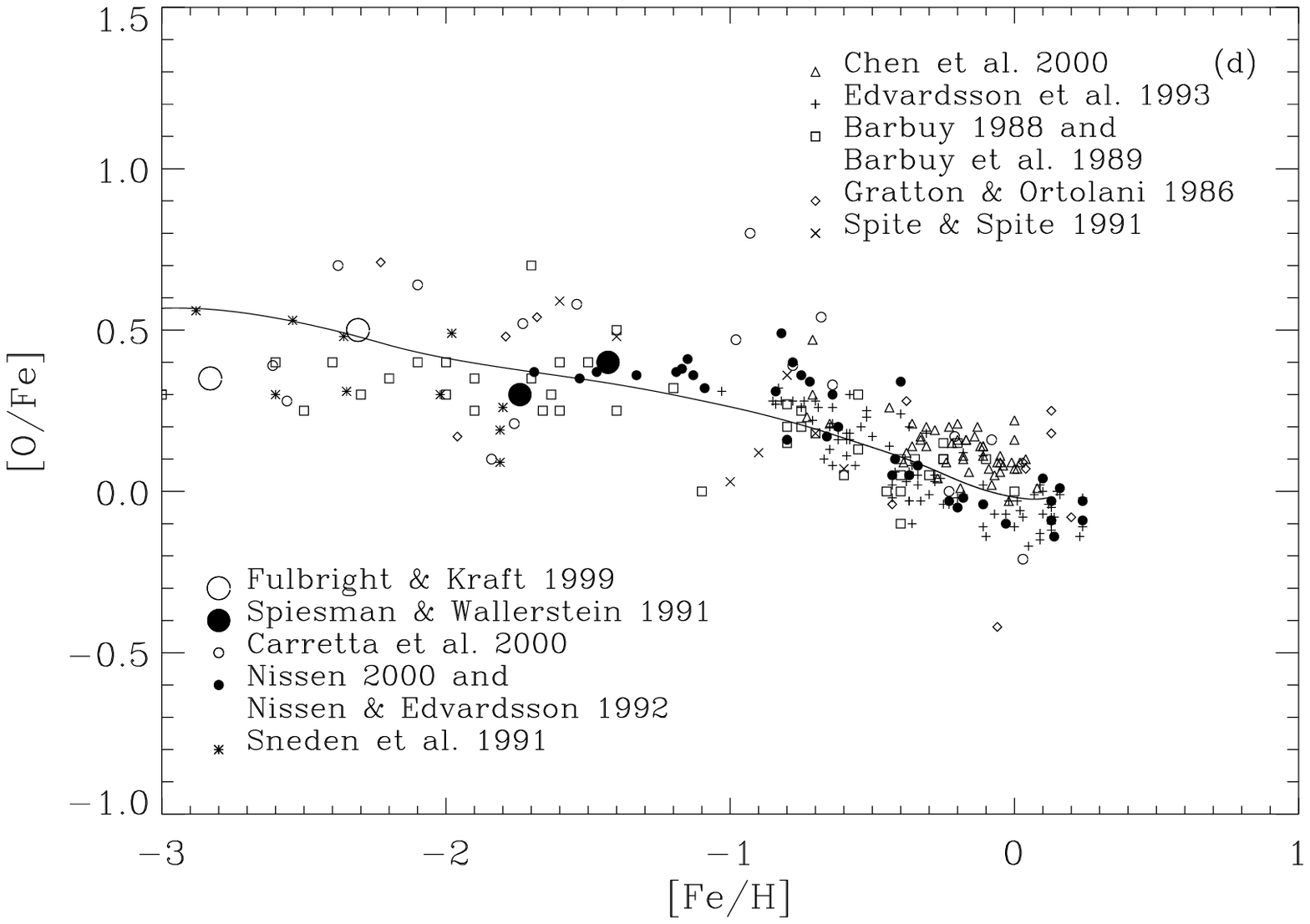}}
\vspace{0cm}
 \caption{Results using the nucleosynthesis yields of VG for ILMS and
          WW for massive stars (the solid lines) (model VG+WW). 
(a) The age-metallicity relation in the solar neighborhood;
(b) The [C/Fe] vs. [Fe/H] relation;
(c) The metallicity distribution of G dwarfs in the solar neighborhood;
(d) The [O/Fe] vs. [Fe/H] relation.
The corresponding observations are shown in the figures.}
\end{figure*}

\subsection{VG+N97}
 
Using the yields of N97 for massive stars and those of VG for ILMS (model VG+N97), 
our GCE results can fit the observational 
age-metallicity relation,
metallicity distribution of G dwarfs and [O/Fe] values (Figs. 2a, c, d). 
But the predicted C abundance is too low to fit
the observations (Fig. 2b). 
Compared to WW, 
the lower C yields calculated by N97 may be mainly caused by 
the choice of a higher rate of  
$^{12}$C$(\alpha,\gamma)^{16}$O reaction. Also, 
the authors of N97 themselves have compared 
the different nucleosynthesis results based on two different 
$^{12}$C$(\alpha,\gamma)^{16}$O reaction rates from 
Caughlan et al. (\cite{CFHZ}) (hereafter CFHZ) 
and Caughlan \& Fowler (\cite{CF88}) (hereafter CF) respectively. 
CFHZ gave a higher reaction rate than CF: CFHZ$\approx$2.3$\times$CF; 
and WW adopted 1.7$\times$CF$\approx$0.74$\times$CFHZ 
in their calculation, which is lower than in N97.
So when [O/Fe] can be matched, [C/Fe] is lower than the observations. 
These results underline the importance of the $^{12}$C$(\alpha,\gamma)^{16}$O reaction rate
to elemental nucleosynthesis.

Certainly, besides the $^{12}$C$(\alpha,\gamma)^{16}$O reaction rate,
convective mechanism and explosion energy can affect the
nucleosynthesis yields (N97).

\begin{figure*}
\input epsf
\vspace{0cm}
\hbox{\hspace{0cm}\epsfxsize=8.8cm \epsfbox{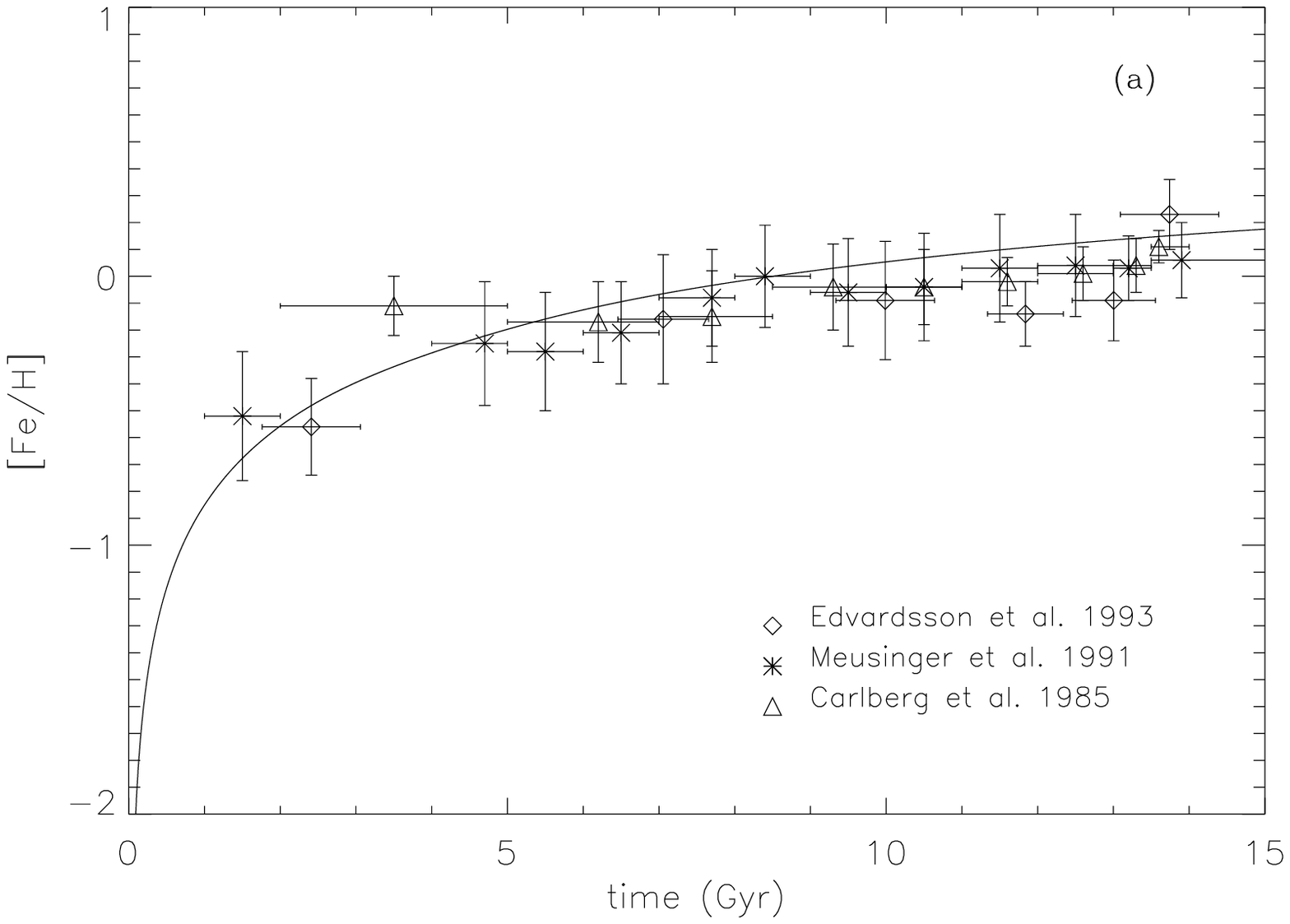}
\epsfxsize=8.8cm \epsfbox{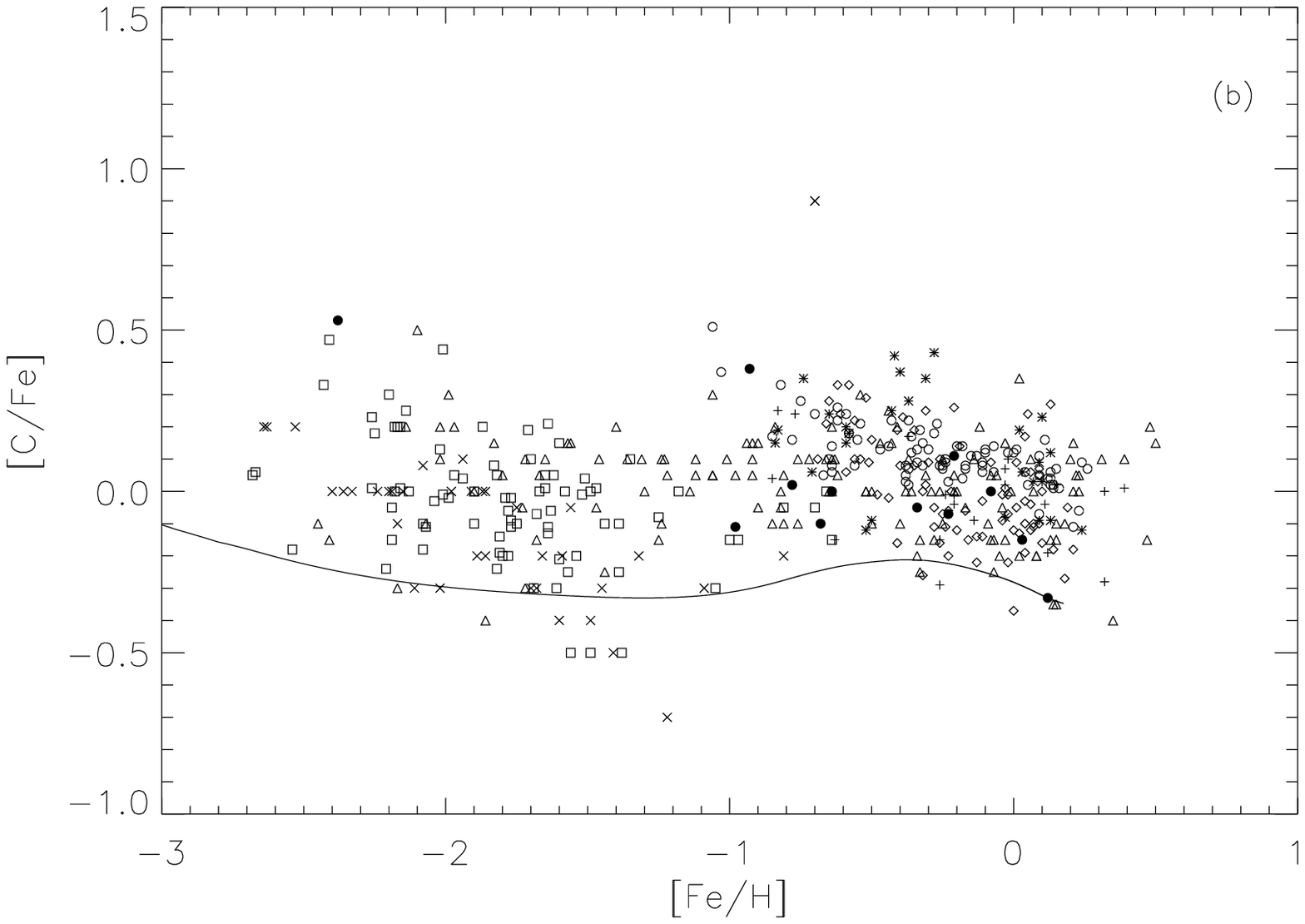}}
\vspace{0cm} 
\hbox{\hspace{0cm}\epsfxsize=8.8cm\epsfbox{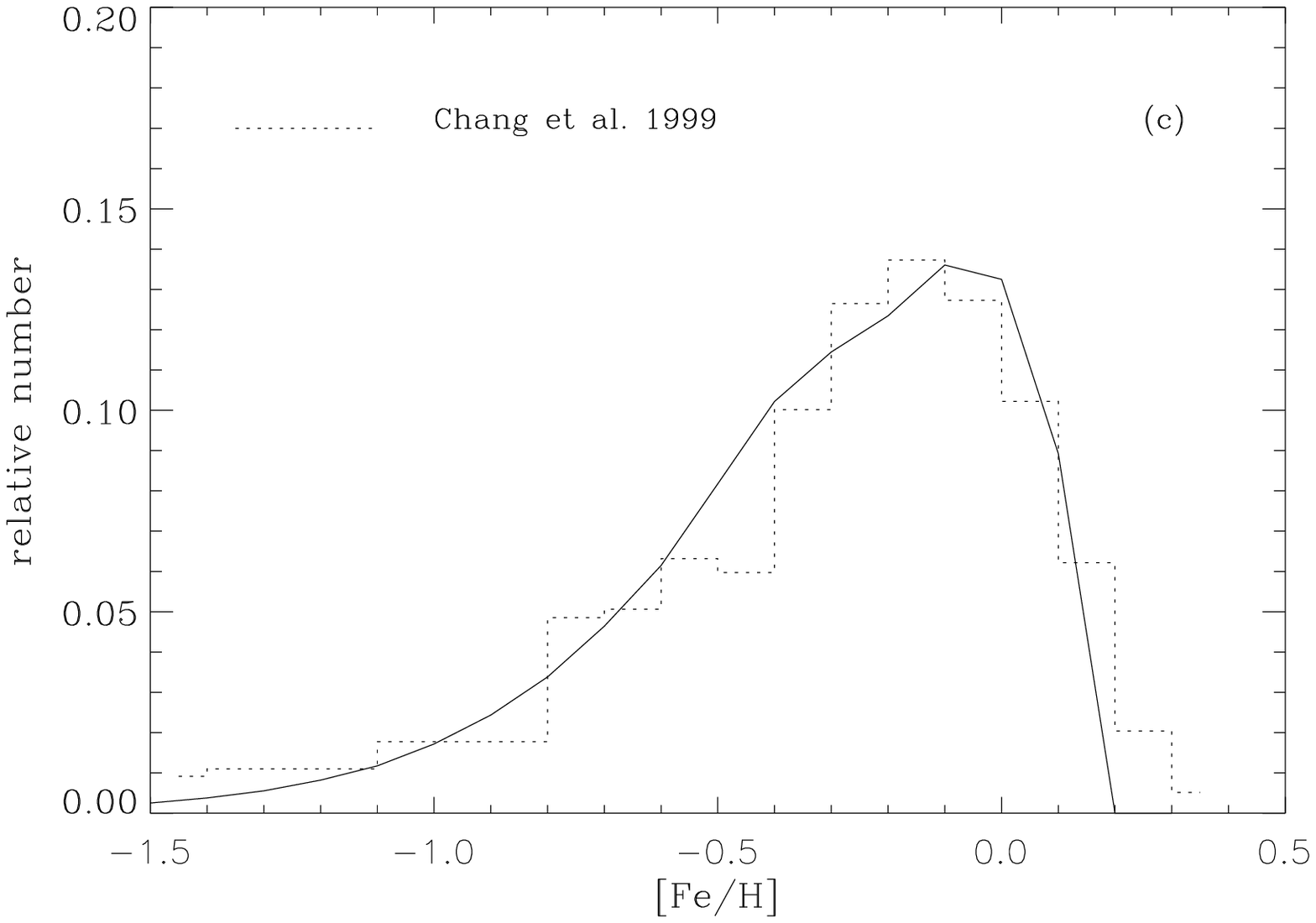}
\epsfxsize=8.8cm \epsfbox{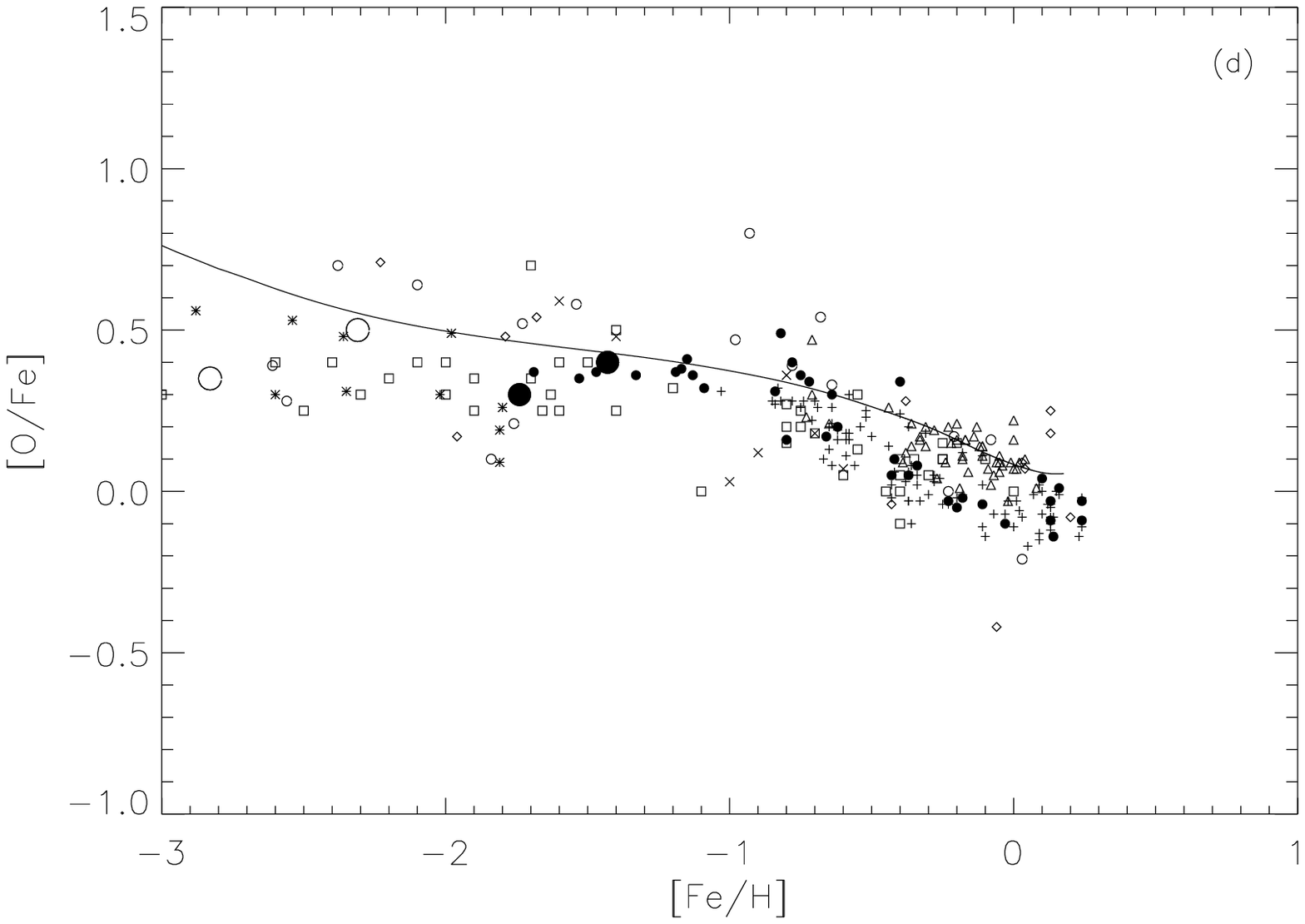}}
\vspace{0cm} 
 \caption{Same as Fig. 1, but with the nucleosynthesis yields of VG for ILMS
and N97 for massive stars (model VG+N97).}
\end{figure*}

\subsection{VG+M92}

The results predicted by model VG+M92 can fit the observed age-metallicity 
relation,
metallicity distribution of G dwarfs  and [O/Fe] vs. [Fe/H] (Figs. 3a, c, d).
There is a strong increase in the calculated [C/Fe] for [Fe/H] $>-1.2$ (Fig. 3b).
What is the main contributor?
ILMS? Stellar winds of W-R stars? Or a combined contribution of the two?
Let us try to identify the main source.

\begin{figure*}
\input epsf
\vspace{0cm}
\hbox{\hspace{0cm}\epsfxsize=8.8cm \epsfbox{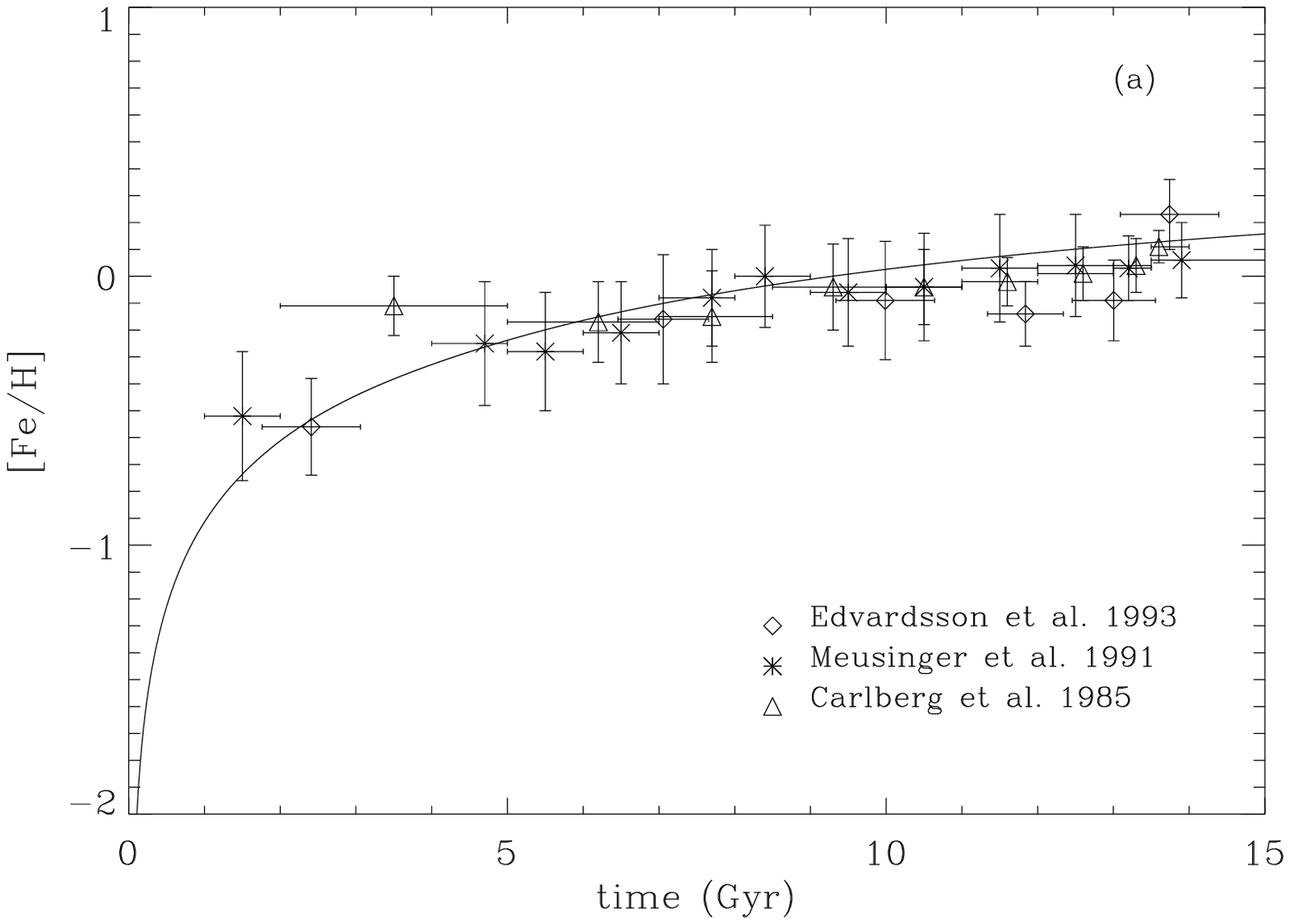}
\epsfxsize=8.8cm \epsfbox{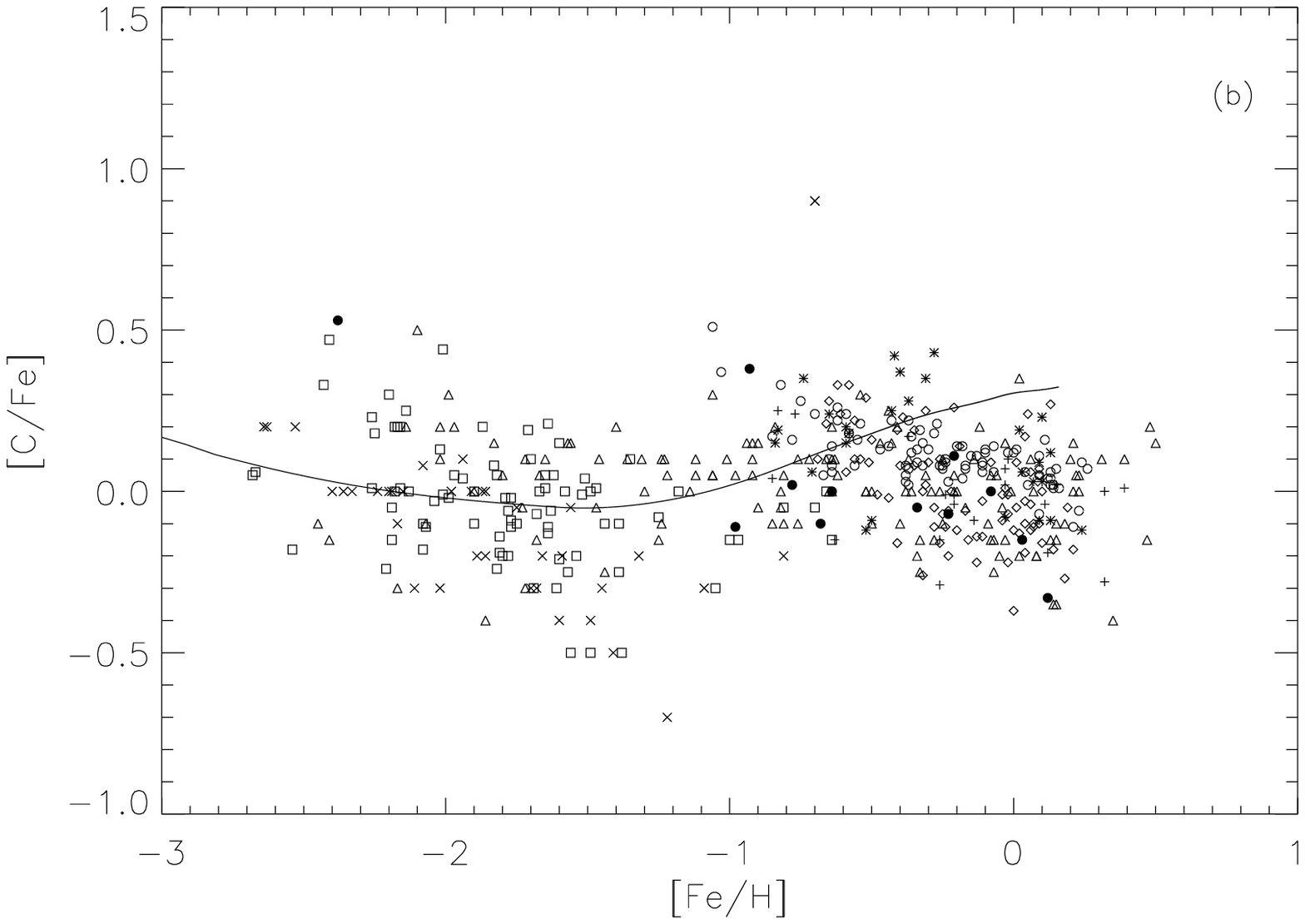}}
\vspace{0cm} 

\hbox{\hspace{0cm}\epsfxsize=8.8cm\epsfbox{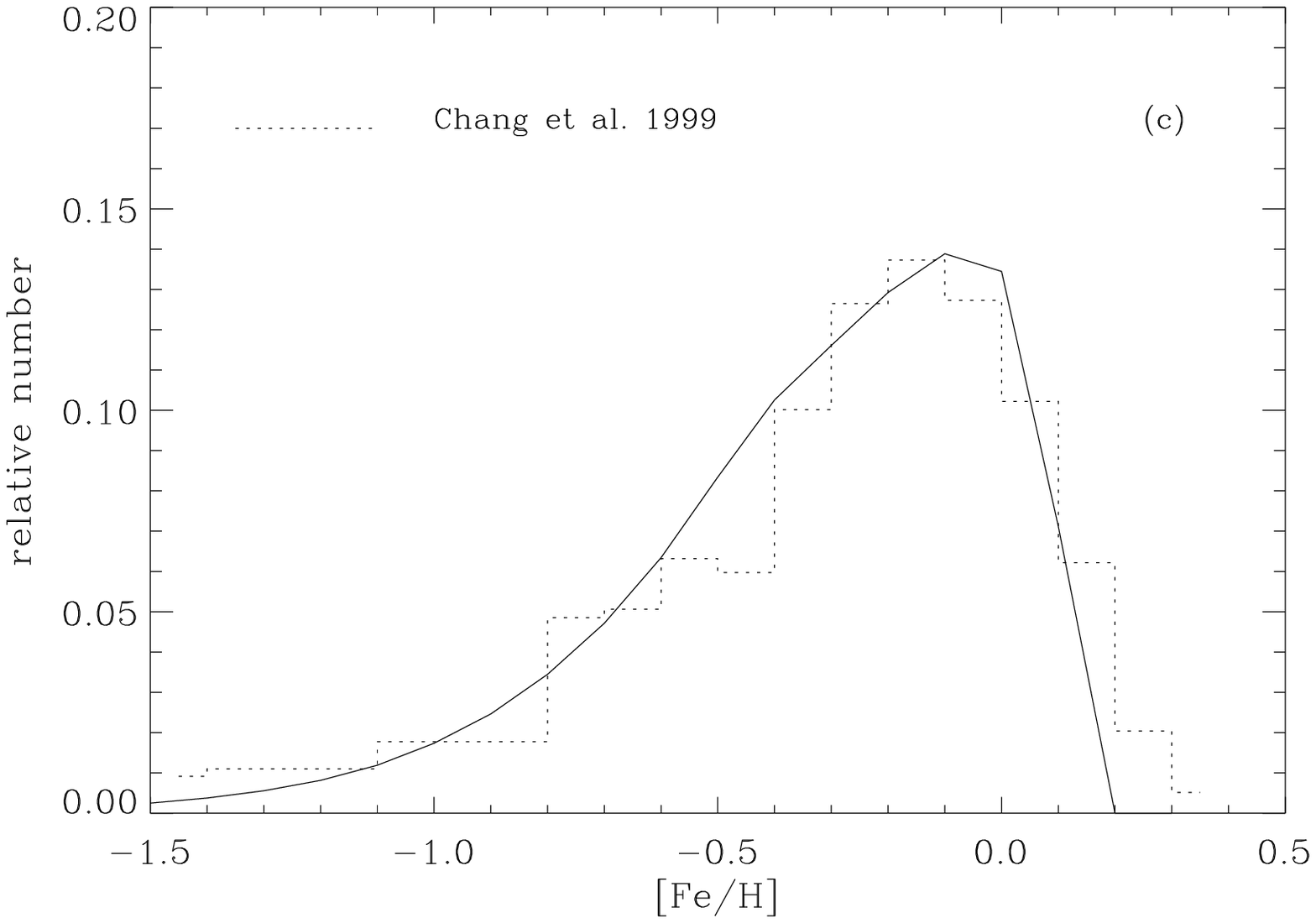}
\epsfxsize=8.8cm \epsfbox{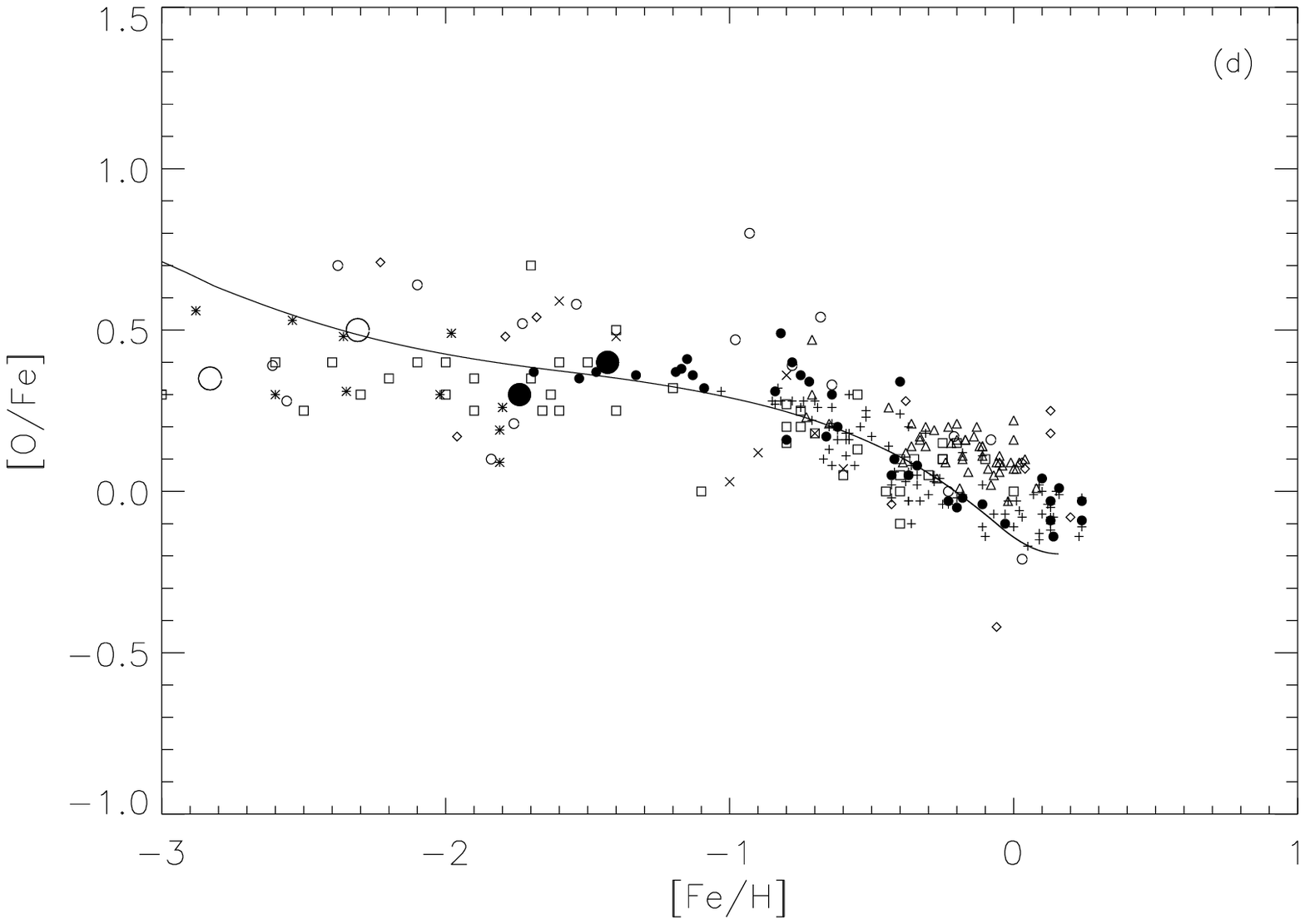}}
\vspace{0cm} 

\hbox{\hspace{0cm}\epsfxsize=8.8cm \epsfbox{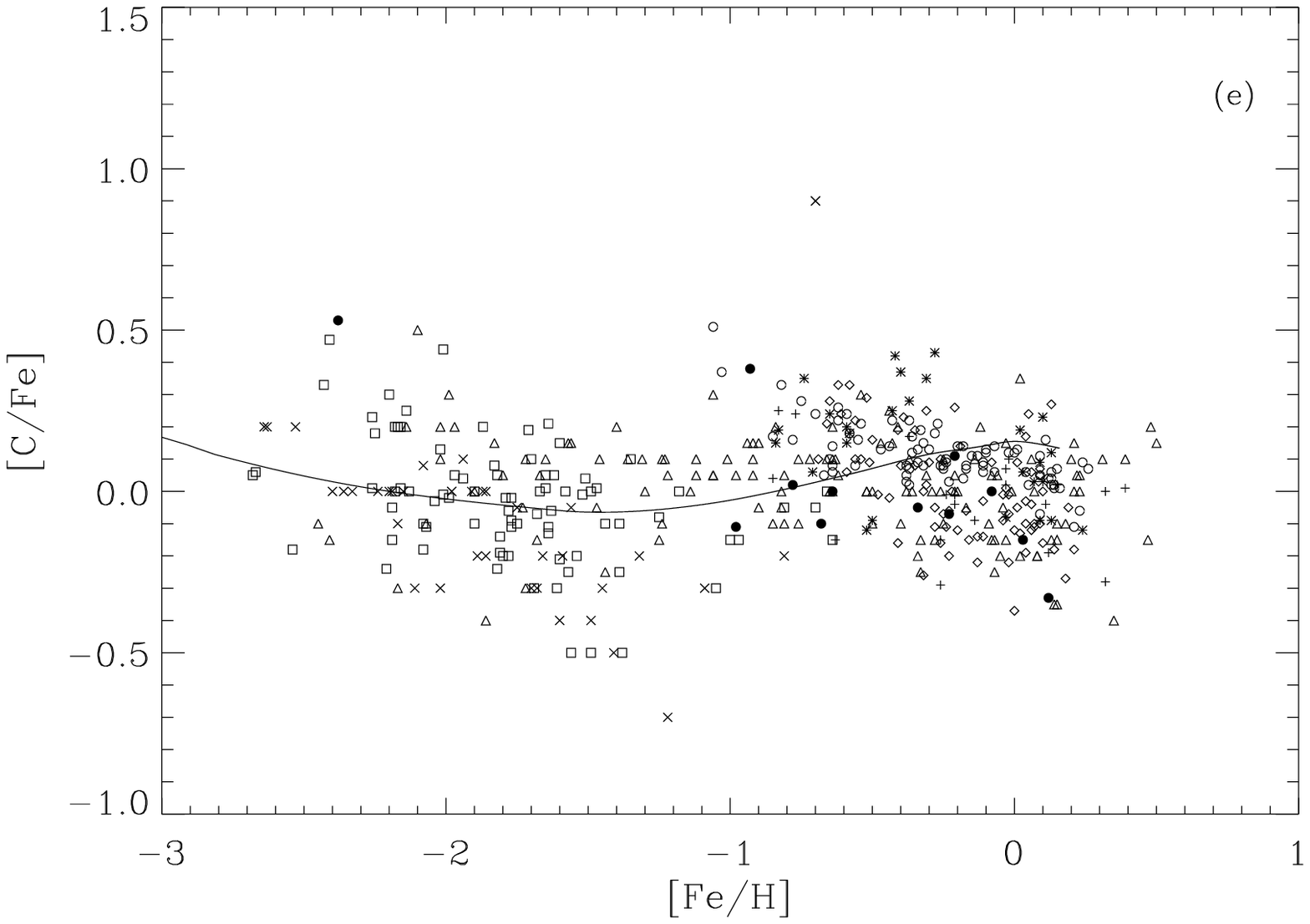}
\epsfxsize=8.8cm \epsfbox{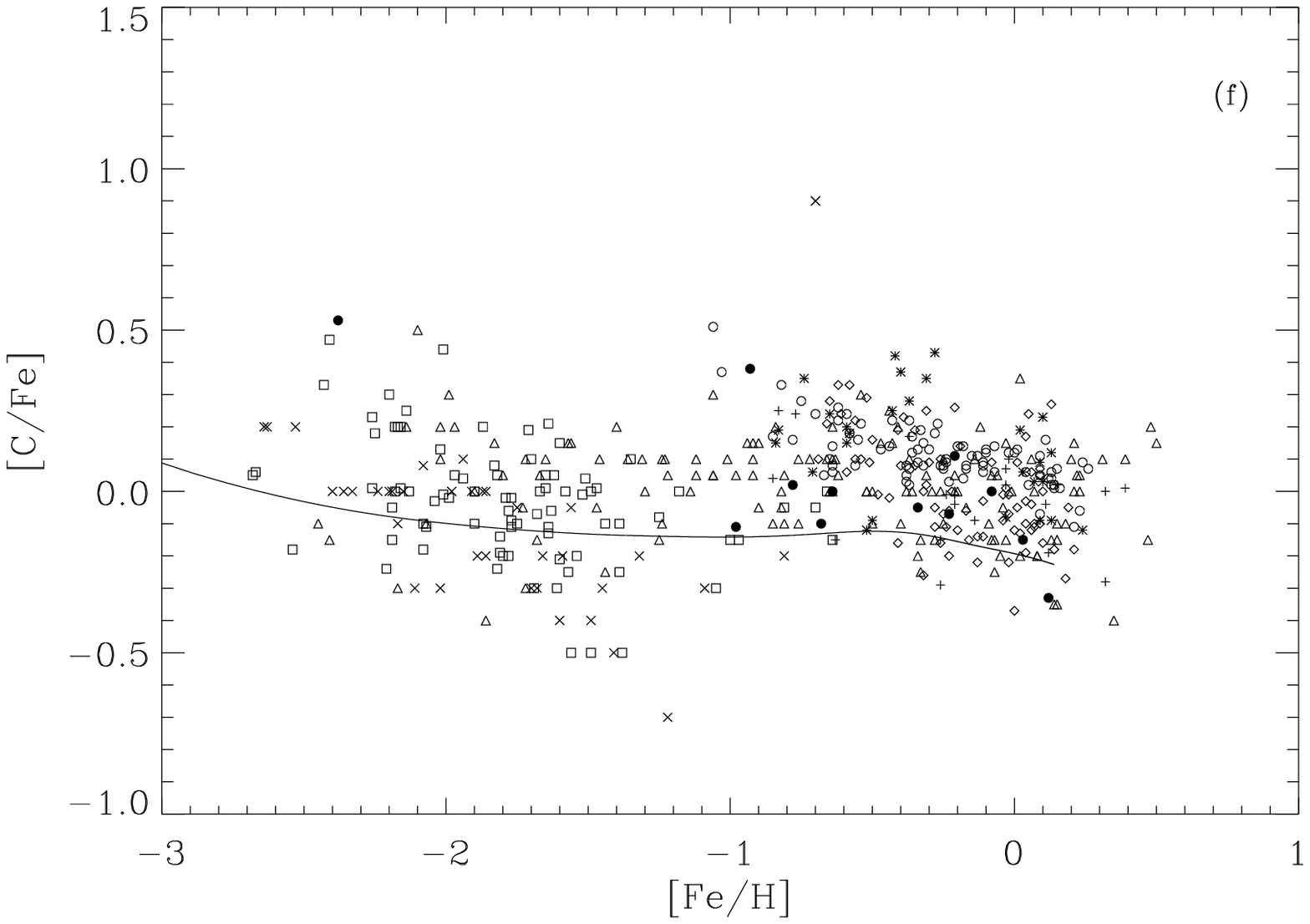}}
\vspace{0cm}
\label{fighgw}
 \caption{(a), (b), (c) and (d) are same as in Fig. 1, 
but with the nucleosynthesis yields of VG for ILMS and M92 for massive stars (model VG+M92). 
(e) The calculated [C/Fe] vs. [Fe/H] only considering
    the C yields of massive stars of M92; 
(f) The calculated [C/Fe] vs. [Fe/H] considering C yields of VG for ILMS and
    C yields of only $M\leq 40M_{\odot}$ stars from M92 for massive stars.}
\end{figure*}

Fig. 3e gives the result for only the yields of massive stars of M92.
It shows the predicted [C/Fe] increases with metallicity 
in metal-rich region. 
The calculated [C/Fe] using the yields of ILMS (VG) and only massive stars with 
$M\leq 40M_{\odot}$ (M92) is
presented in Fig. 3f: here, the calculated [C/Fe] 
is slightly lower than the observations in metal-rich region. 
The difference between Fig. 3e and Fig. 3f 
shows that the contribution to carbon 
from stellar wind of high metallicity W-R stars calculated by 
M92 is greater than the contribution from ILMS given by VG. 
The  
$M>40M_{\odot}$ massive stars with higher metallicity, the W-R stars,
can eject significant amounts of carbon, which
causes the [C/Fe] value to increase with increasing 
metallicity in the range of [Fe/H] $>-1.0$ in Fig. 3e.

\subsection{RV+WW}

Using results of RV as the yields of ILMS and WW for massive stars (model RV+WW),
we calculated the corresponding age-metallicity relation, metallicity distribution
of G dwarfs, [O/Fe] vs. [Fe/H] and [C/Fe] vs. [Fe/H] (Figs. 4a, c, d, b). 
When these three
results match the observations, [C/Fe] first increases slightly and then decreases,
and is slightly higher than 
the prediction of model VG+WW in metal-rich region. 
These results show that the RV calculation gives more carbon than does VG
due to the use of different parameter values.
(see M2K, Liang \& Zhao \cite{liang21} for details). 

\begin{figure*}
\input epsf
\vspace{0cm}
\hbox{\hspace{0cm}\epsfxsize=8.8cm \epsfbox{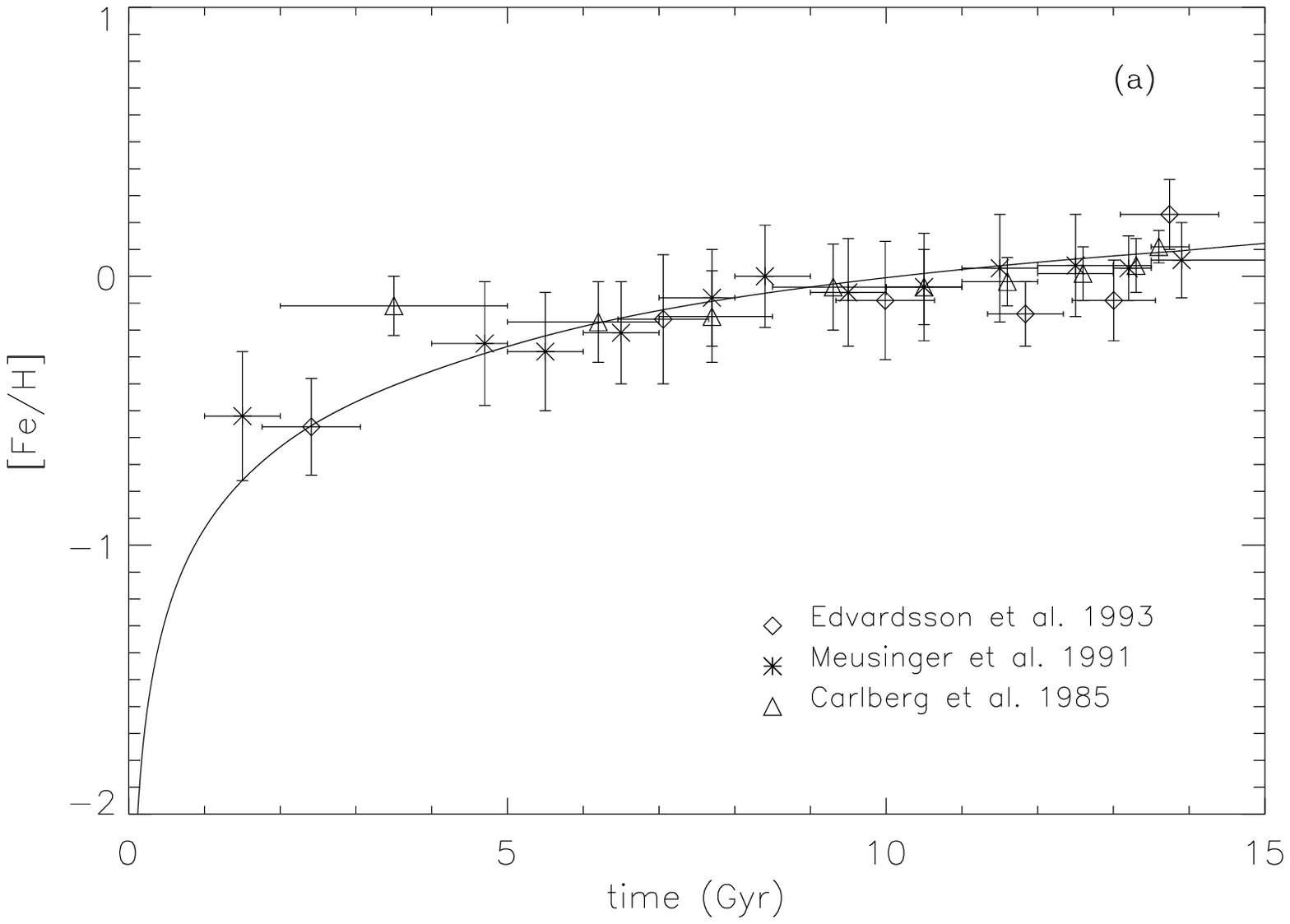}
\epsfxsize=8.8cm \epsfbox{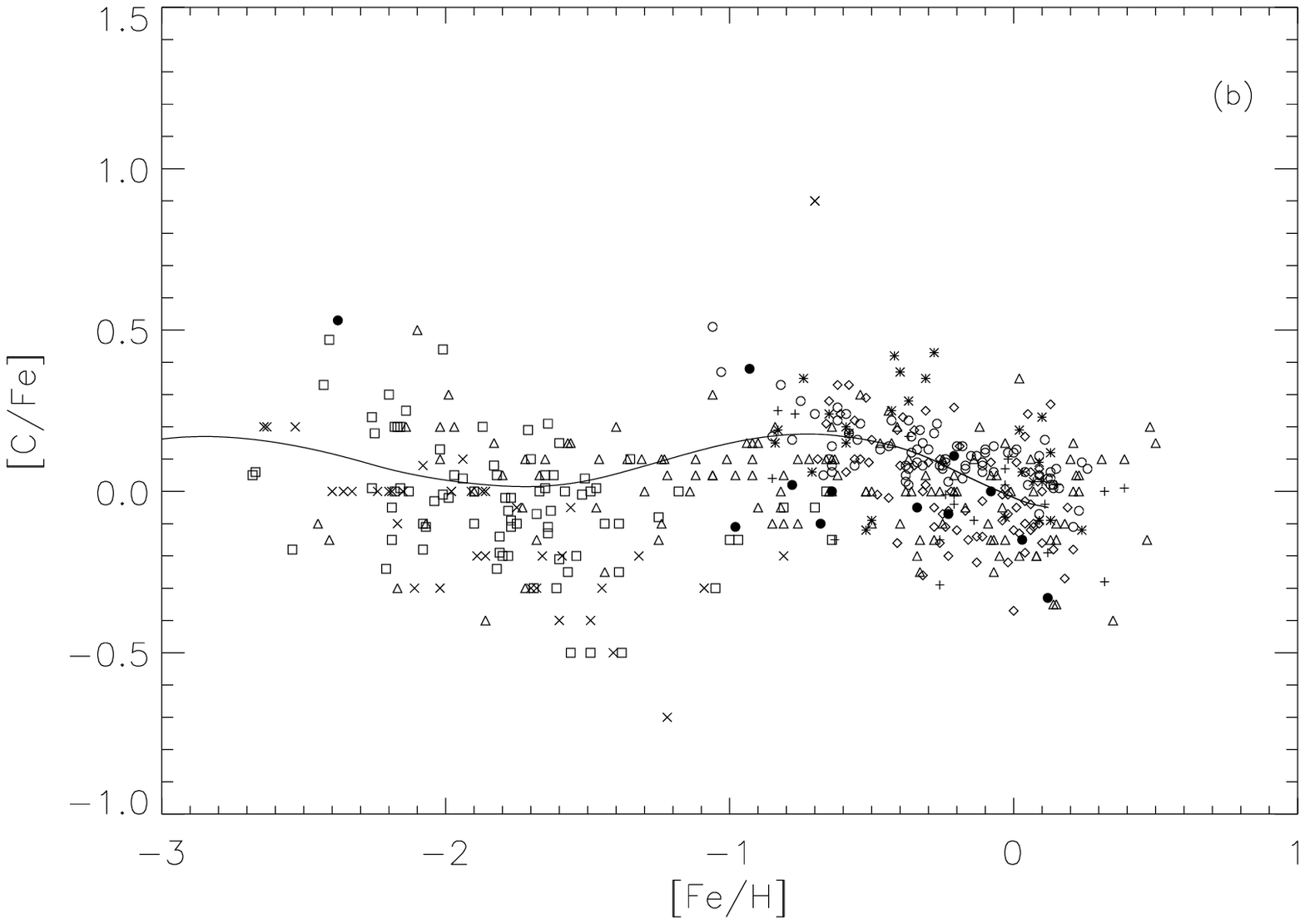}}
\vspace{0cm} 
\hbox{\hspace{0cm}\epsfxsize=8.8cm\epsfbox{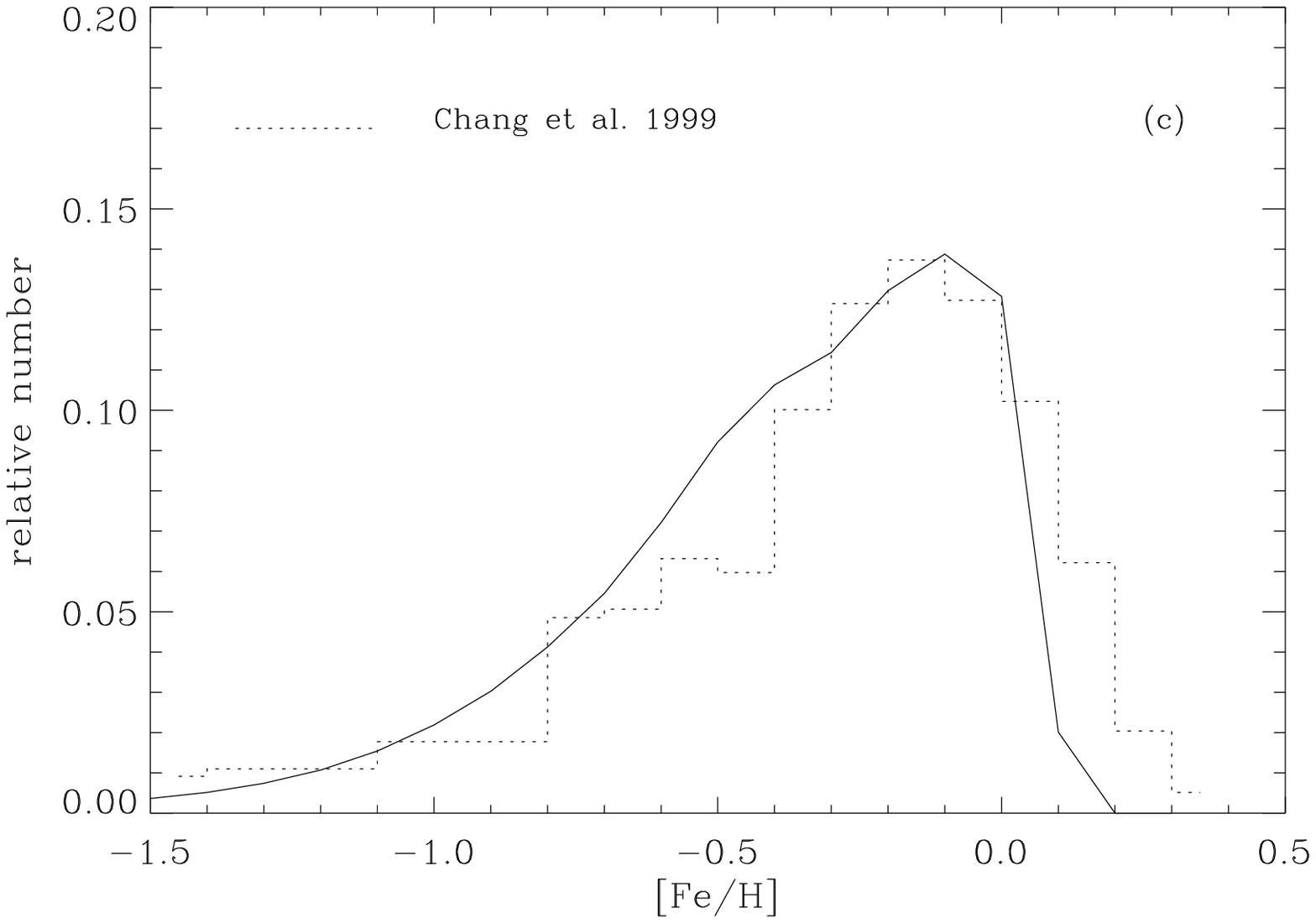}
\epsfxsize=8.8cm \epsfbox{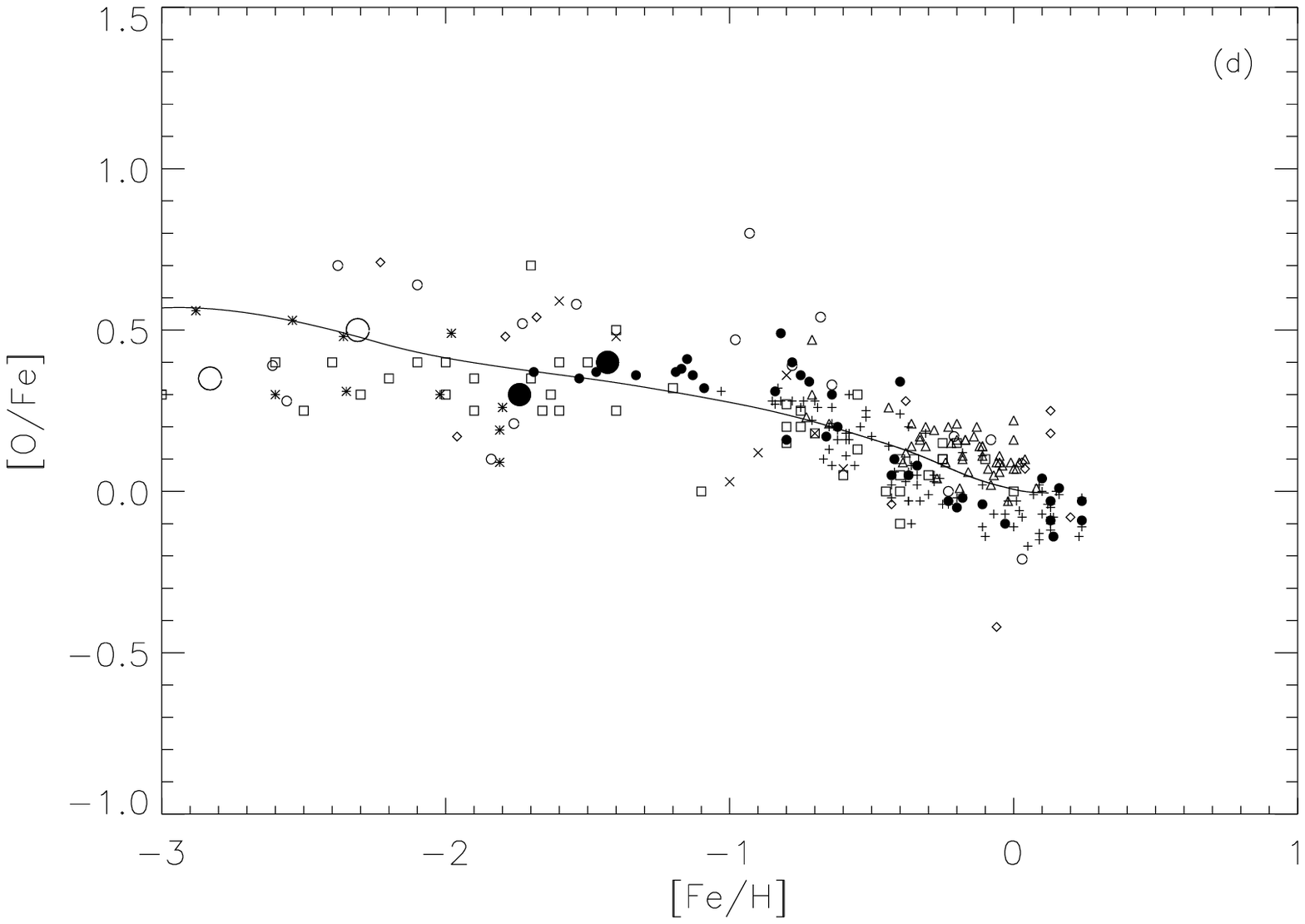}}
\vspace{0cm} 
\label{figrvw}
 \caption{Same as Fig. 1, but with the nucleosynthesis yields of RV for ILMS
and WW for massive stars (model RV+WW).}
\end{figure*}

\subsection{RV+N97}

Figs. 5a, c, d show that the calculated results using model RV+N97 
can fit the observations,
i.e., the age-metallicity relation, the metallicity distribution of G dwarfs
and the [O/Fe] vs. [Fe/H] relation.
The calculated [C/Fe] vs. [Fe/H] then shows that
more carbon is given by RV than VG, and this 
increases the predicted [C/Fe] in metal-rich region (Fig. 5b).
Because N97 gives low carbon yields for the reason of adopting a low 
$^{12}$C$(\alpha,\gamma)^{16}$O
reaction rate, 
[C/Fe] falls below the observations in the whole metallicity range  (Fig. 5b).

\begin{figure*}
\input epsf
\vspace{0cm}
\hbox{\hspace{0cm}\epsfxsize=8.8cm \epsfbox{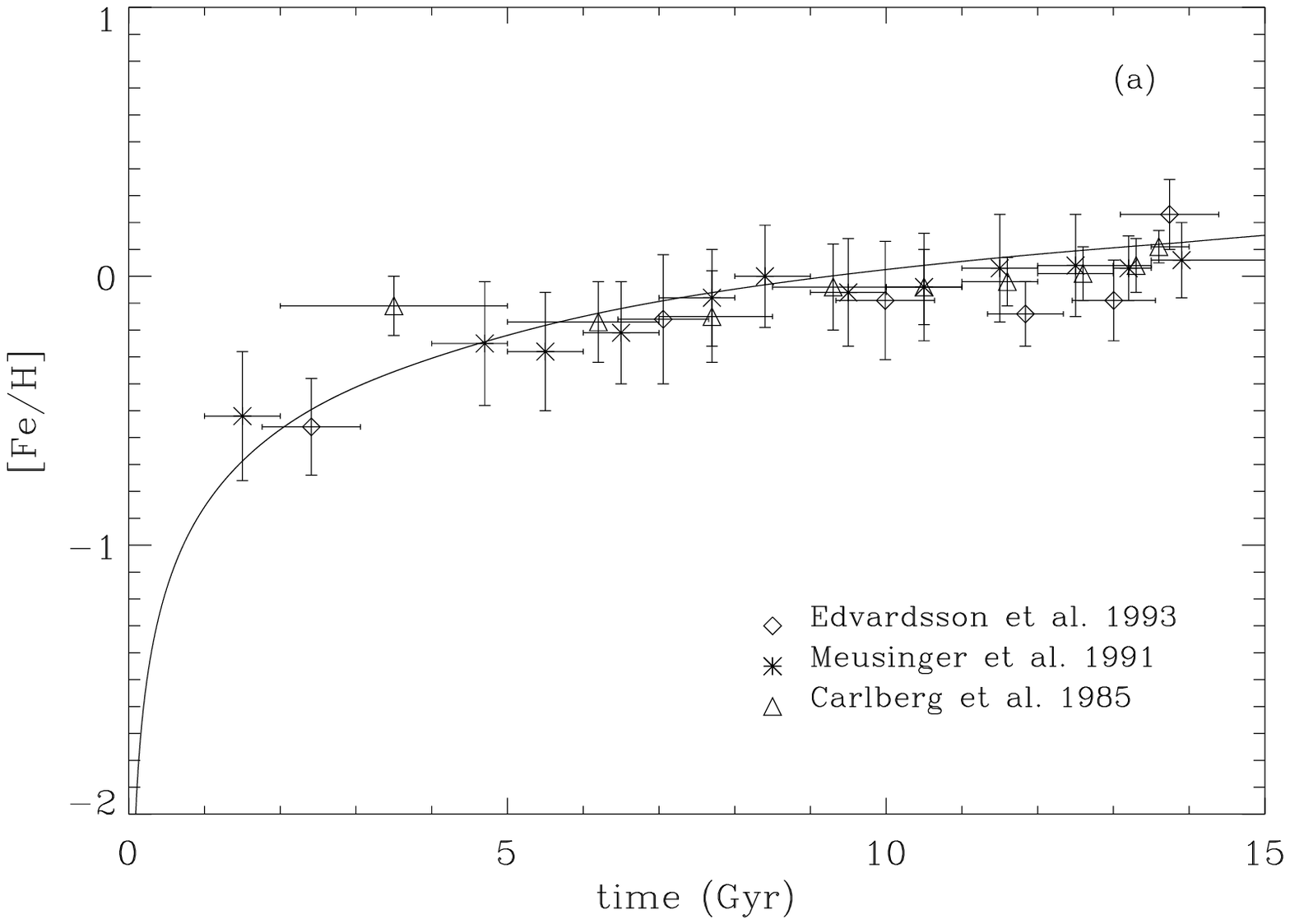}
\epsfxsize=8.8cm \epsfbox{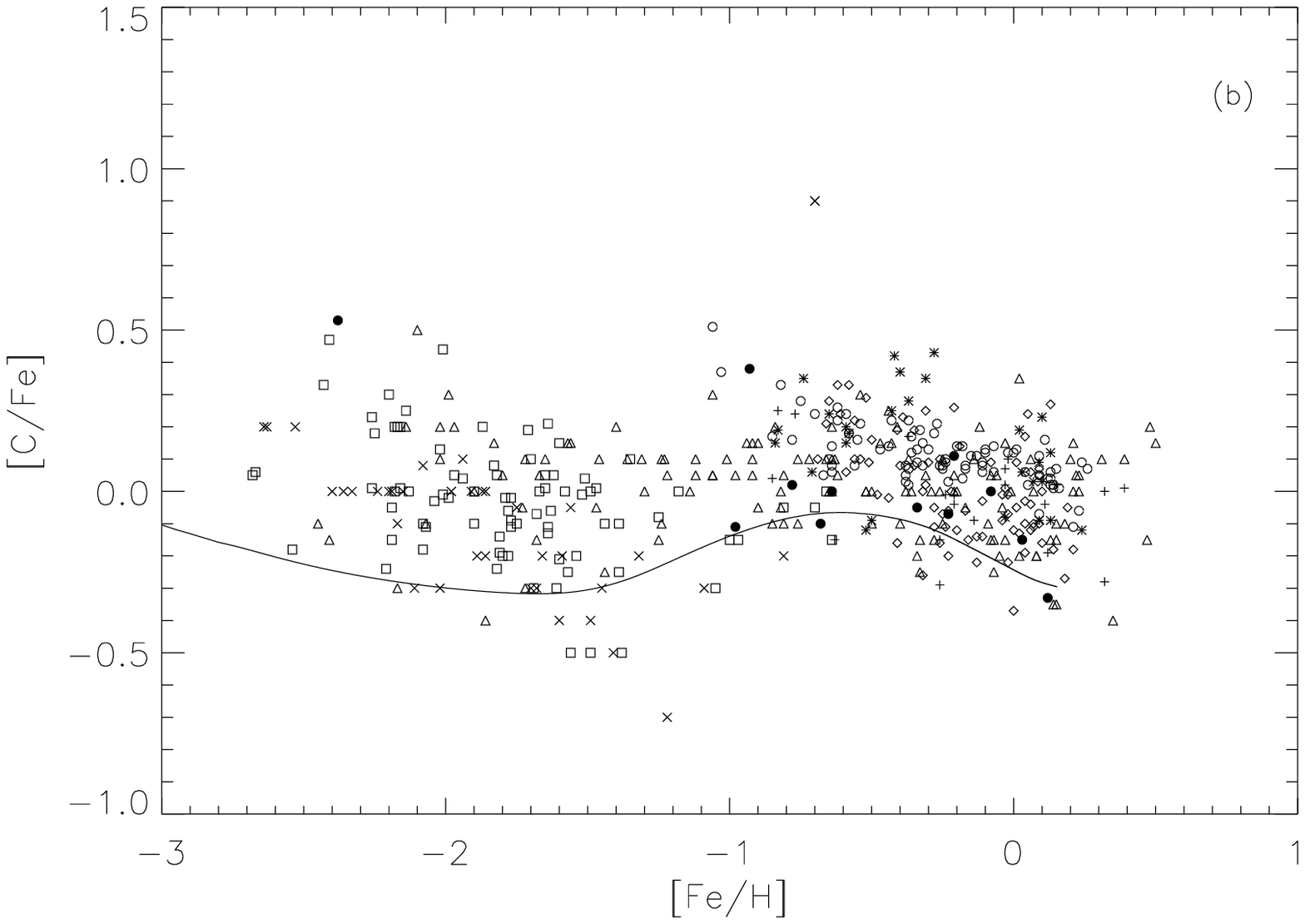}}
\vspace{0cm} 
\hbox{\hspace{0cm}\epsfxsize=8.8cm\epsfbox{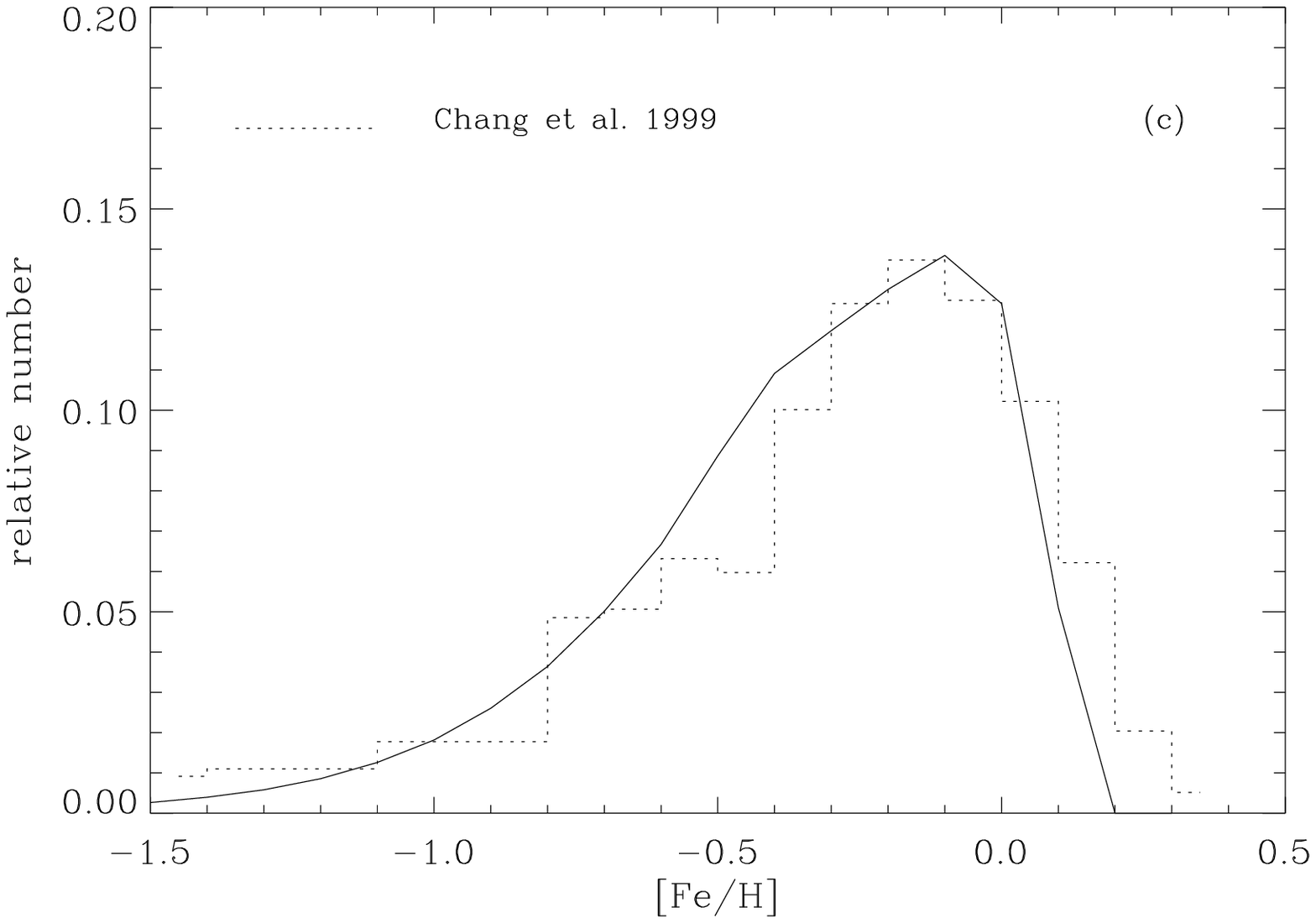}
\epsfxsize=8.8cm \epsfbox{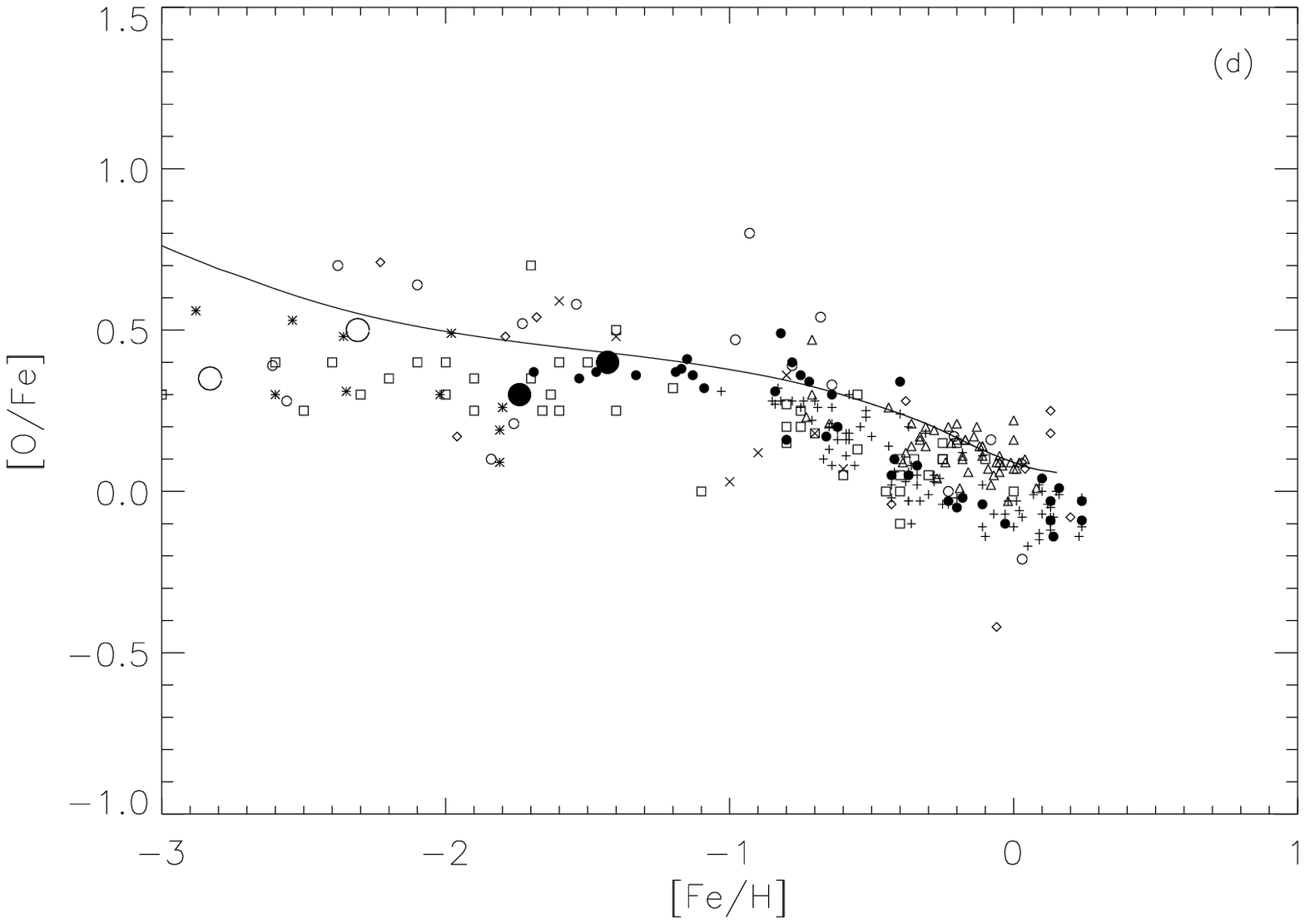}}
\vspace{0cm}
 \caption{Same as Fig. 1, but with the nucleosynthesis yields of RV for ILMS
and N97 for massive stars (model RV+N97).} 
\end{figure*}

\subsection{ RV+M92}

Figs. 6a$-$d display the results based on the yields of RV for ILMS 
and of M92 for massive stars (model RV+M92).
Figs. 6a, c, d show the calculations can fit the corresponding observations, while Fig. 6b shows that 
[C/Fe] increases with metallicity in metal-rich region (Fig. 6b). 
There are no apparent difference here between this model and 
the model VG+M92 (Fig. 3b);
this can be 
understood because the C contribution 
from W-R stars given by M92 is 
higher than that from ILMS given by RV or VG.
 
Fig. 6e shows the [C/Fe] vs. [Fe/H] relation based only on the 
yields of massive stars
from M92; with no contribution from ILMS, 
this is still higher than the observations in metal-rich region.
Fig. 6f exhibits the predicted [C/Fe] using the yields of RV for ILMS and those of M92 
for massive stars with $M\leq 40M_{\odot}$; it  
fits the observations, and shows a slightly higher [C/Fe]
than is given by model VG+M92 in metal-rich region (Fig. 3b).
This result shows that RV gives higher C yields
than does VG with a different choice of the parameters. 

\begin{figure*}
\input epsf
\vspace{0cm}
\hbox{\hspace{0cm}\epsfxsize=8.8cm \epsfbox{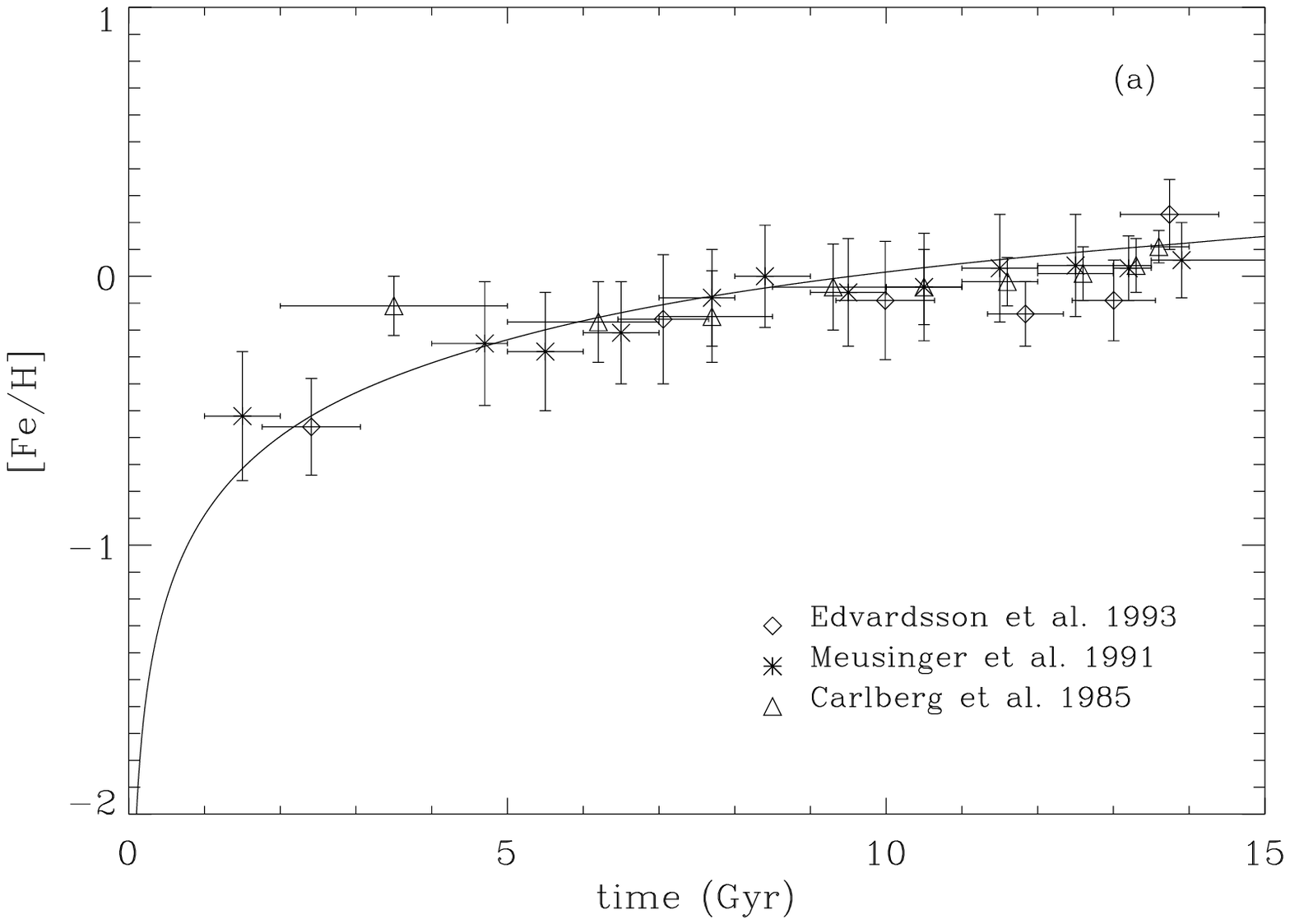}
\epsfxsize=8.8cm \epsfbox{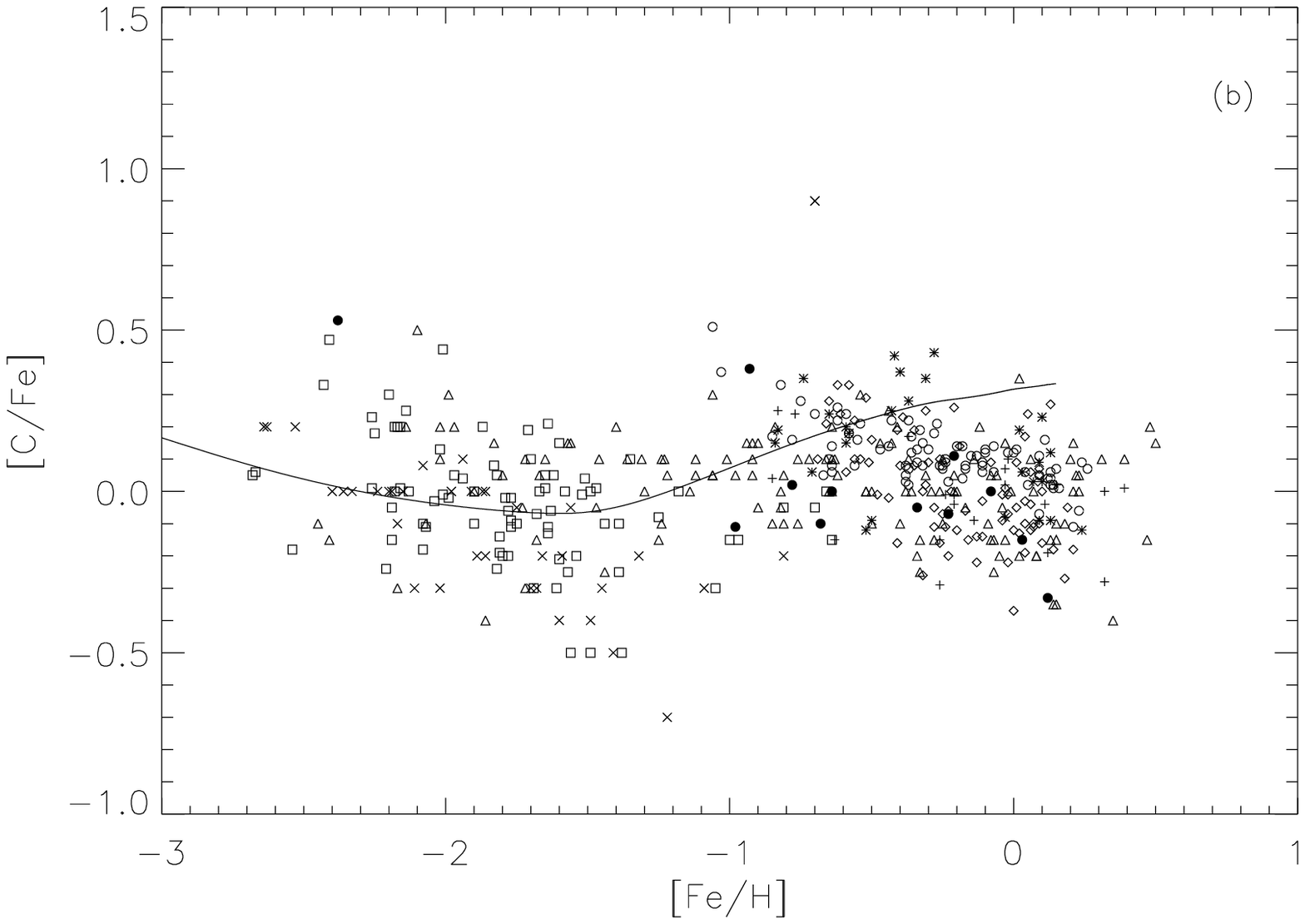}}
\vspace{0cm} 
\hbox{\hspace{0cm}\epsfxsize=8.8cm\epsfbox{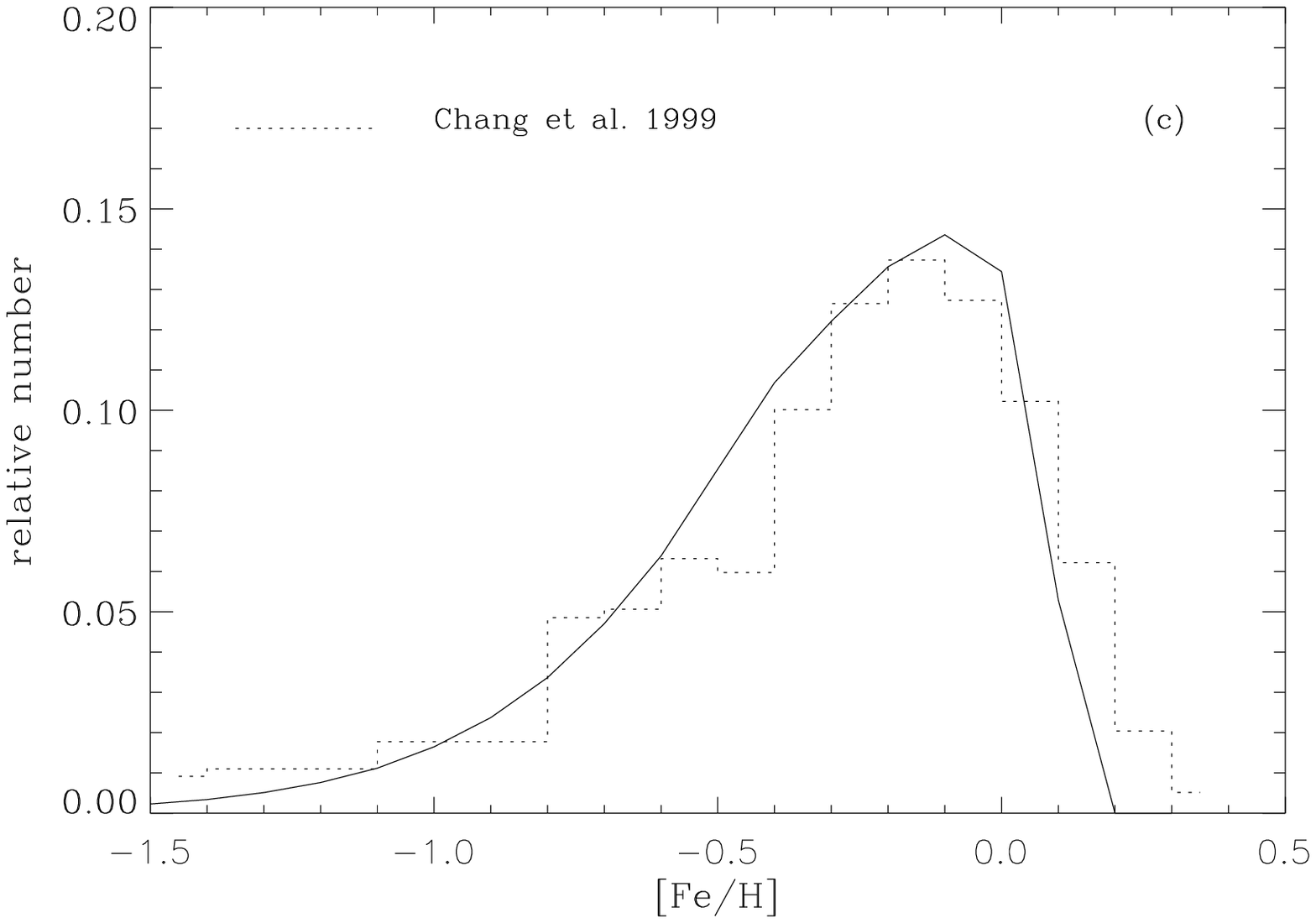}
\epsfxsize=8.8cm \epsfbox{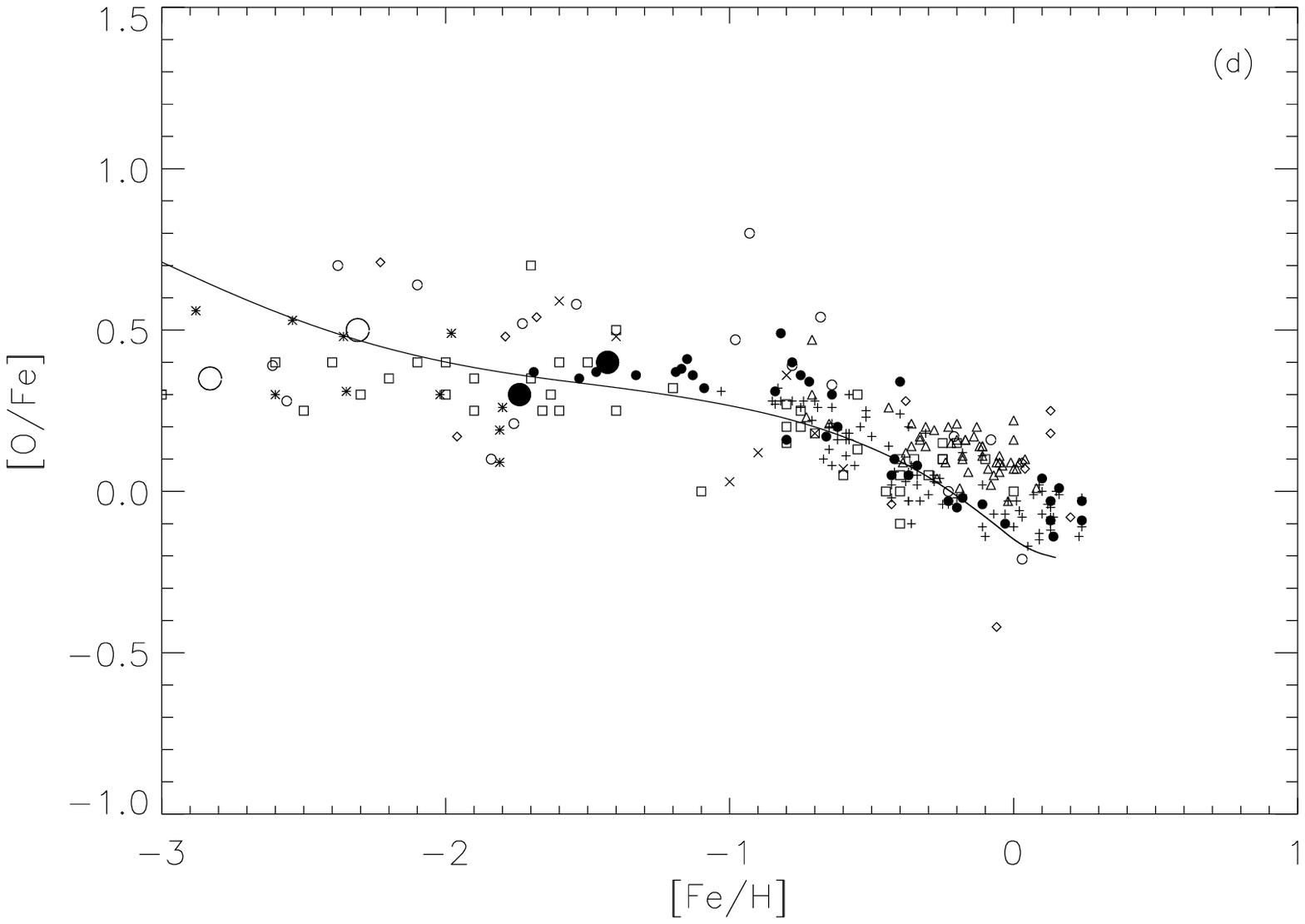}}
\vspace{0cm} 
\hbox{\hspace{0cm}\epsfxsize=8.8cm\epsfbox{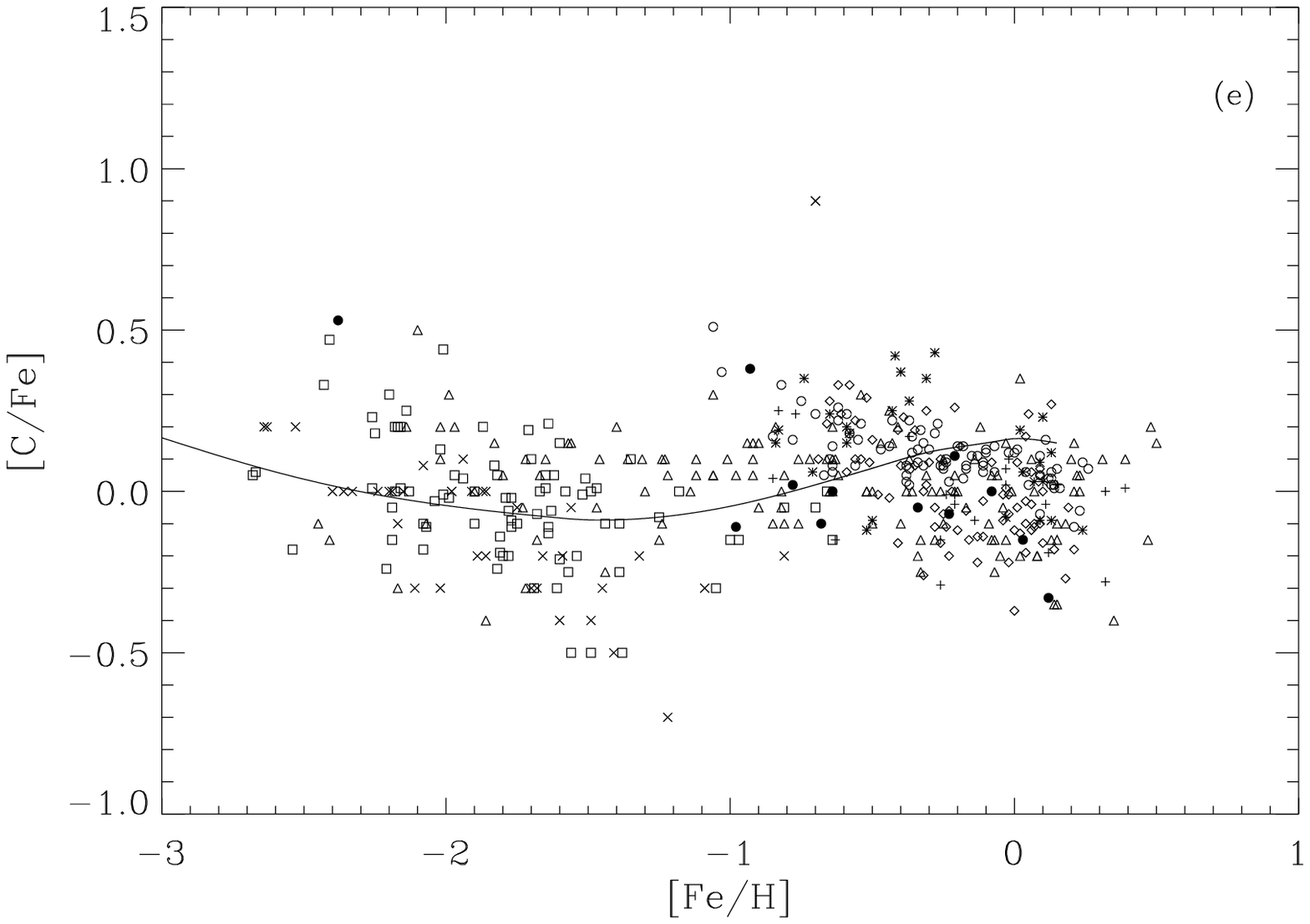}
\epsfxsize=8.8cm \epsfbox{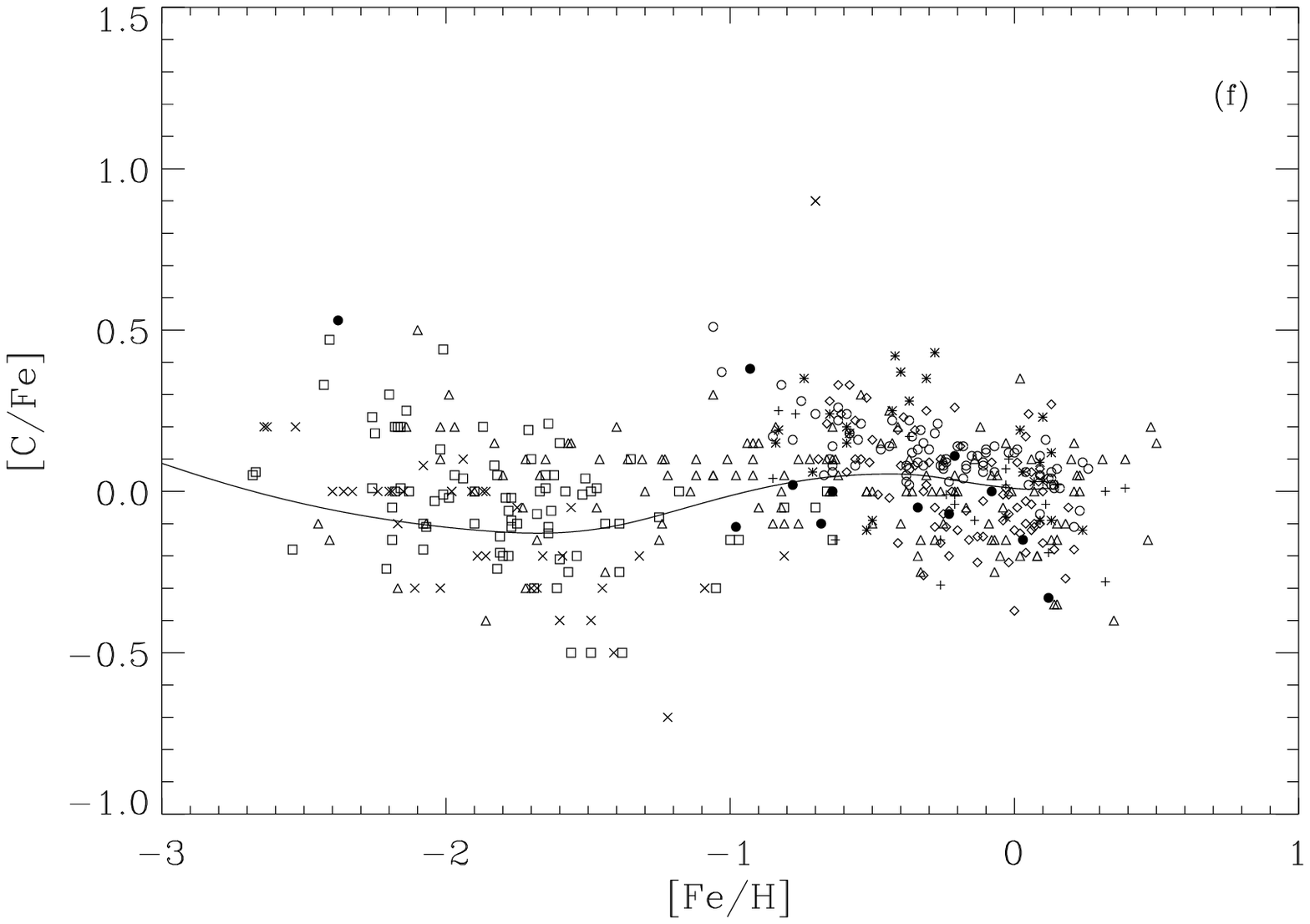}}
 \caption{Same as Fig. 3, but with nucleosynthesis yields of RV for ILMS 
           and M92 for massive stars (model RV+M92).} 
\end{figure*}

\subsection{MBC+PCB}

It is proper to combine the yields of MBC and PCB 
together since they used 
the same parameters (Padova group).
They adopted a larger
convective overshooting, so that stars with $M>6M_{\odot}$ rather than
the standard $>$8$M_{\odot}$ would end their lives through SN\,II explosions.
Both the age-metallicity relation and metallicity distribution of G dwarfs
can fit the observations well (Figs. 7a, c).
The O abundance is slightly lower than the observations (Fig. 7d), which may be caused by 
the lower $^{12}$C$(\alpha,\gamma)^{16}$O reaction rate (CF) used by the 
Padova group.  
The predicted [C/Fe] vs. [Fe/H] relation is given in Fig. 7b, which
shows a strong increase of [C/Fe] with increasing metallicity 
from [Fe/H] $\approx -0.7$ on.

Fig. 7e displays the [C/Fe] based only on the contribution of carbon from massive stars; it 
shows that W-R stars with a high metallicity ($Z=0.008, 0.02$) can
eject significant amounts of carbon into the ISM, much more than the lower metallicity stars. 
Thus, [C/Fe] increases steeply starting from [Fe/H] $\approx -0.7$.
The subsequent decrease from [Fe/H] $\approx -0.3$ on is 
the result of Fe contribution from SN\,Ia explosions. 

Fig. 7f shows the case of only ILMS and the $M \leq 40M_{\odot}$ 
part of massive stars; there is basic 
fit to the observations. 
Comparing Figs. 7b, 7e and 7f, we find the main effect of the C yields from metal-rich W-R stars
is the obvious increase of [C/Fe] in metal-rich region shown
in Fig. 7b;---an increase to above the 
observations.

\begin{figure*}
\input epsf
\vspace{0cm}
\hbox{\hspace{0cm}\epsfxsize=8.8cm \epsfbox{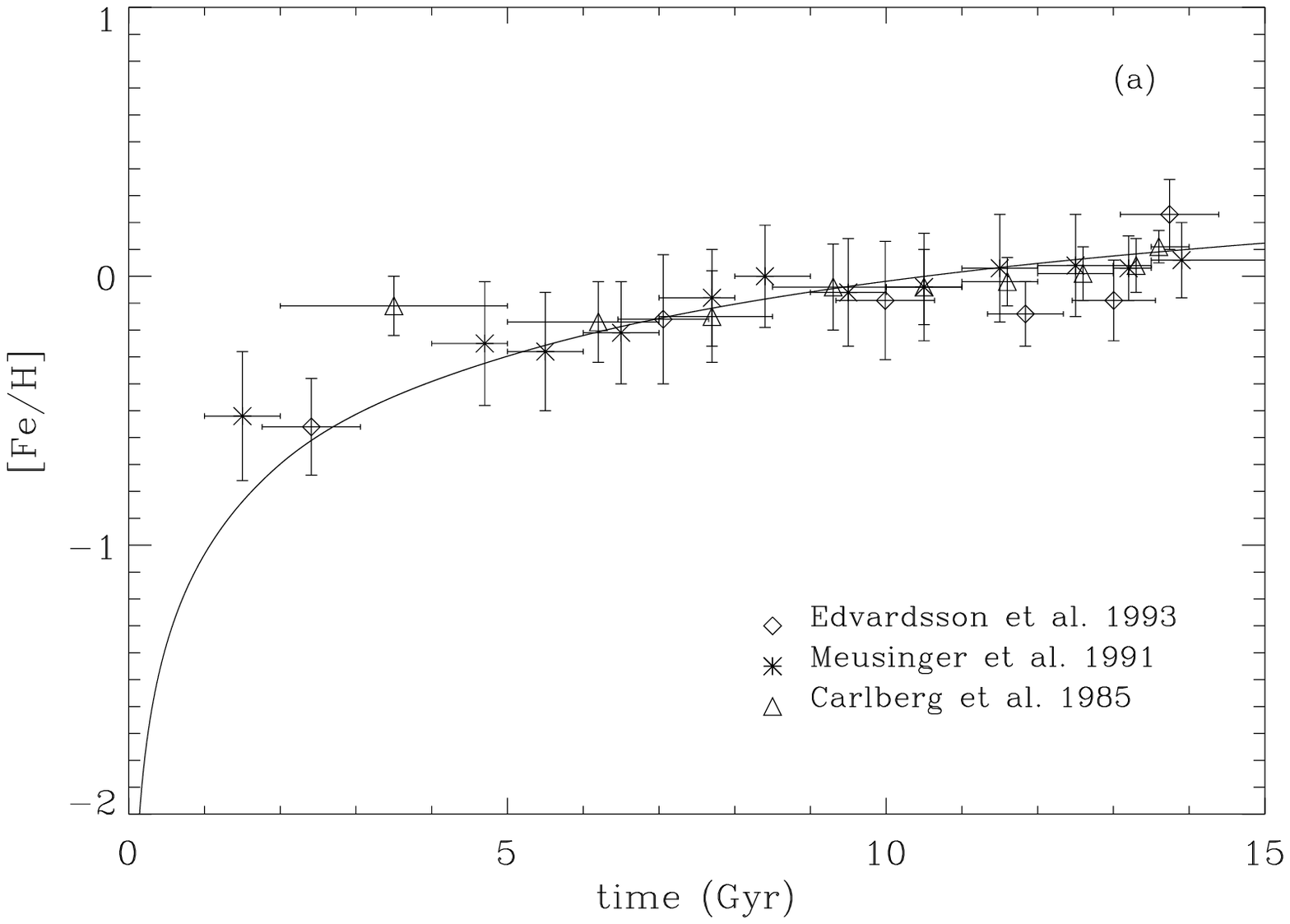}
\epsfxsize=8.8cm \epsfbox{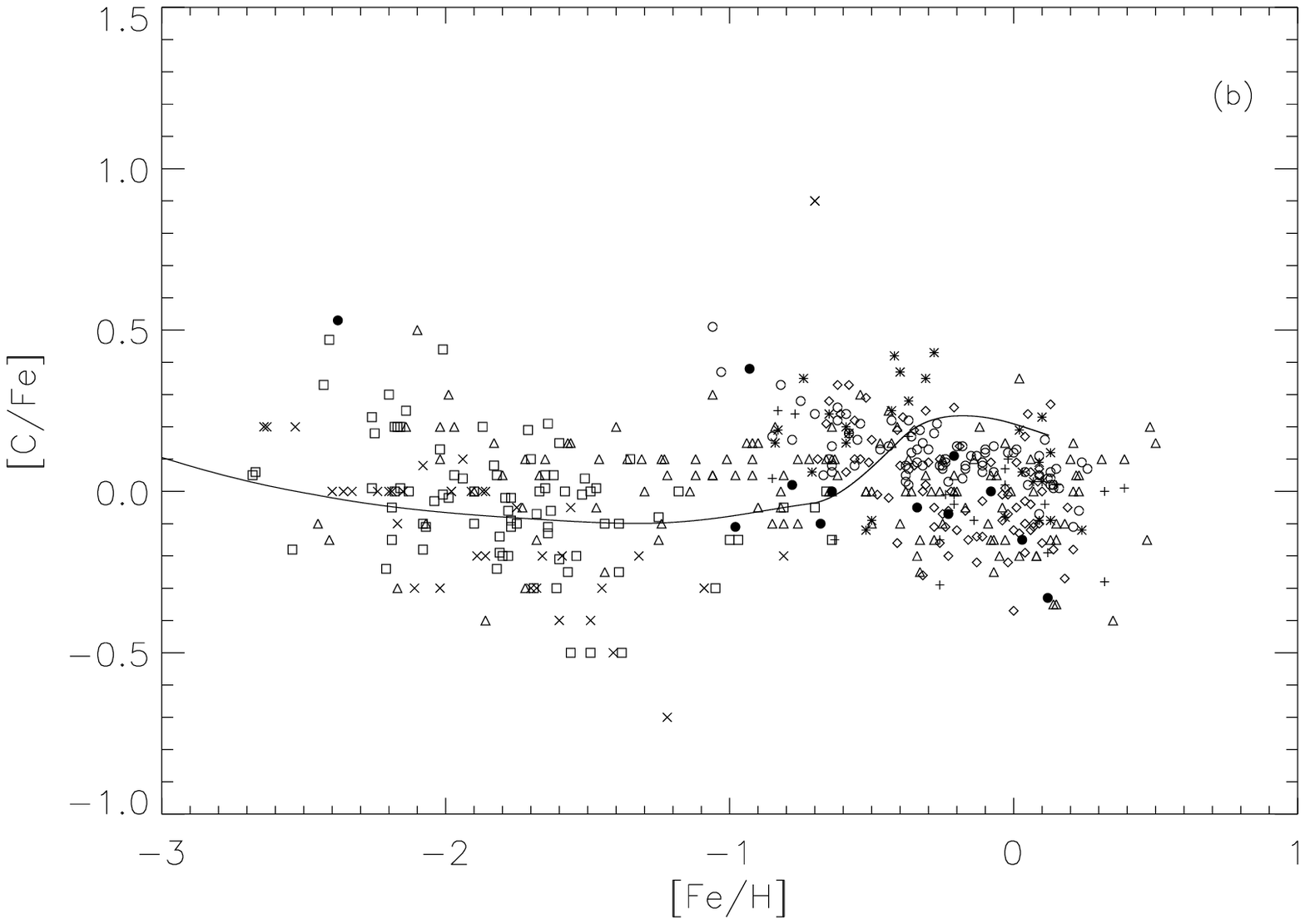}}
\vspace{0cm} 
\hbox{\hspace{0cm}\epsfxsize=8.8cm\epsfbox{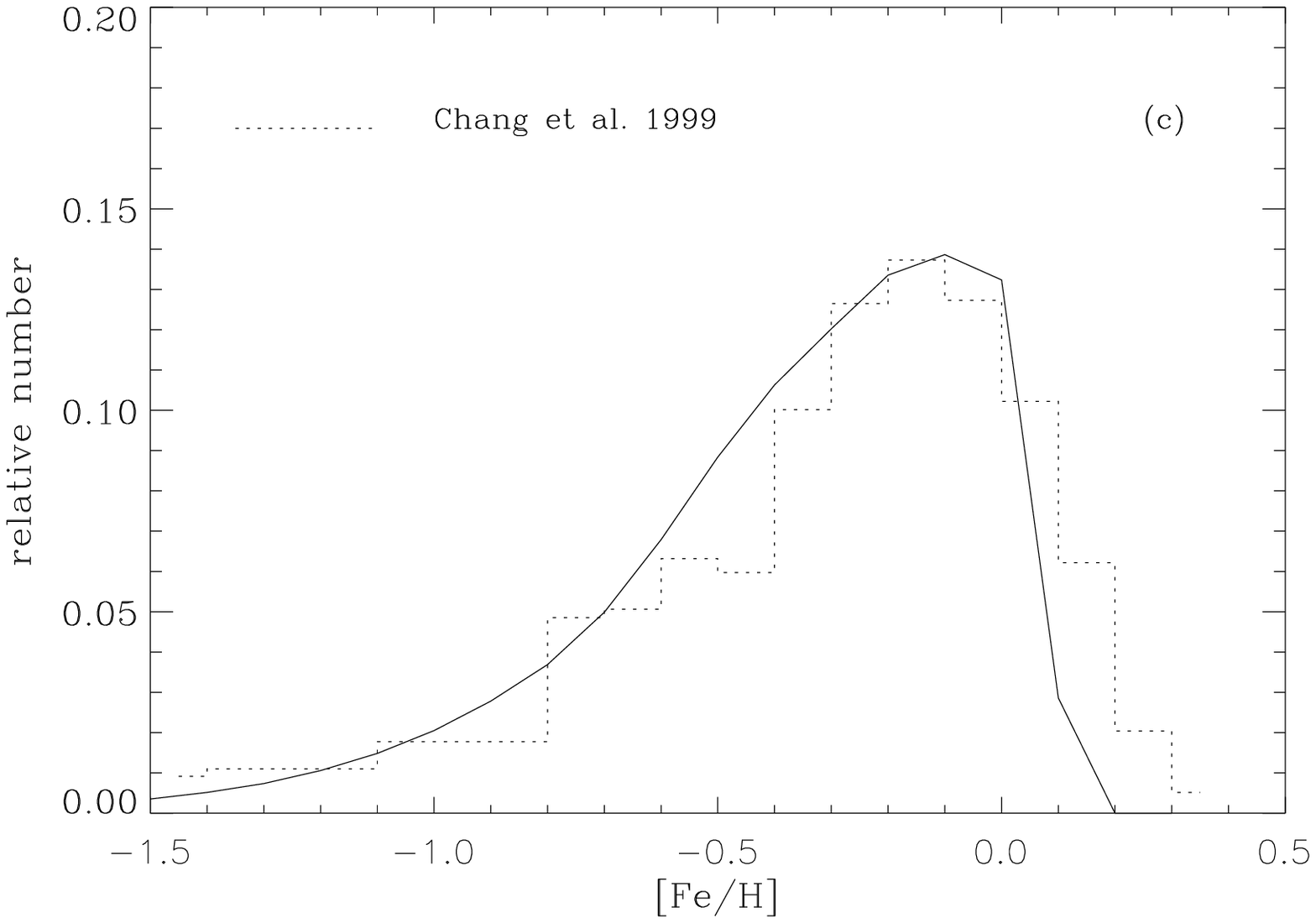}
\epsfxsize=8.8cm \epsfbox{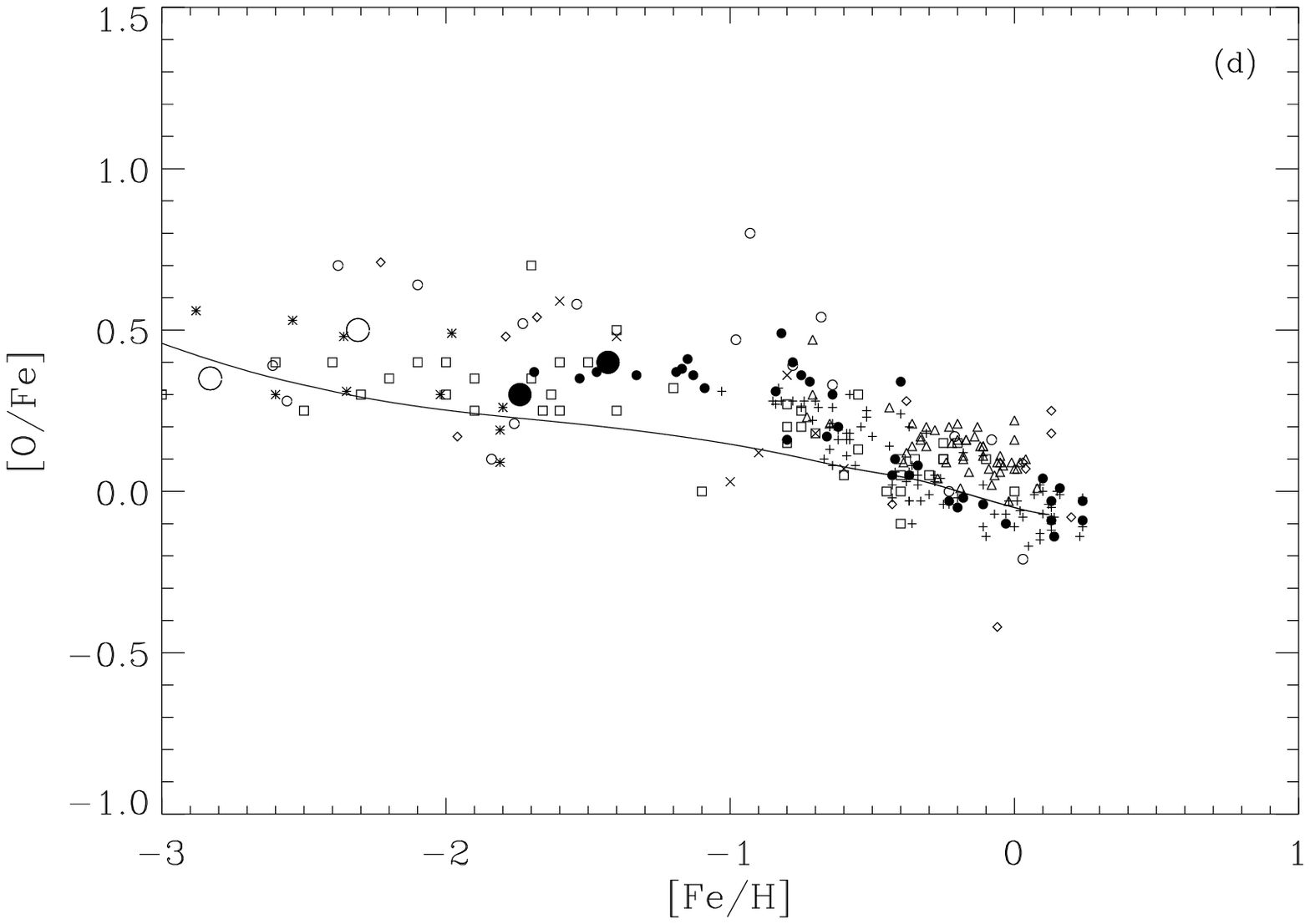}}
\vspace{0cm} 
\hbox{\hspace{0cm}\epsfxsize=8.8cm \epsfbox{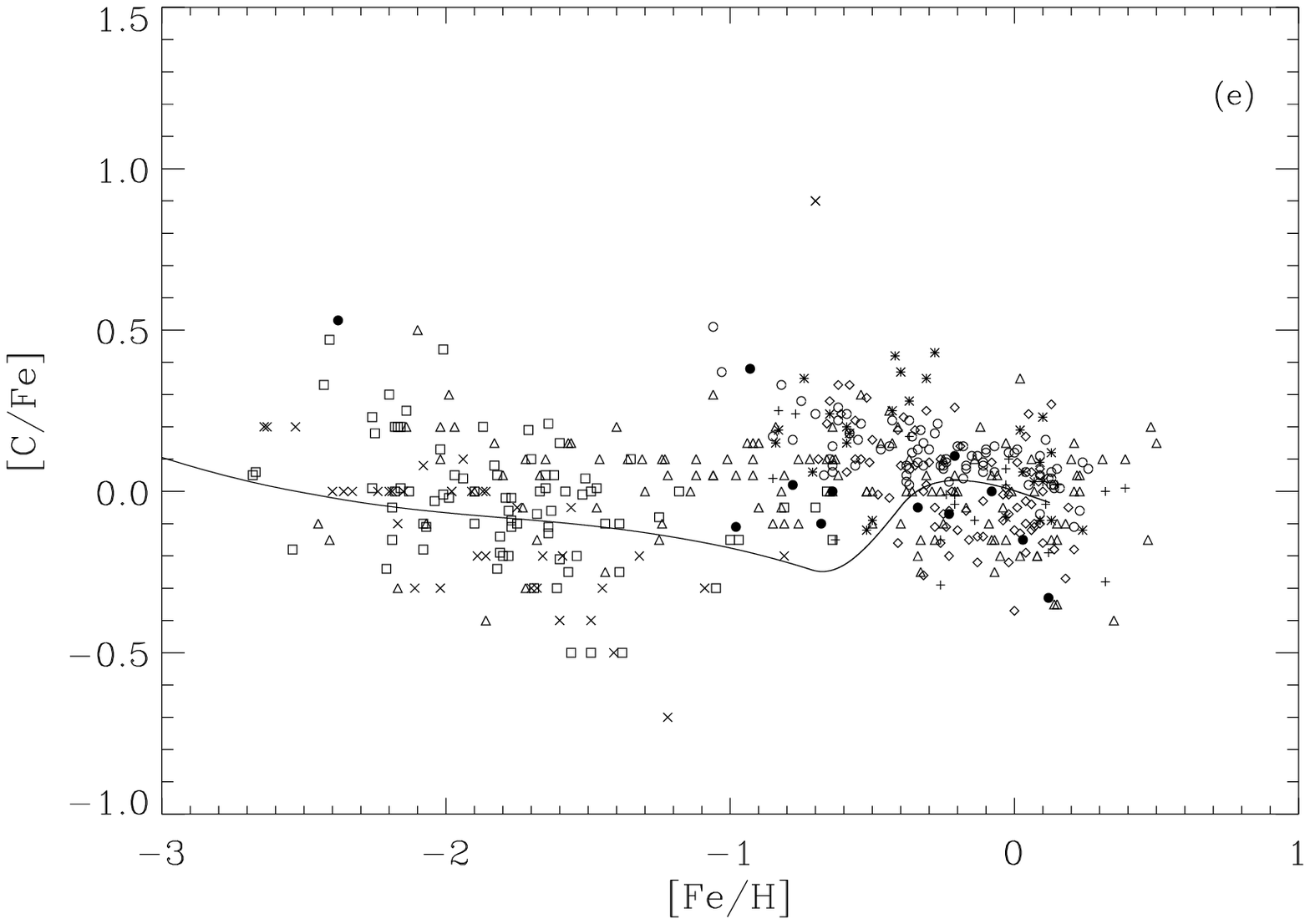}
\epsfxsize=8.8cm \epsfbox{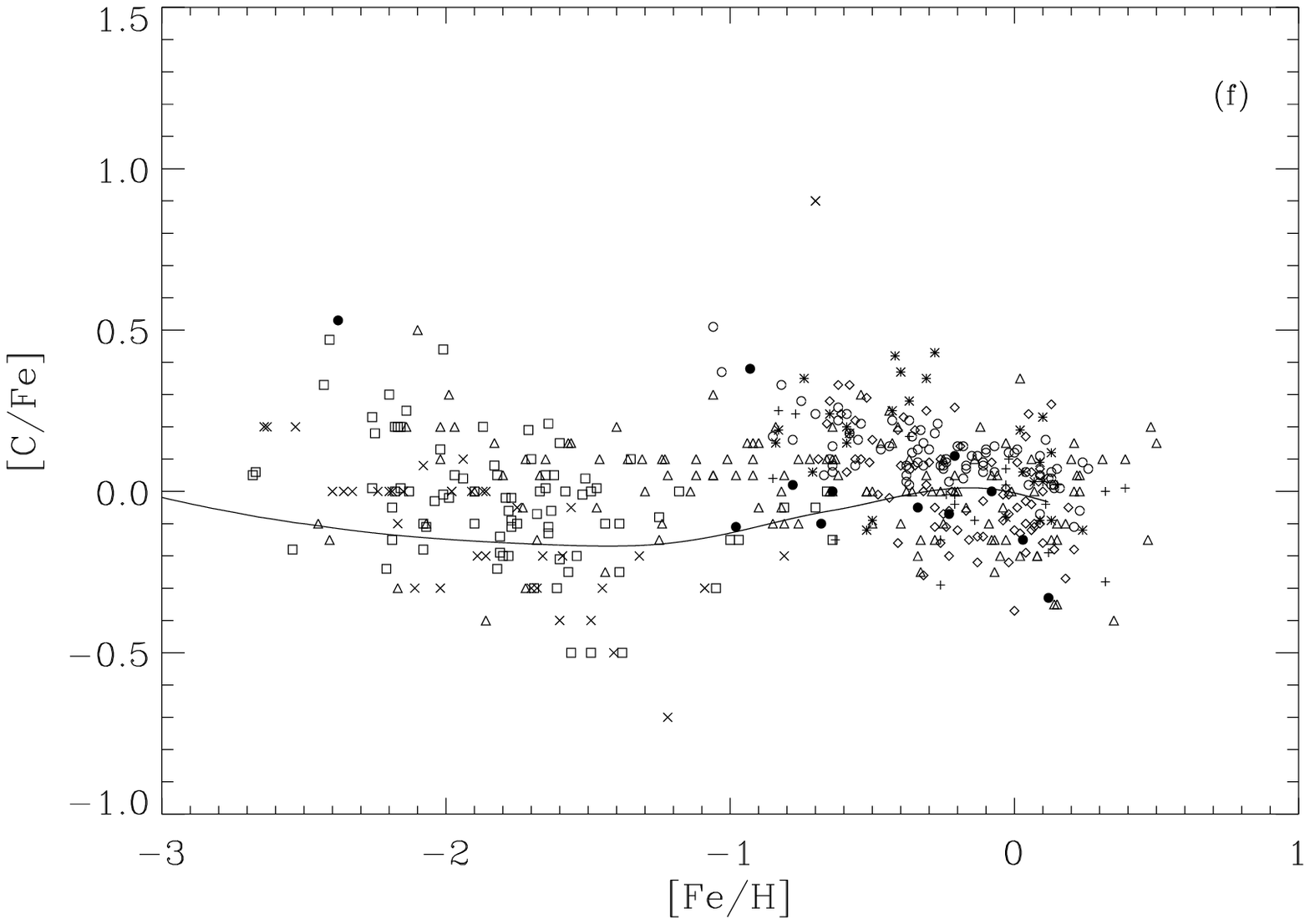}}
 \caption{Same as Fig. 3, but with nucleosynthesis yields of MBC for ILMS 
           and PCB for massive stars (model MBC+PCB).} 
\end{figure*}

\subsection{M2K+PCB}

We use the nucleosynthesis yields of M2K 
in our GCE calculations because this new set of yields 
include the results of 
ILMS with the low metallicity value $Z=0.004$,
which may contain significant information on the ILMS
during the early Galactic stage.

We combine M2K and PCB in our 
model M2K+PCB.
The calculated age-metallicity relation, metallicity distribution of
G dwarfs and [O/Fe] vs. [Fe/H] relation again match the observations, and are similar
to the results of model MBC+PCB, and so are not shown here.
The predicted evolution of [C/Fe] with metallicity is given
in Fig. 8a, which shows 
a strong increase 
with increasing metallicity
from [Fe/H] $\approx -1.5$ on. In order to understand 
this increase, we re-calculate the [C/Fe] vs. [Fe/H] relation using only 
the contribution of massive stars (Fig. 8c),
and only the yields of ILMS and $M\leq 40M_{\odot}$ massive stars (Fig. 8d) 
respectively. 
Comparing Figs. 8c and 7e, and  
Figs. 8d and 7f, we guess that maybe it is the ILMS with $Z=0.004$ that produce a significant
amount of carbon, 
which lead to the strong increase of [C/Fe]
displayed in Fig. 8a. 
Fig. 8e confirms our guess, in which the solid line represents
the calculated [C/Fe] vs. [Fe/H] relation using only the yields of $Z=0.008, 0.02$ 
(the same metallicities as in MBC)
of M2K for ILMS and PBC for massive stars, 
and the dashed line represents
the results of model MBC+PCB (the same line as in Fig. 7b).

The solid and dashed lines in Fig. 8f
represent the corresponding results for nitrogen. 
Figs. 8e, f show that
the nitrogen yields of M2K with $Z=0.008, 0.02$ are very similar to 
the corresponding results of MBC,
though M2K gave more carbon and nitrogen according to the updated parameters,
resulting in the slightly higher [C/Fe] and [N/Fe] values in metal-rich region.
Fig. 8b illustrates the [N/Fe] vs. [Fe/H] relation using the yields of M2K for ILMS; the relation 
is higher than that shown in Fig. 8f.

These results show
that ILMS with low metallicity $Z=0.004$ contribute 
very important amounts of C and N to the ISM (also see Fig. 1 of M2K). 
The lower metallicity ILMS have higher C and N abundances 
because for them the HBB process is more efficient, 
and mass loss is less so. 
The lower mass loss rates correspond to longer TP-AGB lifetimes, 
hence a greater number of the third dredge-up events and a longer duration of the HBB process.
The up-to-date nucleosynthesis calculations (VG) only go as low as 
$Z=0.001$. 
Calculations of ILMS of even lower metallicities are needed; 
these may revise the
trends of elemental abundance evolution, and provide different hints
to the early GCE history.

Though the nucleosynthesis results with $\alpha=1.68$ and 2.00 given by M2K
are different (more nitrogen is produced by intermediate mass stars with $\alpha=2.00$ 
than with $\alpha=1.68$), 
we find 
no obvious differences between these two sets of results.
So we only give the GCE results based on the yields with $\alpha=2.00$ in this paper. 

\begin{figure*}
\input epsf
\vspace{0cm}
\hbox{\hspace{0cm}\epsfxsize=8.8cm \epsfbox{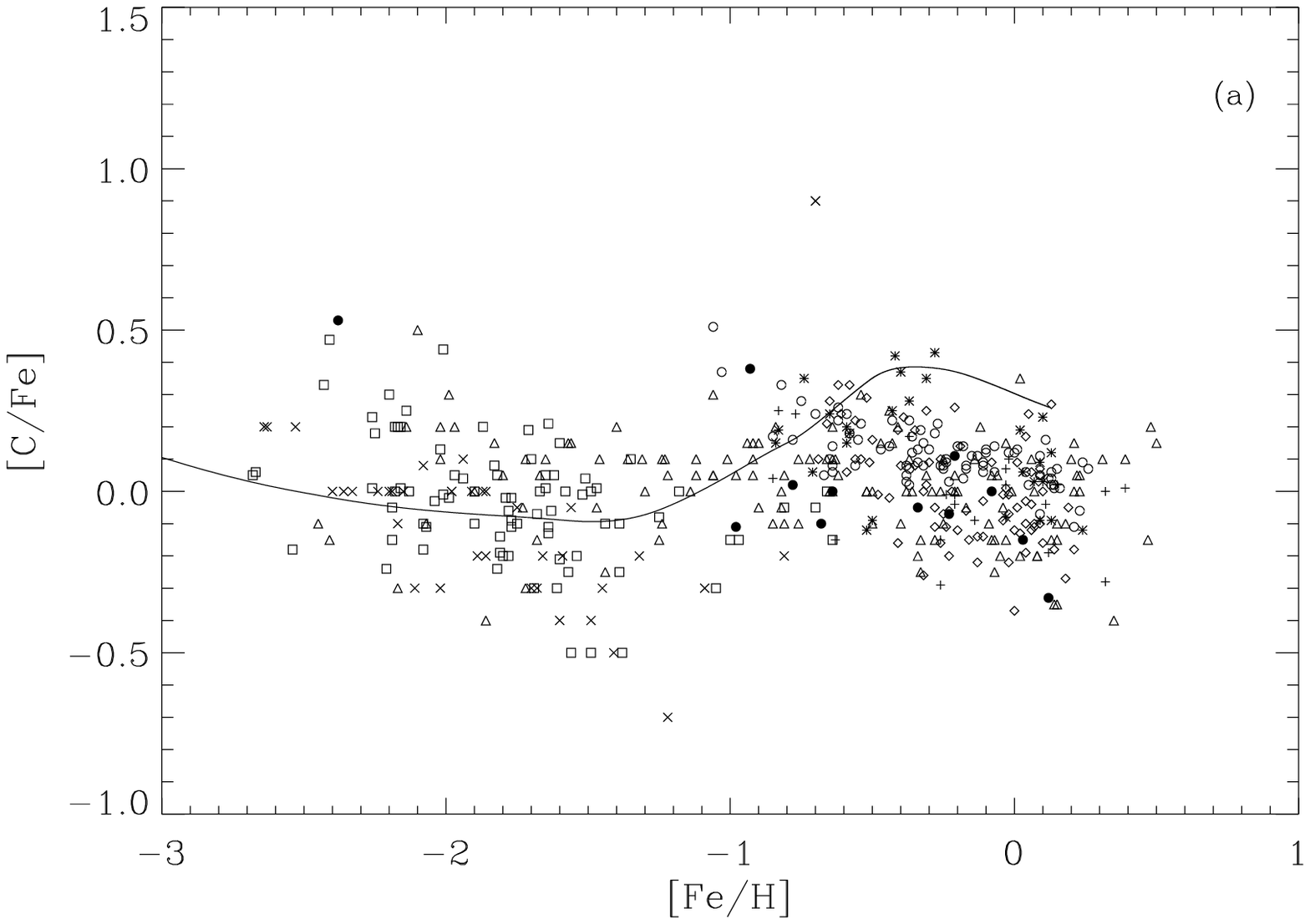}
\epsfxsize=8.8cm \epsfbox{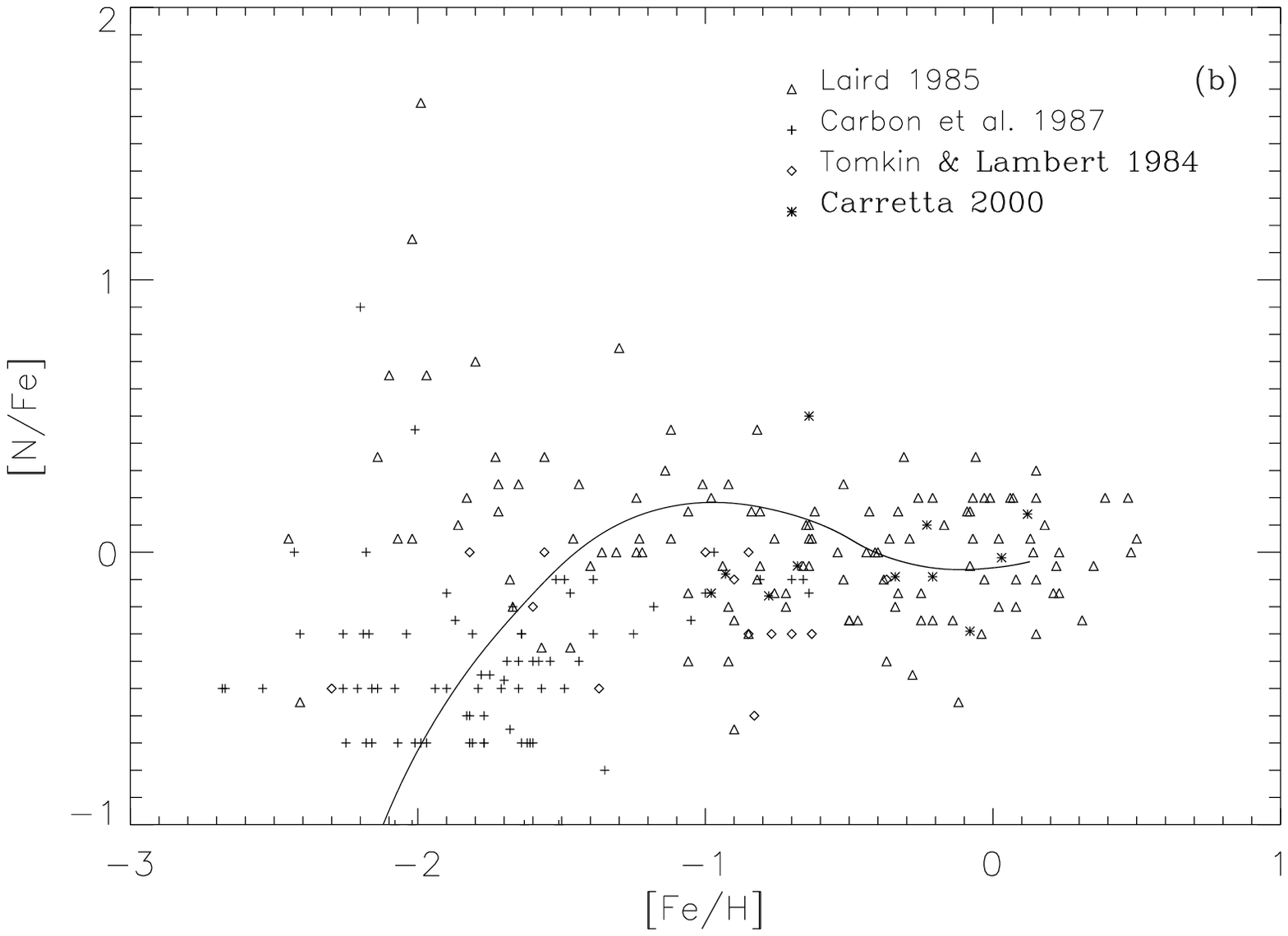}}
\vspace{0cm}
\hbox{\hspace{0cm}\epsfxsize=8.8cm \epsfbox{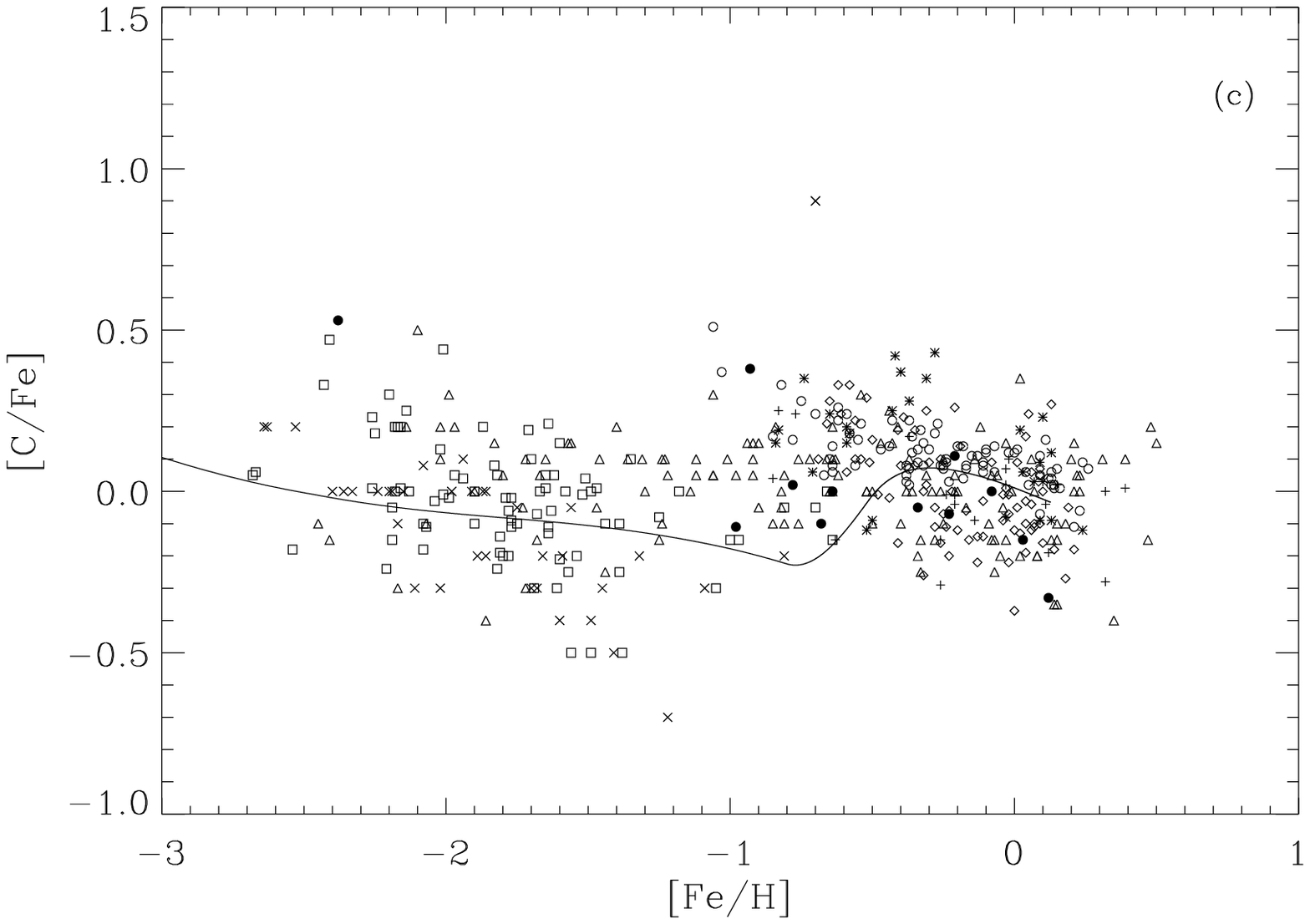}
\epsfxsize=8.8cm \epsfbox{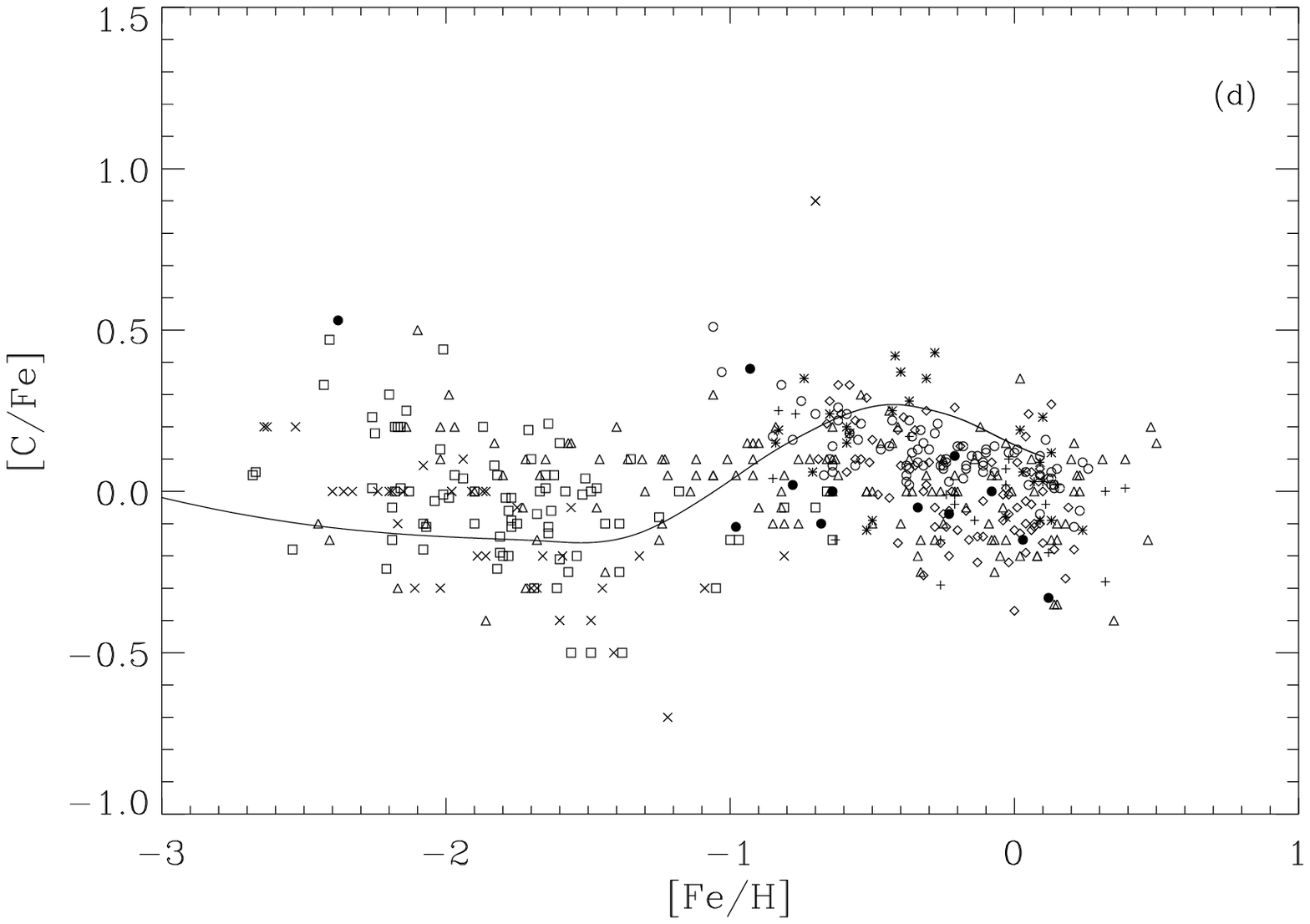}}
\vspace{0cm}
\hbox{\hspace{0cm}\epsfxsize=8.8cm \epsfbox{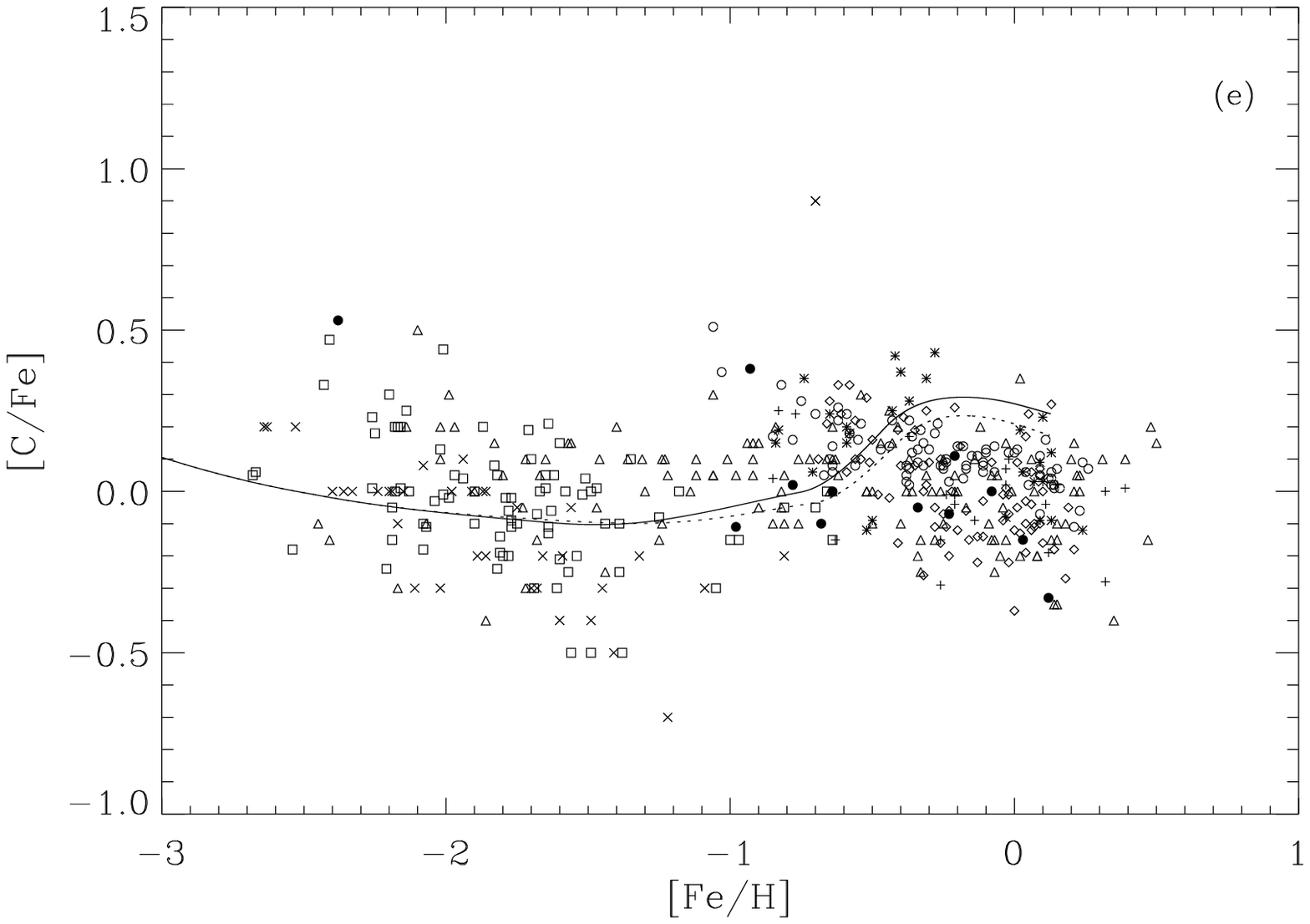}
\epsfxsize=8.8cm \epsfbox{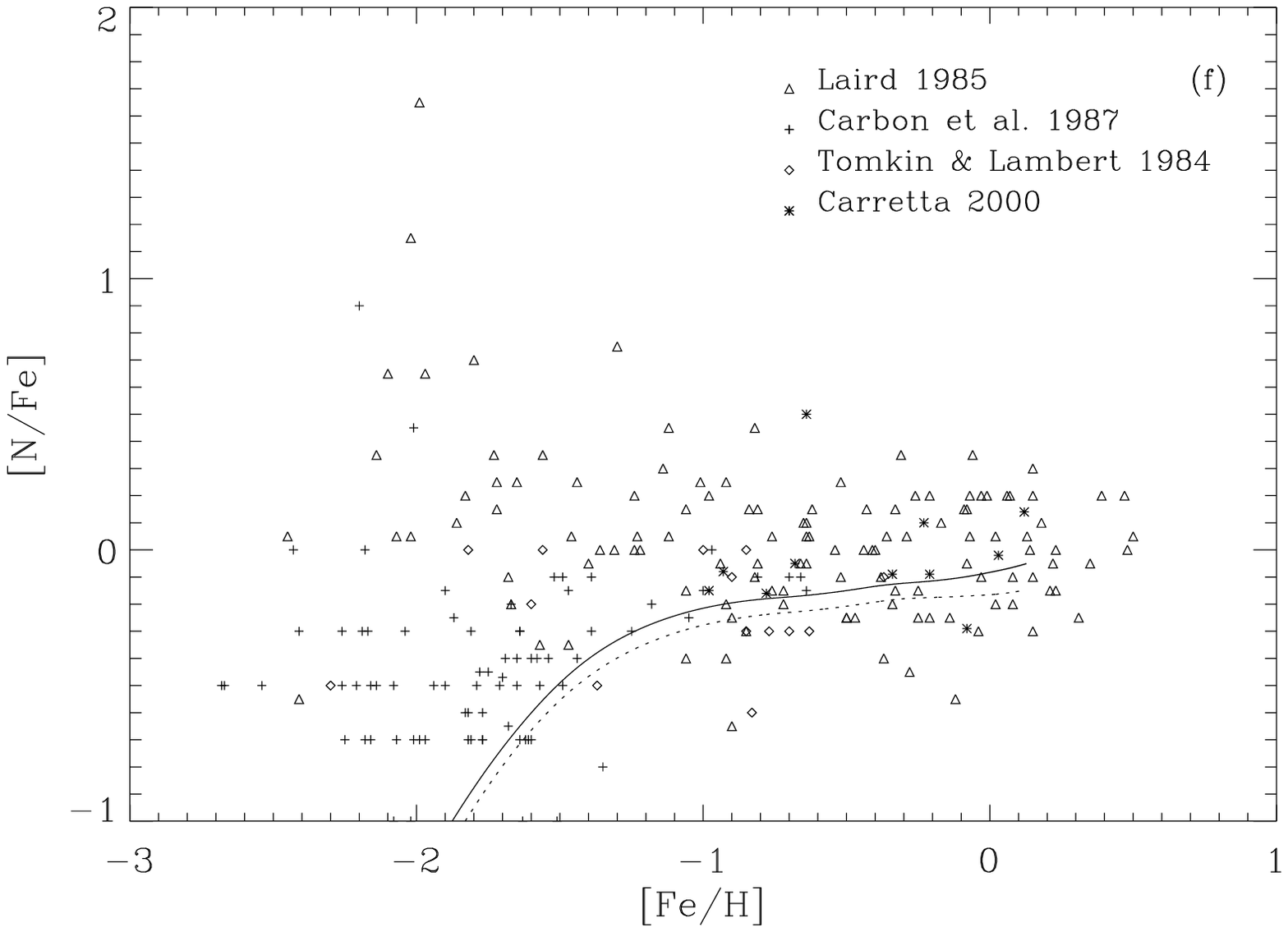}}
\vspace{0cm}
 \caption{(a) The calculated [C/Fe] vs. [Fe/H] using nucleosynthesis yields of M2K
for ILMS and PCB for massive stars (model M2K+PCB); 
(b) Same as (a) but for nitrogen;
(c) The calculated [C/Fe] vs. [Fe/H] considering only yields of massive stars from
PCB without ILMS;
(d) The calculated [C/Fe] vs. [Fe/H] considering yields of M2K for ILMS and only
     $M\leq 40M_{\odot}$ massive stars from PCB;
(e) The solid line represents the [C/Fe] vs. [Fe/H] relation 
considering the yields of PCB for massive stars and M2K
     only with $Z=0.008$ and $Z=0.02$ metallicities for ILMS,
    and the dashed line represents the same result as in Fig. 7b;
(f) Same as (e) but for nitrogen.} 
\end{figure*}

In summary, we give and analyze the calculated abundance evolution of carbon 
using 8 recently published sets of nucleosynthesis yields. 
The parameters of the 8 models are listed in Table 3. 
Here are some of the obvious implications of the parameters.  The higher the $\zeta$ value, 
the more contribution comes from massive stars, and then the predicted O abundance will be 
higher,
especially in low metallicity region. A higher $\beta$ value means
a higher SN\,Ia rate, hence a greater amount of Fe element from SN\,Ia explosion. 
A higher $\nu$ value
represents a higher SFR, more stars form, 
leading to an increase in the O abundance and [Fe/H], 
and the effect is more obvious in metal-poor region. 
At the same time, a higher $\nu$ 
will lead to a higher peak in the metallicity distribution of G dwarfs. 

\begin{table*} 
\centering
 \caption[]{Input parameters of our 8 models}

 {\scriptsize
\begin{tabular}{llll|llll} \hline

model &  $\zeta$  & $\beta$  &   $\nu$   & model &  $\zeta$  &  $\beta$  &  $\nu$ \\ \hline
VG+WW  &  0.51     &   0.037  &    1.0    & RV+WW  &  0.54     &   0.033  &   1.0     \\
VG+N97 &  0.40     &   0.045  &   1.0     & RV+N97 &  0.40     &   0.040  &   1.0      \\ 
VG+M92 & 0.40      &  0.040   &   0.7     & RV+M92 &  0.40     &   0.035  &   0.7       \\  
MBC+PCB&  0.38     &  0.040   &   0.5     &M2K+PCB &  0.38     &   0.040  &  0.5              \\  \hline
\label{tab-p}
\end{tabular}
}
\end{table*} 

\subsection{The $^{12}$C$(\alpha,\gamma)^{16}$O reaction rate}

The $^{12}$C$(\alpha,\gamma)^{16}$O reaction rate is very important to 
stellar nucleosynthesis. It determines the C and O abundances,
hence the abundances of the heavier elements.
Generally, in nucleosynthesis calculations, the 
$^{12}$C$(\alpha,\gamma)^{16}$O reaction rates given by CFHZ and CF are
widely used. CFHZ gives higher values than does CF:
CFHZ$\approx$2.3$\times$CF.
And the higher reaction rate 
leads to a lower C abundance. N97 exhibited this result in their Fig. 7.
They suggested that the actual reaction rate was between  
CFHZ and CF.
WW adopted 1.7$\times$CF ($\approx$0.74$\times$CFHZ) to be the actual reaction rate
in their nucleosynthesis calculations. They gave higher C yields
than did N97. 

For the higher massive stars, stellar wind mass loss strongly affects the C yields,
so the effect caused by the $^{12}$C$(\alpha,\gamma)^{16}$O reaction rate
is obscured. The lower reaction rate of CF has led to  
the lower O yields of PCB.

\subsection {Understanding the W-R stars }

Leitherer (\cite{lei95}) suggested that the most massive stars known in the universe
have masses around 100$M_{\odot}$ and lifetimes of a few Myrs. They are rare: 
in the solar neighborhood only about one such massive star is counted
per $10^5$ to $10^6$ solar-type stars. 
There are three key phases during the evolution of a massive
star: O star, luminous blue variable star and Wolf-Rayet star.
The mass loss of a massive star can be calculated from UV wind lines,
 H$_\alpha$ or radio fluxes. 
The UV P Cygni profile, ubiquitous in O-type stars,
provides a direct indication of stellar wind; 
H$_\alpha$ has been recognized as the prime source of mass-loss
in early-type stars;
and winds in hot stars can be readily
observed at IR-mm-radio wavelengths via the free-free ``thermal" excess
caused by the stellar wind. The mass loss rate scales as 
$Z^{0.5}$ for O stars and early B stars, as $Z^{0.8}$ for mid-B supergiants,
and as $Z^{1.7}$ for A supergiants (Kudritzki \& Puls \cite{KP20}; Crowther
\cite{crowther1}).

W-R stars are easily detectable due to their
intrinsic brightness and the profound effects on the interstellar environment. 
They eject a large amount of material into the ISM via 
the radiative-driven stellar wind. 
The nucleosynthesis calculations show that 
the ejected carbon by massive stars
are very important to the enrichment of the ISM.
In general, the mass loss rate of massive stars adopted
in the stellar evolution calculations are $\dot M \sim M^{2.5}$ and $\dot M\sim Z^{0.5}$
(M92; PCB). 
However, according to recent reviews, our knowledge of the mass loss of massive stars is  not
very clear. 

Meynet \& Maeder (\cite{MeynetM20}) calculated the evolution of stars with rotation, and suggested
that rotation can increase mass loss, also can bring significant surface He$-$
and N$-$ enhancements; the enhancements are the greater for the higher the mass and the rotation.
In his review, Crowther (\cite{crowther1}) found that the 
recent mass-loss rates for Galactic W-R stars indicated a
downward revision of 2$-$4 relative to previous calibrations 
(e.g. Langer \cite{langer89}: $\dot M \sim M^{2.5}$) due to clumping 
(e.g. Schmutz \cite{schm97}). 
Indeed, metallicity affects the yields of carbon.
When mass loss gets very efficient, $^{12}$C will increase because He-burning 
products can be revealed on the surface. If mass loss is extreme, anyway, 
$^{12}$C and $^{16}$O yields may even decrease because most of the mass is rapidly lost
in the wind in the form of $^{4}$He.
But at present, the question of a $Z-$dependent mass loss in W-R stars remains open. 
Crowther (\cite{crowther1}) suggested that
there is a clear distinction among spectral
subtypes in galaxies with different metallicities, which is  qualitatively
explained by evolutionary and spectroscopic models (Crowther \cite{crowther2}).
WN stars in LMC show negligible spectroscopic difference from their  Galactic
counterparts. And the situation is less clear for the SMC, since most W-R stars
are complicated by binarity (Crowther \cite{crowther1}).  

\subsection{Contributions by stars of different masses}

Comparing our calculated results using different nucleosynthesis yields 
(Sect. 5.1$-$5.8),
considering the uncertainty of stellar wind mass loss 
of massive stars (see Sect. 5.10),
and assuming that the actual 
$^{12}$C$(\alpha,\gamma)^{16}$O reaction rate to be between the values of 
CFHZ and CF, 
we choose to use 
the results of model VG+WW 
in an analysis of the contributions of carbon by stars of different masses.

Here, the reason of choosing VG rather than RV for ILMS is
because of suggestions that the yields of RV are not successful 
in predicting the observed C/O and N/O ratios in stars and planetary nebulae, 
or in reproducing the carbon stars of relatively low luminosities 
in the Magellanic Clouds (Gustafsson et al. \cite{Gus99}).

Fig. 9 shows the fraction of carbon contributed by massive stars  
based on the WW yields for 
massive stars and the VG yields for ILMS, as a function of the 
Galactic age.  
In the early Galactic stage, almost all the carbon is produced by massive stars.
As the age advances, ILMS contribute more and more,
and an overtake the massive stars 
from about 1.65 Gyr on. 
At the present epoch, many low mass stars have also evolved and ejected matter into the ISM, and the relative contribution by the ILMS is further increased.

\begin{figure}
\input epsf
\vspace{0cm}
\epsfxsize=8.8cm \epsfbox{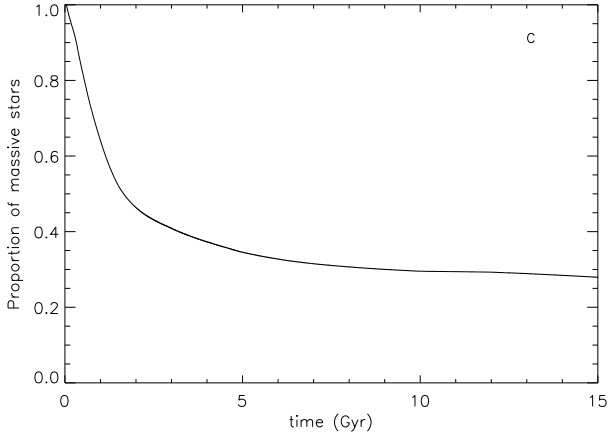}
\vspace{0cm}
 \caption{The relative contribution of massive stars 
 to carbon production at different Galactic ages using the results of WW 
 for nucleosynthesis yields of
 massive stars and VG for ILMS.} 
\end{figure}

\section{Nitrogen Abundance}

The element nitrogen includes a $primary$ and a $secondary$ component.  
If nitrogen production depends on the initial CNO abundance, 
the [N/Fe] would be dependent on [Fe/H], and this part of nitrogen is
$secondary$;
however, if nitrogen production is chiefly
$primary$, then [N/Fe] would be independent of the value of [Fe/H].

Laird (\cite{laird85}) and Carbon et al. (\cite{carbon87}) 
found the observed [N/Fe] to be 
constant over the entire metallicity range, although 
the precise level is uncertain. 
So, it seems that the observations require a 
$primary$ nitrogen source from massive stars. 
However, if the observed constant nitrogen in low
metallicity region is correct, then the present yields by massive stars cannot
supply the needed $primary$ nitrogen source (Timmes et al. \cite{timmes}; 
Chiappini et al. \cite{CMG97}).  
Our results, based on the recently published nucleosynthesis yields
of massive stars, also point to an absence of a $primary$ nitrogen source.
Figs. 10a, b, c show the calculated results of [N/Fe] vs. [Fe/H] 
using our GCE models VG+WW, VG+N97 and VG+M92 respectively. 
Those models with the RV yields of ILMS give similar trends 
to these with the VG yields; 
so we do not show them here. The [N/Fe] values of 
model MBC+PCB and M2K+PCB will be given
in Sect. 5.8.

In these figures, the solid lines represent the results with 
the N yields of
both ILMS and massive stars,
and the dashed lines represent the results with only the contribution from massive stars.
These results show that ILMS are the dominant donor of nitrogen,
while 
massive stars have provided only a small part to the ISM 
during the whole Galactic age up to the present.
However, in the range of $-3<$ [Fe/H] $<-2$, [N/Fe] increases steeply 
(solid lines in Figs. 10a, c), 
due to 
intermediate mass stars as source of secondary nitrogen.
As metallicity increases up to [Fe/H] $\approx -1.0$, 
[N/Fe] increases gradually,---typical behaviour of 
secondary nitrogen.  
As [Fe/H] increases further,  
[N/Fe] approaches the solar value, which shows that the primary 
nitrogen from LIMS plays  
an important role in the production of nitrogen since that time.

From another aspect, comparison between the results shown in Fig. 10a and Fig. 10c
confirms further that the ILMS are the main source of nitrogen.
These two figures are based the models VG+WW and VG+M92 which use  different N yields
from massive stars, yet the results are very similar, so we infer that
it is the N yields of ILMS that provide the bulk of nitrogen to the ISM. 
In addition, 
we should note that M92 gives only the N yields of stellar wind 
and does not include contribution from 
SN\,II explosions, so the dashed line is higher 
in Fig. 10a than in Fig. 10c.
Again, the predicted N abundance evolution by the models MBC+PCB and M2K+PCB
also support the statement that ILMS are the dominant N contributor (see Figs. 8b, f).  
The model VG+N97 predicts high [N/Fe] in early Galactic stage, though 
not as high as the solar abundance. A possible reason is that N97 only gives the 
yields of solar metallicity massive stars, 
which eject more nitrogen than do low metallicity stars.  
With increasing metallicity, 
Fe from SN\,Ia explosions 
plays an increasingly more important role in the enrichment of the ISM, 
and this leads to a [N/Fe] decrease with the Galactic evolution 
(dashed line in Fig. 10b).
 
In general, the relative weight of the secondary and primary components on the 
theoretical yield of
$^{14}$N of ILMS depends on the interplay between the secondary enrichment 
caused by the first and second
dredge-up episodes, and the primary contribution given by the CNO-cycle during the 
envelope burning.
$^{13}$C and $^{14}$N are regarded as secondary elements in the sense 
that they are formed from $^{12}$C and $^{16}$O originally present in the star 
at the time of its formation. Typical in this sense is the secondary production 
of $^{13}$C and $^{14}$N during the first dredge-up, 
and the secondary production of $^{14}$N during the second dredge-up. However, 
when the third dredge-up process operates in conjunction with the HBB process, 
primary $^{13}$C and $^{14}$N
are produced. Meanwhile the primary $^{12}$C, dredged up after each He-shell flash is 
later converted in part into primary $^{13}$C and $^{14}$N by the burning at 
the base of the convective envelope during the interpulse phase. 

Is it sure that the observed [N/Fe] is near 0.0 
in metal-poor region?
Carbon et al. (\cite{carbon87}) found two stars with high N abundances,
HD\,74000 with [N/Fe] $\sim$ +0.9, HD\,25329 with [N/Fe] $\sim$ +0.45.
Laird (\cite{laird85})
obtained high N abundances
for four stars, HD\,74000 with [N/Fe] $\sim$ +1.15 and [Fe/H] $=-2.02$, 
HD\,97916 with [N/Fe] $\sim$ +0.75 and [Fe/H] $=-1.30$, 
HD\,160617 with [N/Fe] $\sim$ +1.65 and [Fe/H] $=-1.99$,
HD\,166913 with [N/Fe] $\sim$ +0.70 and [Fe/H] $=-1.80$. 
The possible reasons for
the very high N abundance given by the authors are: (1) internal mixing;
(2) mass transfer in a binary system;
(3) primordial N enhancement. 
Beveridge \& Sneden (\cite{BS94}) reanalyzed
the chemical abundances of two N-rich dwarf stars, namely HD 74000 and HD 25329.
Their results showed that all the very heavy elements synthesized through neutron-capture s-process
were enhanced in these two stars. 
It is likely that the 
enrichment of N and the very heavy elements originated in material dredged up from the
helium burning shell of former AGB stars.
Also, Carbon et al. (\cite{carbon87}) pointed out that,
omitting the N-rich objects, the observed [N/Fe] vs. [Fe/H] plot showed a negative
correlation with [Fe/H].
In this sense, our results can basically fit the observations. 
More explorations of metal-poor N-rich objects are needed. 

We now 
follow Sect. 5.11 and calculate the contribution by massive stars
to nitrogen using the model VG+WW. 
Fig. 10d shows
the fractional contribution as a function of 
the Galactic age.
We note that, except at the extreme early stage before the intermediate mass stars 
began to evolve,
the relative contribution from massive stars 
is very low in the early stage, 
and that it gradually increases with
the evolution of the Galaxy, then slightly decreases in the late stage. 
This trend means that the yield of nitrogen varies strongly with the metallicity.
However, the proportion of nitrogen contributed by massive stars is generally very low,
even the peak is less than 20\%. 
WW calculated a very low N yield for stars with zero metallicity, 
on the order of $10^{-6}M_{\odot}$.
The N yields of ILMS with low metallicities  
are higher than WW, for example, the newly formed and ejected nitrogen
from 8$M_{\odot}$ stars with $Z=0.001$ is about $10^{-2}M_{\odot}$,
thus, the intermediate mass stars
have been making important contributions since they began to evolve 
and eject materials into the ISM. 
With increase of metallicity, the N yield of massive stars also increases, 
as is shown by rising part of the curve 
of Fig. 10d.
The eventual decrease in the curve 
at the late stage is probably caused by the higher N yield from the  ILMS with higher metallicities.

\begin{figure*}
\input epsf
\vspace{0cm}
\hbox{\hspace{0cm}\epsfxsize=8.8cm \epsfbox{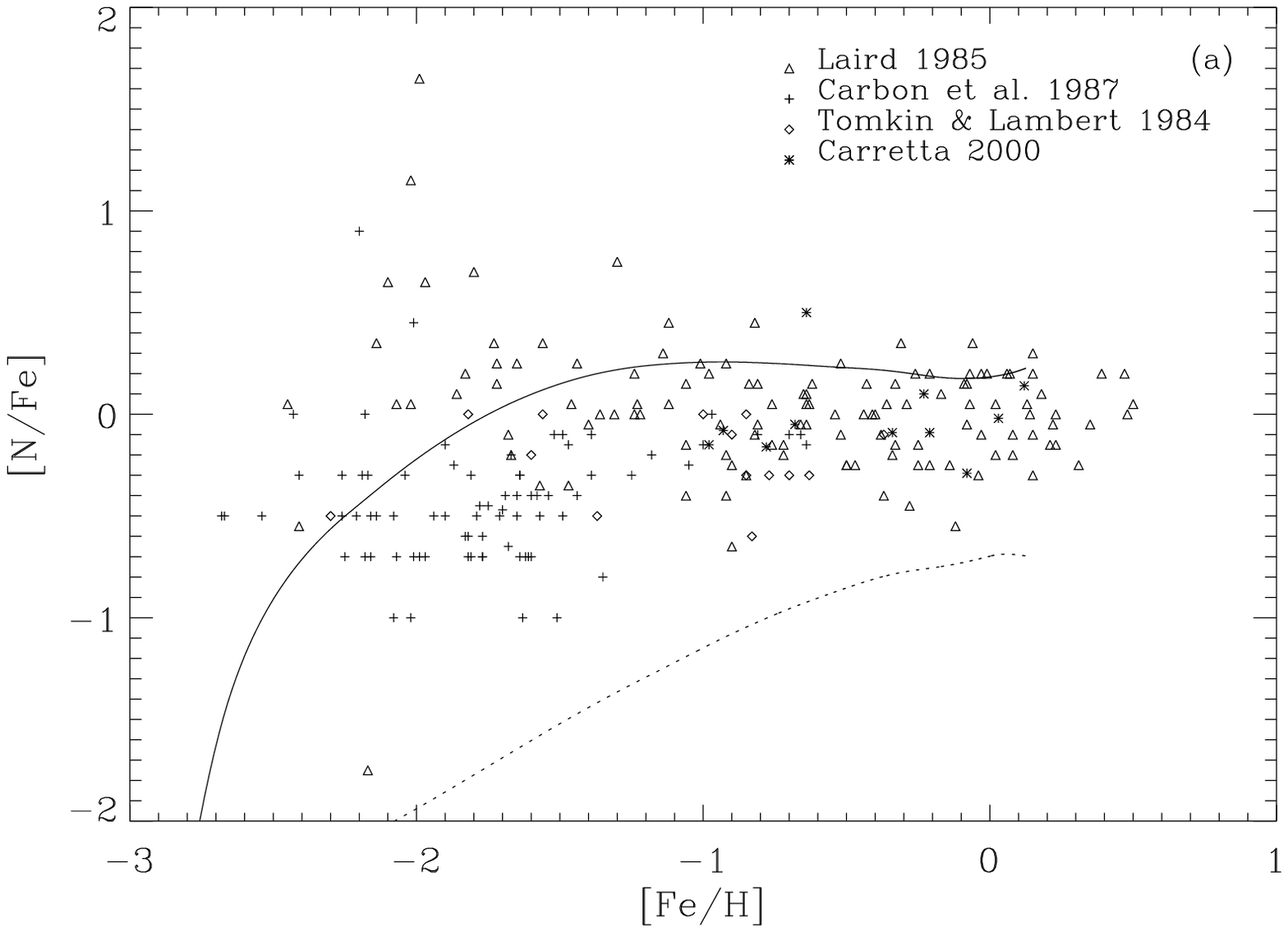}
\epsfxsize=8.8cm \epsfbox{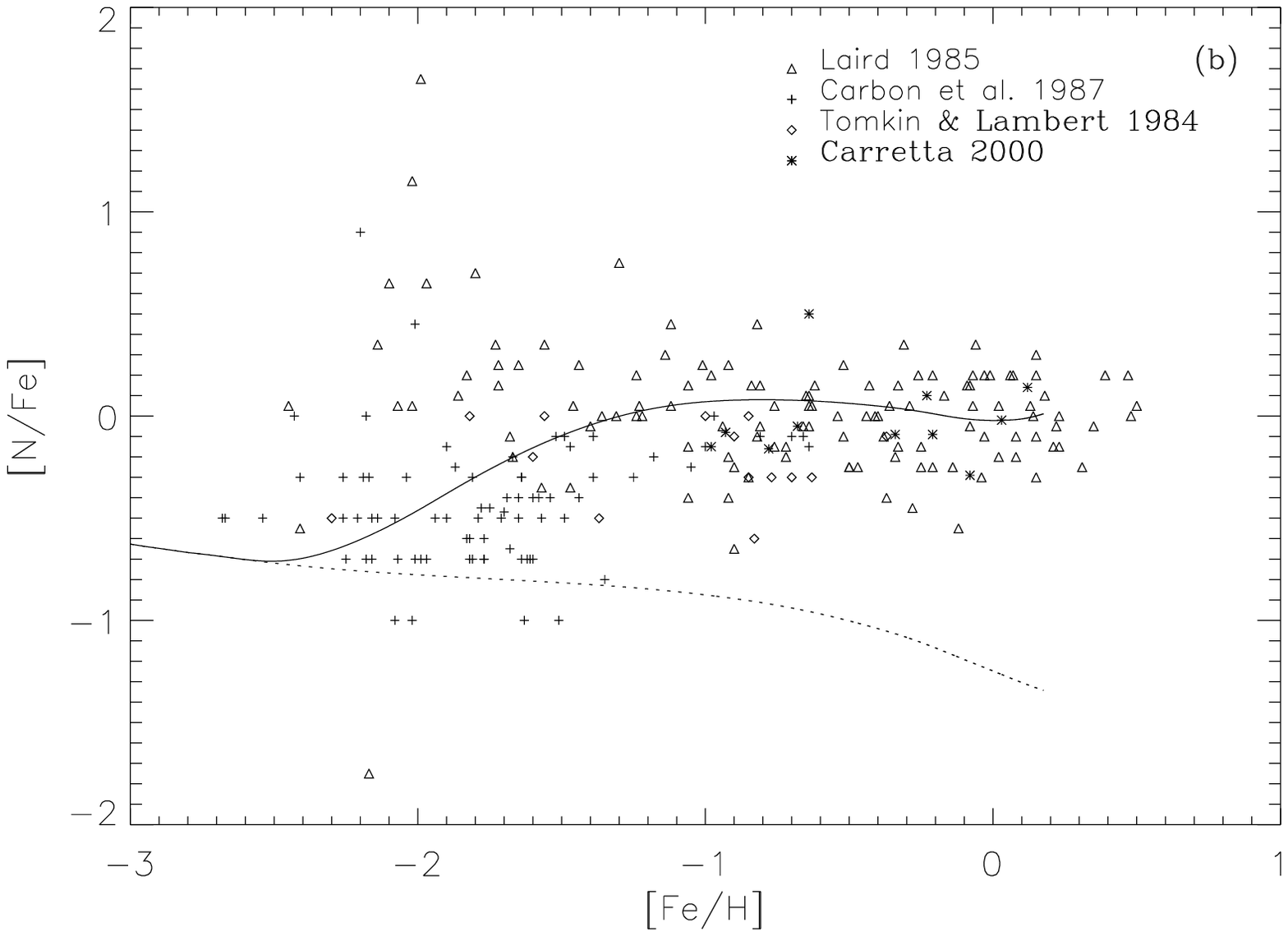}}
\vspace{0cm}
\hbox{\hspace{0cm}\epsfxsize=8.8cm \epsfbox{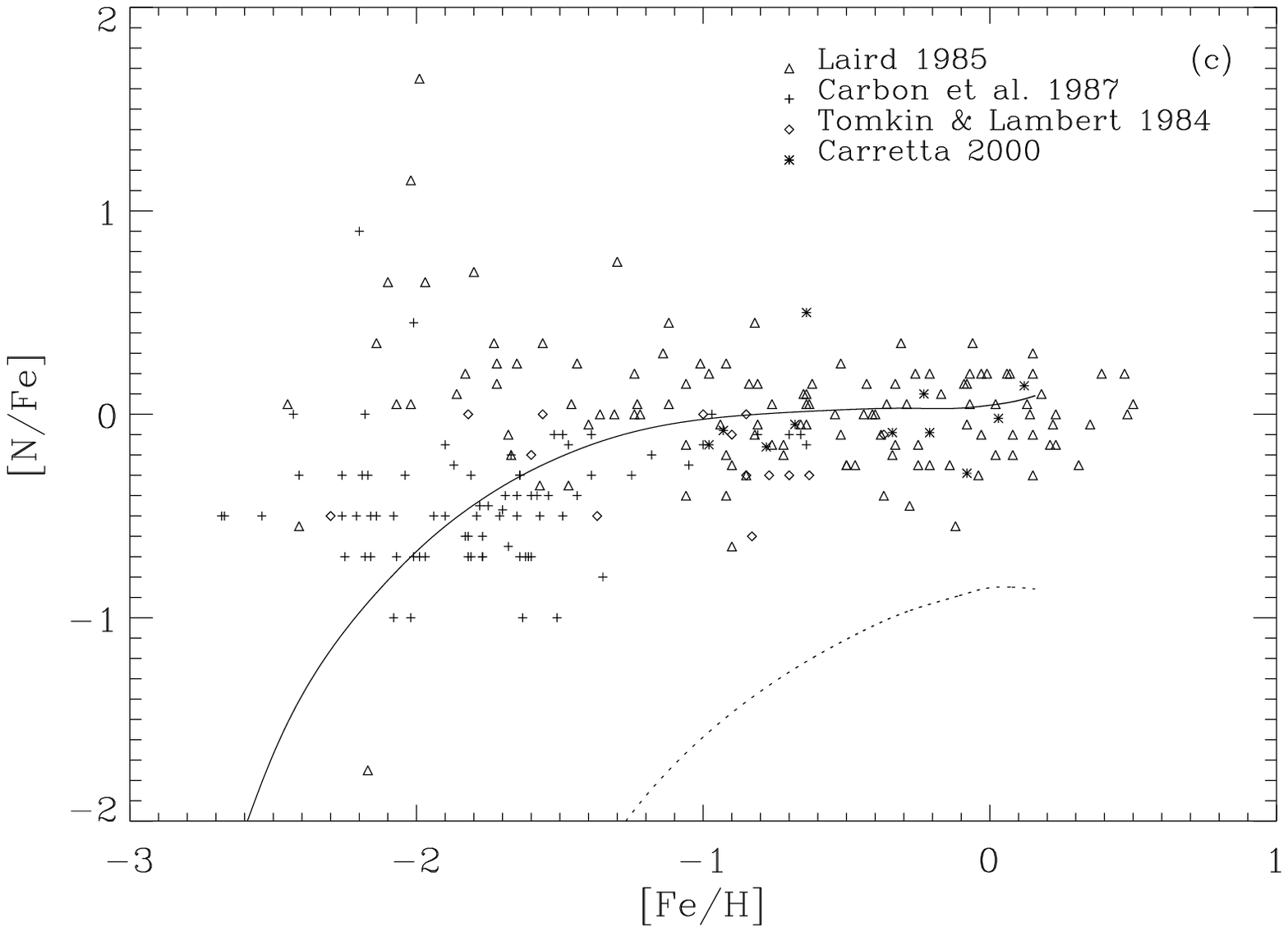}
\epsfxsize=8.8cm \epsfbox{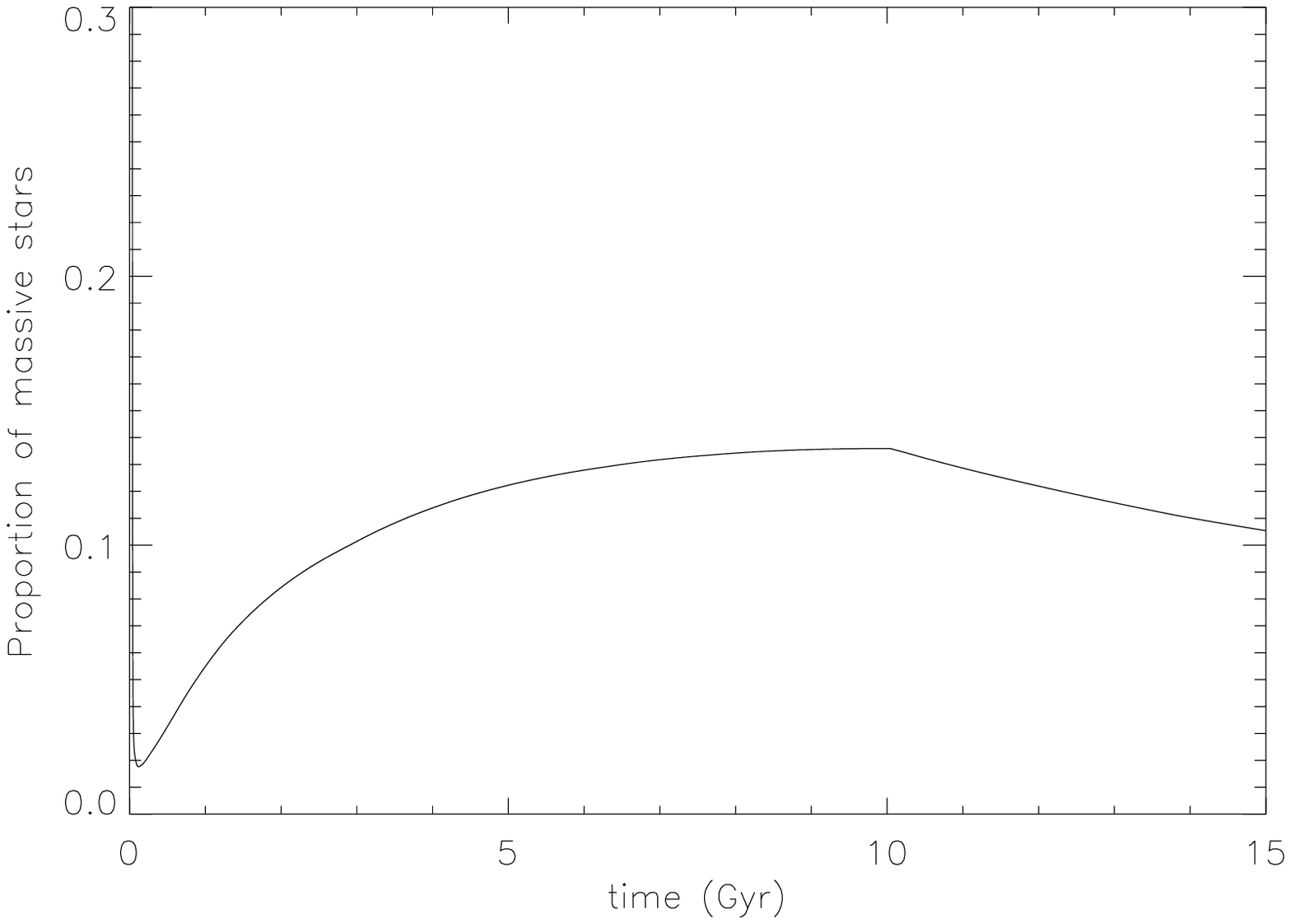}}
\vspace{0cm}
 \caption{(a) The calculated [N/Fe] vs. [Fe/H] relation using nucleosynthesis
    yields of VG for ILMS and WW for massive stars. The solid line represents
    the result with both the contributions of 
    ILMS and massive stars, and the dashed line represents
    the result with only massive stars. 
   (b) Same as (a) but with the yields of N97 for massive stars.
   (c) Same as (a) but with the yields of M92 for massive stars.
   (d) The relative contribution of massive stars to nitrogen production 
with the Galactic 
ages predicted by using the yields of VG for ILMS and WW for massive stars. } 
\end{figure*}

\section{Oxygen Abundance}

The source of oxygen is generally clear, namely, it is 
produced mainly in
short-lived, massive stars and ejected through SN\,II explosions. 
As our Galaxy evolves, more and more Fe is produced through long lifetime SN\,Ia explosions, 
and the ratio [O/Fe] decreases.
This trend is shown in all of our calculated results given in Sect. 5.

Conti et al. (\cite{conti67}) presented the first data indicating 
the existence of an oxygen overabundance in metal-poor stars. 
Many authors continued to explore the evolution of [O/Fe],
and confirmed the overabundance of oxygen in metal-poor stars (see Sect. 2).
Edvardsson
et al. (\cite{E93}) and Chen et al. ({\cite{chen}) separately obtained 
the oxygen abundances of a large
sample of disk stars, both showed that [O/Fe]   
decreases 
with increasing metallicity in Galactic disk stars. 
For halo stars, however, available observations exhibit an apparent contradiction,
namely,
the [O/Fe] values of metal-poor giants derived from the [\ion{O}{i}] forbidden 
line (+0.4) 
are lower
than those of dwarfs from the \ion{O}{i} triplet at $\lambda$\,7770 \AA $~$and OH lines 
(+1.0) (see Fig. 11).
And using the $\lambda$\,6300 \AA $~$forbidden line, similar abundance values to the giants 
have been obtained for some metal-poor dwarfs, 
though the line 
is very weak in these stars (Nissen et al. \cite{NPA20}).

The difference can be understood from two aspects.
In the first aspect, we believe that oxygen has a similar behavior to other 
$\alpha$ elements (Mg, Si, S, Ca), that is, 
we believe that the lower [O/Fe] in metal-poor region
is true, and the relatively higher ratio derived from the \ion{O}{i} triplet or OH line is wrong.
The reasons are as follows.
The \ion{O}{i}
triplet has a very high excitation feature (9.15 eV) and is quite sensitive to the
temperature structure of the adopted model atmosphere, and NLTE effects
cause the higher O abundance derived from it.
And the OH lines in the UV are 
very sensitive to the temperature structure of the stellar atmosphere and/or
to NLTE effects too.  
According to this argument,
there should be no obvious change in the O abundance 
as the dwarfs evolve to the giants. 

On the other hand, if we believe that the higher [O/Fe] derived 
from the \ion{O}{i} triplet and OH line reflects the true O abundance, 
then we have to reconsider the abundance evolution
of oxygen, hence the GCE result. 
Maybe the top-heavy IMF is true, which favours the formation of 
massive stars in the early Galactic stage,
hence a high O abundance.
This assumption would imply a corresponding high [$\alpha$/Fe] in metal-poor dwarfs, which is not found in 
present available data.
Could the difference between [O/Fe] and [$\alpha$/Fe]
mean that their behaviors are {\it not\/} similar?
Perhaps NLTE effects can revise the present [$\alpha$/Fe] value.
Recently, Zhao \& Gehren (\cite{ZG20}) reported that NLTE
effects can increase the [Mg/Fe] abundance, and the corrections are larger 
for the lower metallicities of the metal-poor stars. 
Thus, maybe NLTE corrections can also 
increase the abundances of other $\alpha$ elements at the early Galactic stage. 

\begin{figure}
\input epsf
\vspace{0cm}
\epsfxsize=8.8cm \epsfbox{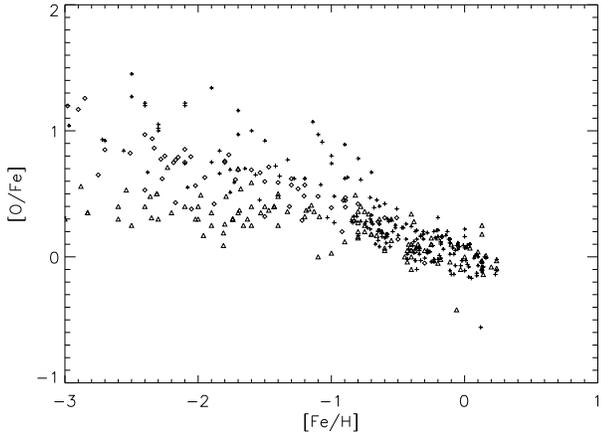}
 \caption{The observed [O/Fe] vs. [Fe/H] data.  
        Oxygen abundances derived from [O I] 6300 \AA $~$forbidden line
        are indicated by triangles, from O I triplet by pluses, 
        and from OH lines in UV region by diamonds.} 
\end{figure}

Goswami \& Prantzos (\cite{GP20}) pointed out three reasons for the higher [O/Fe]
value
in low metallicity region: (1) The Fe produced by SN\,Ia explosions 
entered the galactic
scene as early as [Fe/H] $\sim -3$, instead of [Fe/H] $\sim -1$ in the ``standard"
scenario; (2) the O yields from massive stars are, for some reasons, metallicity dependent;
(3) there is the possibility of  
the yields of Fe and all elements heavier than oxygen being metallicity-dependent.

\section{ Conclusions}

Using the standard GCE model, we have attempted an exploration of  
the questions of the sources of carbon
and the abundance evolution of CNO elements.
Because the yield of stellar nucleosynthesis is a very important parameter
of the GCE model, and different yields give different evolutionary
behaviors of the elements, we chose the recently published different 
nucleosynthesis yields: 
the yields given by RV, VG, MBC and M2K for intermediate-,
low-mass single stars; the results of the classical W7 model of 
Nomoto et al. (\cite{nomoto2}) for binary stars through SN\,Ia explosions; the 
yields given by WW, N97, M92 and PCB for massive single stars.
Then we set up 8 specific models for the combinations, VG+WW, VG+N97, VG+M92, RV+WW, RV+N97, RV+M92, 
MBC+PCB and M2K+PCB.

Our results show that, after fitting the predicted age-metallicity relation, 
metallicity distribution of G dwarfs and [O/Fe] vs. [Fe/H] relation
to the observations,  very different [C/Fe] vs. [Fe/H] relations 
are given by the different models with different yield combinations.

The $^{12}$C$(\alpha,\gamma)^{16}$O reaction rate is very important to 
stellar nucleosynthesis calculations. Hence, it affects the results of the GCE calculations.
A higher rate of this reaction 
will produce less carbon, and a lower rate, more.  
A suitable choice is between the values given by CFHZ and CF.

Generally speaking, in the early stage
of our Galaxy, massive stars are the main contributor of carbon.
As our Galaxy evolves to the 
late stage, the long lifetime ILMS begins to play an important role
in the enrichment of the ISM.  
At the same time, 
the high metallicity W-R stars eject significant amounts of
carbon into the ISM 
by radiative-driven stellar wind; the metallicity influences the outer
opacities and therefore the mass loss rate by stellar winds in massive stars.
However, on the present published nucleosynthesis yields, 
our results still cannot distinguish the source of carbon in metal-rich Galactic stage. It may be that 
just the massive stars ($M>8M_{\odot}$) 
are sufficient to account for the observed [C/Fe] values (see Figs. 3e, 6e); or it may be just the ILMS and 
$M\leq 40M_{\odot}$ massive stars without the W-R stars 
are enough to explain the observed carbon in metal-rich stars
(see Figs. 1b, 7f).

However, we can at least distinguish the different contributions of 
ILMS and massive stars to carbon from our calculations.
Having comparing the GCE results shown in Sect. 5.1$-$5.8,
and considered the undetermined characters of stellar wind  mass loss 
of massive stars,
and assuming the actual 
$^{12}$C$(\alpha,\gamma)^{16}$O reaction rate to be between CFHZ and CF, we calculated 
the respective contributions to carbon from massive stars and ILMS in the model WW+VG.
Fig. 9 shows that, 
in the early stage, almost all carbon is produced by massive stars;
with increase of the Galactic age, the contribution from ILMS becomes  more and more and eventually exceeds that of the massive stars. 
The latter contribution is increased when the contribution from W-R stars is included.

Henry et al. (\cite{henry20}) gave an excellent analysis on the source of C and N. 
But they only gave 
detailed results for the yields of VG for ILMS and M92 
for massive stars. And they increased the yields of M92 in their calculations
(see their Table 3), which increased the contribution of carbon
from massive stars. 

Carigi (\cite{cari20}) compared different set of yield, and suggested that massive stars
with stellar wind are the main source of carbon.
Compared to their work, we use the abundances of a large sample of dwarf stars as 
one of the most important observational constraints  
rather than just the stellar age. The reason is that 
few halo stars have determined
ages, while the results reported by Edvardsson et al. (\cite{E93}), Gustafsson
et al. (\cite{Gus99}) and Chen et al. (\cite{chen}) 
are only for disk stars,
so the element abundance in the early stage of the Galaxy cannot be tested by observations.
In addition, we divided the massive stars into two ranges, $M\leq 40M_{\odot}$ and 
$M>40M_{\odot}$, so that we can compare the carbon contributions 
given by W-R stars and ILMS more clearly.
And we use the new set of nucleosynthesis yields of 
M2K, which contain important information 
on low metallicity ILMS.

Our results confirm that most of nitrogen is produced by ILMS, especially
by intermediate mass stars. And the present yields given by massive stars
cannot supply the needed primary N source. 
Massive stars produce the main part of oxygen in the universe.
But the observed O abundances derived from different atomic 
lines lead to 
obvious differences in the [O/Fe] determination in metal-poor stars. 
Maybe we should check the procedure and many corresponding 
parameters to obtain abundance of oxygen. 
Also, perhaps this difference
hints at some new information on Galactic chemical evolution, especially, 
it is necessary to understand the evolution of our Galaxy in the early stage 
carefully.

  \begin{acknowledgements}
  We thank Dr. B. Edvardsson for very valuable suggestions 
on the original manuscript. 
We thank Dr. T. Kiang and Dr. P. Podsiadlowski for spendind much time and energy
to help us to improve the English expression.  
Thank Dr. L. C. Deng
 for the interesting discussions about massive stars, 
and thank Zhang Yanxia and Fan Yingjie for their friendly help.
  This research work is supported by the National Natural Science Foundation of
  China under grant No. 19725312 
and NKBRSF G1999075406. 
  \end{acknowledgements}

\end{document}